\documentclass[twocolumn]{aastex62}

\usepackage{graphicx}
\usepackage{amssymb}
\usepackage{xspace}

\newcommand{\Msun}{M$_\odot$\xspace}
\newcommand{\sfrunit}{M$_\odot$ yr$^{-1}$\xspace}
\newcommand{\unitlpco}{K\,km\,s$^{-1}$\,pc$^2$\xspace}
\newcommand{\unitxco}{$M_{\sun}$ (\unitlpco)$^{-1}$\xspace}
\newcommand{\lpco}{${L^{\prime}}_{\rm\!\!CO}$\xspace}
\newcommand{\lhcn}{${L^{\prime}}_{\rm\!\!HCN}$\xspace}
\newcommand{\lhco}{${L^{\prime}}_{\rm\!\!HCO+}$\xspace}
\newcommand{\kms}{km s$^{-1}$\xspace}
\newcommand{\meanu}{$\langle U \rangle$\xspace}

\begin{document}
\graphicspath{{./}{figures/}}

\title{The State of the Molecular Gas in Post-Starburst Galaxies}

\author[0000-0002-4235-7337]{K. Decker French}
\affiliation{Department of Astronomy, University of Illinois, 1002 W. Green St., Urbana, IL 61801, USA}  

\author[0000-0003-2599-7524]{Adam Smercina}
\affiliation{Astronomy Department, University of Washington, Seattle, WA 98195, USA}

\author[0000-0001-7883-8434]{Kate Rowlands}
\affiliation{AURA for ESA, Space Telescope Science Institute, 3700 San Martin Drive, Baltimore, MD, USA}
\affiliation{Department of Physics and Astronomy, Johns Hopkins University, Baltimore, MD 21218, USA}

\author[0000-0002-6582-4946]{Akshat Tripathi}
\affiliation{Department of Astronomy, University of Illinois, 1002 W. Green St., Urbana, IL 61801, USA}  

\author[0000-0001-6047-8469]{Ann I. Zabludoff}
\affil{Steward Observatory, University of Arizona, 933 N Cherry Ave, Tucson, AZ 85721, USA}

\author[0000-0003-1545-5078]{John-David T. Smith}
\affiliation{Ritter Astrophysical Research Center, University of Toledo, Toledo, OH 43606, USA}

\author[0000-0002-7064-4309]{Desika Narayanan}
\affiliation{Department of Astronomy, University of Florida, 211 Bryant Space Science Center, Gainesville, FL 32611, USA}

\author[0000-0003-3078-2763]{Yujin Yang}
\affiliation{Korea Astronomy and Space Science Institute, 776 Daedeokdae-ro, Yuseong-gu, Daejeon 34055, Republic of Korea}

\author{Yancy Shirley}
\affil{Steward Observatory, University of Arizona, 933 N Cherry Ave, Tucson, AZ 85721, USA}

\author[0000-0002-4261-2326]{Katey Alatalo}
\affiliation{Space Telescope Science Institute, 3700 San Martin Dr., Baltimore, MD 21218, USA}
\affiliation{Department of Physics and Astronomy, Johns Hopkins University, Baltimore, MD 21218, USA}

\vspace{1 cm}
\begin{abstract}

The molecular gas in galaxies traces both the fuel for star formation and the processes that can enhance or suppress star formation. Observations of the molecular gas state can thus point to when and why galaxies stop forming stars. In this study, we present ALMA observations of the molecular gas in galaxies evolving through the post-starburst phase. These galaxies have low current star formation rates, regardless of the SFR tracer used, with recent starbursts ending within the last 600 Myr. We present CO (3--2) observations for three post-starburst galaxies, and dense gas HCN/HCO$^+$/HNC (1--0) observations for six (four new) post-starburst galaxies. The post-starbursts have low excitation traced by the CO spectral line energy distribution (SLED) up to CO (3--2), more similar to early-type than starburst galaxies. The low excitation indicates that lower density rather than high temperatures may suppress star formation during the post-starburst phase. One galaxy displays a blueshifted outflow traced by CO (3--2). MaNGA observations show that the ionized gas velocity is disturbed relative to the stellar velocity field, with a blueshifted component aligned with the molecular gas outflow, suggestive of a multiphase outflow. Low ratios of HCO$^+$/CO, indicating low fractions of dense molecular gas relative to the total molecular gas, are seen throughout post-starburst phase, except for the youngest post-starburst galaxy considered here. These observations indicate that the impact of any feedback or quenching processes may be limited to low excitation and weak outflows in the cold molecular gas during the post-starburst phase.

\end{abstract}

\section{Introduction}

Multiwavelength observations of galaxies across cosmic time are revealing a detailed picture of how galaxies grow, evolve, and ultimately become quiescent. While some galaxies in the local Universe have gradually ended star formation over many Gyr, others show signs of a sudden end to star formation, having undergone rapid evolution from starbursting to quiescent \citep[e.g.,][]{Schawinski2014}. Such post-starburst galaxies display substantial populations of young A stars, yet little emission line flux from HII regions around O or B stars, indicating a recent starburst that has since ended \citep{Dressler1983,Couch1987}. Post-starburst galaxies provide evidence for fast evolution, likely driven by recent mergers \citep{Zabludoff1996, Pawlik2015, Sazonova2021}. At higher redshifts, large fractions of quiescent galaxies show signs of being post-starburst, suggesting that this process of rapid evolution is more common \citep{Wild2009,Snyder2011,Whitaker2012,Wild2016,Rowlands2017,Belli2019, Wild2020,DEugenio2020}. In order to match the high mass, quiescent end of the galaxy population, simulations must add in feedback from active galactic nuclei (AGN) to limit and ultimately end star formation \citep[e.g.,][]{DiMatteo2005,Croton2006}. Yet the process of how and if AGN feedback operates in its different regimes across galaxy type, and the contribution of stellar feedback, is not well understood.

Galaxies evolving through the post-starburst phase are laboratories for understanding how and when star formation ends, the role of feedback processes, and the connection between the evolution of galaxies and their supermassive black holes (see \citealt{French2021} for a recent review). In order to understand the changes in star formation, we must look to the molecular gas properties of evolving galaxies to study the potential fuel for that star formation. Previous work has uncovered high molecular gas fractions traced by CO (1--0) single dish observations in multiple samples of post-starburst galaxies at $z\sim0.1$ \citep{French2015, Rowlands2015, Alatalo2016b}, contrary to expectations that these galaxies would already be devoid of gas. 

The observation of large CO-traced molecular gas reservoirs remaining in post-starburst galaxies raises the question of why they have become quiescent and what prevents the CO-traced molecular gas from forming stars. If large gas reservoirs exist, but are in a relatively diffuse state, this would explain why star formation is no longer occurring at high rates. In previous work, we presented the non-detection of dense gas tracers in two post-starburst galaxies, with upper limits consistent with the low SFRs of these galaxies, indicating that these galaxies have stopped forming stars due to a lack of dense gas \citep{French2018}. This absence of detected dense gas raises the new question of what physical properties in the diffuse molecular gas are preventing its collapse into forming stars: Is the gas heated? Is it kinematically disturbed? Is there evidence of energy injection from AGN? In order to address these questions, we require detections (rather than limits) of the dense gas tracers in post-starburst galaxies.

In this work, we aim to explore the molecular gas state of post-starburst galaxies by studying their CO excitation and by using multiple tracers of the dense molecular gas from observations with the Atacama Large Millimeter/submillimeter Array (ALMA). We present observations of CO (3--2) for three post-starburst galaxies with previous CO (1--0) and (2--1) observations, as well as observations of dense gas tracers HCN (1--0), HCO$^+$ (1--0), and HNC (1--0) for four new post-starburst galaxies, which we combine with the previous sample of two measurements from \citet{French2018b}. When needed, we assume a flat cosmology with $h=0.7$ and $\Omega_m=0.3$.

\section{Observations}

\subsection{Sample Selection}
\label{sec:sample}

Our targets are selected from two parent samples of post-starburst galaxies with previous molecular gas detections, both of which were originally selected using SDSS spectroscopy. Galaxies from \citet{French2015, Smercina2018} are selected to have low H$\alpha$ emission (H$\alpha$ equivalent width $<3$ \AA\ in emission) and strong Balmer absorption (Lick $H\delta_{\rm A}$ at least $1\sigma$ above 4 \AA; H$\delta_{\rm A}$ $-$ $\sigma$(H$\delta_{\rm A}$) $>$ 4). Galaxies from \citet{Rowlands2015} were selected using a principal component analysis (PCA) from \citep{Wild2007, Wild2009}. These PCA components correspond roughly to $D_n4000$ and H$\delta$. The PCA-selected post-starburst galaxies identify galaxies earlier in the post-starburst phase and with higher SFRs than post-starburst galaxies selected using a cut against H$\alpha$ emission \citep{Wild2009,French2018}.

We summarize our targets, their coordinates and redshifts, post-burst ages from \citet{French2018}, and ALMA dataset numbers in Table \ref{tab:targets}. We select our samples from previous studies of CO-traced molecular gas in post-starburst galaxies. We select two samples of galaxies for observing CO (3--2) (rest-frame $\nu = 345.7960$ GHz) in ALMA band 7 and for observing HCN (1--0) (Hydrogen Cyanide; rest-frame $\nu = 88.6316$ GHz),  HCO$^+$ (1--0) (Formylium; rest-frame $\nu = 89.1885$ GHz), and HNC (1--0) (Hydrogen Isocyanide; rest-frame $\nu = 90.6636$ GHz) in ALMA band 3. The CO (3--2) sample is a subset of the HCN/HCO$^+$/HNC sample. 

For the CO (3--2) observations, we select galaxies from the \citet{French2015, Smercina2018} sample with CO (1--0) and CO (2--1) detections, declinations observable with ALMA, and redshifts such that the CO (3--2) line is not in regions of high atmospheric absorption. Of the sample of four proposed galaxies, three were observed. Following the naming convention from \citet{French2015}, these are H02, H03, and S02. 

For the HCN (1--0), HCO$^+$ (1--0), and HNC (1--0) observations, we select galaxies from \citet{French2015, Smercina2018} as well as from the sample of younger post-starburst galaxies from \citet{Rowlands2015}. We require the galaxies to have declinations observable with ALMA and redshifts low enough to observe HCN (1--0) before it redshifts out of ALMA Band 3 ($z\lesssim 0.06$). We select galaxies to span the full range of post-starburst age. The inclusion of two galaxies from the \citet{Rowlands2015} sample allows for a wider post-starburst age baseline for this sample. The galaxies R02 and R05 are the second and fifth galaxies in the \citet{Rowlands2015} sample, respectively. From the \citet{French2015} sample, the galaxies H02 and S05 were previously observed with ALMA in various dense gas tracers, as reported in \citet{French2018b}. We add an additional two galaxies from this sample: H03 and S02, for a total sample of six post-starburst galaxies with dense gas line measurements HCN (1--0), HCO$^+$ (1--0), and HNC (1--0). 

\subsection{ALMA Observations}
\label{sec:almaobs}

Observations of CO (3--2) for three post-starburst galaxies were taken during Cycle 5 (program 2017.1.00930, PI French). We use the Band 7 receiver and place the redshifted CO (3--2) line in a 1875 MHz wide spectral window with 3840 channels of width 1129 kHz. This corresponds to roughly 1700 km/s with channels of width 1 km/s for our objects. The requested spatial resolution was chosen to be $\sim 0.3$ arcsec, in order to obtain $\sim9$ resolution elements per galaxy, assuming the galaxies had similar size to the CO (2--1) measurements of a partially-overlapping sample from \citet{Smercina2022}. The data were pipeline calibrated using the CASA pipeline indicated in Table \ref{tab:redux}. Briggs weighting was used with the robust values chosen to best match the requested beam size and sensitivity.

Observations of several dense gas tracers were taken of four post-starburst galaxies during Cycles 5 and 6 (programs 2017.1.00935 and 2018.1.00948, PI French). We use the Band 3 receiver and three spectral windows to observe HCN (1--0), HCO$^+$ (1--0), and HNC (1--0). We use wide spectral windows ($\sim 1600-6600$ km/s), depending on the closeness of the lines to the edges of the Band 3 frequency range, and channel widths $\sim 2$ km/s. The requested spatial resolution was chosen to be 1 arcsecond, in order to match the total spatial extent of the dense gas emission estimated from CO (2--1) observations \citep{Smercina2022} and dense gas observations in LIRGs. This resolution scale is larger than the typical size of dense gas emitting clumps, which we do not expect to measure individually. The data were pipeline calibrated using the CASA pipeline indicated in Table \ref{tab:redux}. Imaging parameters are chosen to best match the requested sensitivity and beam size. For the cases in which no lines are detected (HNC in R05 and all lines in S02), we manually re-image the data using updated continuum regions and natural weighting (robust = 2), but are unable to detect any additional emission. We also re-imaged these datasets using a UV taper of 3 arcseconds and are still unable to detect any additional emission.  Two of the datasets (H03 and R02) were manually re-imaged by the ALMA pipeline scientists to refine the continuum subtraction. 

Moment maps and extracted spectra for the both datasets are shown in Appendix \ref{sec:appendix}. We extract spectra from a 1.5 arcsec radius centered on the moment 0 map centroid. We integrate the flux density between $\pm3\sigma_{gauss}$ from the CO (1--0) line measurements in \citet{French2015} and \citet{Rowlands2015}. The continuum regions outside of the line regions are used to determine the uncertainty on the flux measurements. The integrated line fluxes are shown in Table \ref{tab:fluxes}. In cases where the signal to noise ratio is less than 3, we provide a 3$\sigma$ upper limit on the possible line flux.

\begin{table*}[]
    \centering
    \begin{tabular}{lllllll}
    \hline
    Galaxy & R.A. & Decl. & $z$ & Post-burst Age & HCN/HCO$^+$/HNC Dataset & CO (3--2) Dataset  \\
     & (deg) & (deg) & & (Myr) & & \\
    \hline
    H02 & 141.580  & 18.6781 & 0.0541 & 201 & (2016.1.00881; [1]) & 2017.1.00930\\
    H03 & 222.067  & 17.5517 & 0.0449 & 381 & 2017.1.00935 & 2017.1.00930\\
    S02 & 49.2288  & -0.0420 & 0.0231 & 522 & 2017.1.00935 & 2017.1.00930\\
    S05 & 146.112  & 4.49912 & 0.0467 & 259 & (2016.1.00881; [1]) & \\
    R02 & 228.951 & 20.0224 & 0.0363 & -4 &  2018.1.00948 & \\
    R05 & 244.398 & 14.0523 & 0.0338 & 10 & 2018.1.00948 & \\
    \hline
    \end{tabular}
    \caption{Post-starburst galaxy targets. Names match those in \citet{French2015, Smercina2018, French2018b, Smercina2022}. Galaxy coordinates and redshifts are from the SDSS main spectroscopic survey \citep{Strauss2002}. Post-burst ages measure the time since the starburst ended, taken from \citet{French2018}. Galaxies with negative ages are those where the best-fit model is a still-declining burst. [1] \citet{French2018b}.}
    \label{tab:targets}
\end{table*}

\begin{table*}[]
    \centering
    \begin{tabular}{lllll}
    \hline
    Target & Dataset & Beam size (\arcsec)& CASA Pipeline & robust \\
    \hline
    H03 HCO$^+$ (1--0) & 2017.1.00935 & 1.26 $\times$ 1.06 & 5.1.1-5 & 0 \\
    S02 HCO$^+$ (1--0) & 2017.1.00935 & 0.97 $\times$ 0.88 & 5.4.0-68 & 2 \\
    R02 HCO$^+$ (1--0) & 2018.1.00948 & 1.03 $\times$ 0.86 & 5.4.0-68 & 1 \\
    R05 HCO$^+$ (1--0) & 2018.1.00948 & 1.16 $\times$ 0.80 & 5.4.0-68 & 2\\
    H02 CO (3--2) & 2017.1.00930 & 0.39 $\times$ 0.38 & 5.1.1-5 & 0.5 \\
    H03 CO (3--2) & 2017.1.00930 & 0.39 $\times$ 0.34 & 5.4.0-68 & -0.5 \\
    S02 CO (3--2) & 2017.1.00930 & 0.26 $\times$ 0.23 & 5.4.0-68 & 2 \\
    \hline
    \end{tabular}
    \caption{ALMA Observations. Beam sizes and robust values for Briggs weighting were chosen to best match the requested beam size and sensitivity.}
    \label{tab:redux}
\end{table*}

\begin{table*}[]
    \centering
    \begin{tabular}{lllllll}
    \hline
    Galaxy &  HCN (1--0) &  HCO$^+$ (1--0) &  HNC (1--0) &  CO (3--2) \\
    & (Jy km/s) & (Jy km/s) & (Jy km/s) & (Jy km/s) \\
    \hline
    H02 & $<0.13$ & $<0.13$ & & $18.87 \pm 0.55$ \\
    H03 & $0.98 \pm 0.02$ & $0.32 \pm 0.02$ & $0.25 \pm 0.02$ & $71.52 \pm 0.45$ \\
    S02 & $<0.08$ & $<0.09$ & $<0.09$ & $20.17 \pm 0.22$  \\
    S05 & $<0.05$ & $<0.05$ & & \\
    R02 & $0.21 \pm 0.03$ & $0.51 \pm 0.03$ & $0.13 \pm 0.03$ & \\
    R05  & $0.22 \pm 0.03$ & $ 0.51 \pm 0.03$ & $<0.09$ & \\
    \hline
    \end{tabular}
    \caption{Measured Integrated Line Fluxes $S_\nu dv$.}
    \label{tab:fluxes}
\end{table*}

\subsection{Archival Data}
\label{sec:sdss}

We use H$\alpha$-based star formation rates (SFRs) for the post-starburst galaxies from \citet{French2015} and \citet{Rowlands2015} as this measurement is the least biased and most available SFR for our post-starburst sample. The H$\alpha$ fluxes are from the SDSS \citep{Strauss2002} MPA-JHU {\tt galSpec} catalogs \citep{Brinchmann2004, Tremonti2004}, and have been corrected for extinction using the Balmer decrement. For the two \citet{Rowlands2015} galaxies (R02, R05), the SFRs are additionally corrected for the contribution from AGN contamination using the method from \citet{Wild2010}, as these galaxies were selected without a cut against significant H$\alpha$ emission (see \S\ref{sec:sample}). We correct for aperture bias using the {\tt galSpec} SFR aperture bias corrections, which use photometry outside of the fiber aperture to estimate the required correction \citep{Salim2007}, and range from 3.5--6.3$\times$ corrections for the galaxies considered here. The impact of using different SFR indicators is discussed in S\ref{sec:sfr}. For the sample of post-starburst galaxies with both H$\alpha$ observations and {\it Spitzer} [NeII] 12.8$\mu$m and [NeIII] 15.6$\mu$m observations, we find no evidence that the H$\alpha$ observations are missing star formation due to dust obscuration. If there was significant dust obscuration for this sample, we would expect to see Ne-based SFRs systematically above the H$\alpha$-based SFRs. Instead, we see no systematic shift between the two tracers. Thus, we use H$\alpha$-based SFRs throughout the main body of this work, and present the results if infrared tracers are used in Appendix \ref{sec:appendix_ir}.

\section{Results}

\subsection{Gas Excitation}

\begin{figure*}
\includegraphics[width=\textwidth]{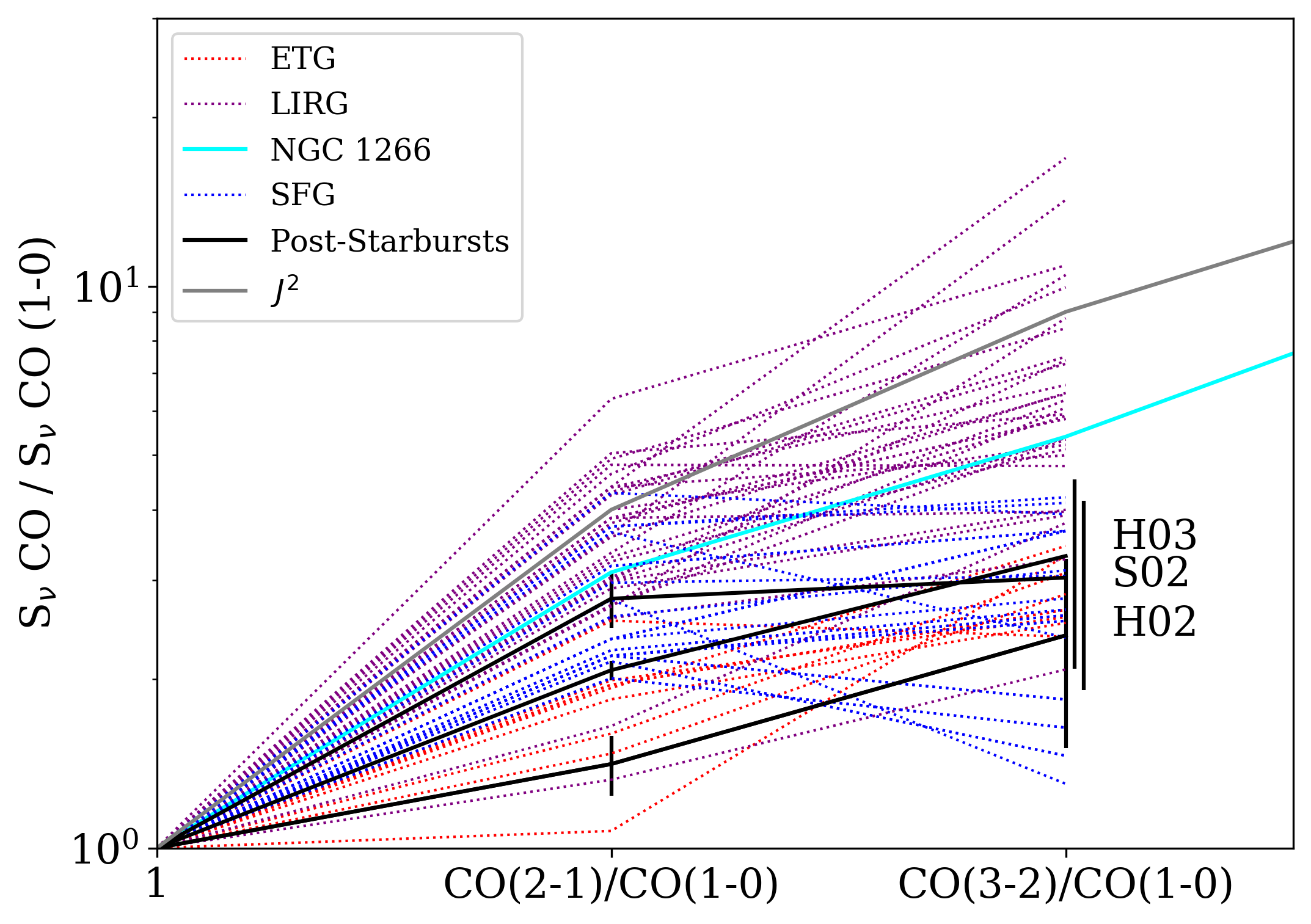}
\caption{CO Spectral Line Energy Density (CO SLED) for the three post-starburst galaxies observed in CO (3-2) (with data for CO (2-1) and CO (1-0) from \citealt{French2015}), their likely progenitors (LIRGs, \citealt{Papadopoulos2012}), star forming galaxies \citep{Leroy2021}, and their likely descendants (early type galaxies, \citealt{Crocker2012, Bayet2013}). We also include the shocked post-starburst galaxy NGC 1266 \citep{Pellegrini2013}. The post-starburst galaxies presented here have excitation states more similar to the early type galaxies than the LIRGs, with lower excitation than observed for NGC 1266. (Note: Several of the LIRGs have CO SLEDs above the expectation for thermalized gas, likely due to optical depth effects \citep{Papadopoulos2012}.) }
\label{fig:co_sled}
\end{figure*}

Observations of multiple $J$ CO lines trace the excitation of the molecular gas; the shape of the CO spectral line energy density (SLED) depends on the density, kinetic temperature, and observed column density of the gas \citep{Weiss2007,Carilli2013,Narayanan2014,Bournaud2015,Kamenetzky2018}. We combine the CO (3--2) measurements of three post-starburst galaxies from this survey with CO (2--1) and CO (1--0) measurements from the IRAM 30m from \citet{French2015}. Resolved studies of CO (2--1) with ALMA of a subset of these galaxies from \citet{Smercina2022} show that the molecular gas in post-starburst galaxies is compact, on $\lesssim 1$ arcsec scales, such that the ALMA observations are not resolving out extended flux, and the single dish measurements are comparable to the ALMA measurements. We compare the ALMA CO(2--1) flux measurements from 
\citet{Smercina2022} to the IRAM 30m measurements from \citet{French2015} to estimate the uncertainties introduced from combining the measurements. For H02 and H03, the ALMA and IRAM CO (2--1) measurements differ by $\sim37$\% (consistent within 1.5$\sigma$ of the combined uncertainty). For H02, the IRAM flux is larger than the ALMA flux; for H03 the ALMA flux is larger than the IRAM flux. We do not have ALMA CO (2--1) measurements for S02, so we assume it will have a similar uncertainty. These errorbars are reflected in Figure \ref{fig:co_sled}.

We compare the CO SLEDs of the three post-starburst galaxies with CO (3--2) measurements with other galaxy samples in Figure \ref{fig:co_sled}. The post-starburst galaxies have low CO excitation, consistent with the population of early type galaxies \citep{Crocker2012,Bayet2013}, slightly below the population of star forming galaxies \citep{Leroy2021}, and below most of the LIRGs \citep{Papadopoulos2012}. Even considering the uncertainties from combining ALMA and IRAM measurements, the post-starburst galaxies have low excitation.

Observations of CO (1--0) and CO (2--1) in post-starburst galaxies show low star formation efficiencies (SFEs), $\rm SFE \propto SFR/$\lpco \citep{French2015,Rowlands2015,Alatalo2016b,Smercina2018,Smercina2022}. The excitation of the CO-traced gas can be used to distinguish between two possible mechanisms that could suppress star formation in the molecular gas, leading to these low SFEs. The first possibility is high kinetic temperatures in the gas paired with low gas densities; in this case, we would expect to see high gas excitation throughout the post-starburst phase. High excitation could also arise from both high temperatures and high densities like those in starburst galaxies, but we would expect to see starburst-like SFRs if the densities were also high. The second possibility for why the post-starburst galaxies have low SFE is if the gas densities are low. In this case, we would expect to see low gas excitation, similar to early type galaxies, throughout the post-starburst phase. The observations of the post-starburst excitation presented here favor the second possibility, that the post-starburst galaxies have low densities and temperatures\footnote{These low temperatures are indicative of the state of the CO, but are not sensitive to the large scale motions that appear to increase the velocity dispersion of the gas. \citet{Smercina2022} infer a source of turbulent heating from high velocity dispersions of CO (2--1) observations, yet this does not preclude the low CO excitation observed here.}. The low density may explain the low SFEs.

We explore these possibilities more quantitatively by modelling the CO SLEDs with RADEX \citep{radex}. RADEX is a non-LTE (non local thermodynamic equilibrium) code that solves for the radiative transfer of a given molecular species, assuming a geometry. Here, we assume a uniform sphere geometry\footnote{Assuming alternate geometries of an expanding sphere or plane parallel slab will not provide substantially different results for this application \citep{Krips2011}. We address the case where the galaxy contains multiple components of gas later in this section.}. We use a grid of logarithmically-spaced values in temperature ($T$; $10-300$ K), density ($n$; $10^1-10^7$ cm$^{-3}$), and column density of CO ($N$; $10^{13}-10^{21}$ cm$^{-2}$), similar to that used by \citet{Krips2011}. We generate the expected flux ratios of the CO (1--0), CO (2--1), and CO (3--2) lines for this grid of parameters and compare them to the observed line ratios. 

\begin{figure*}
\includegraphics[width=0.5\textwidth]{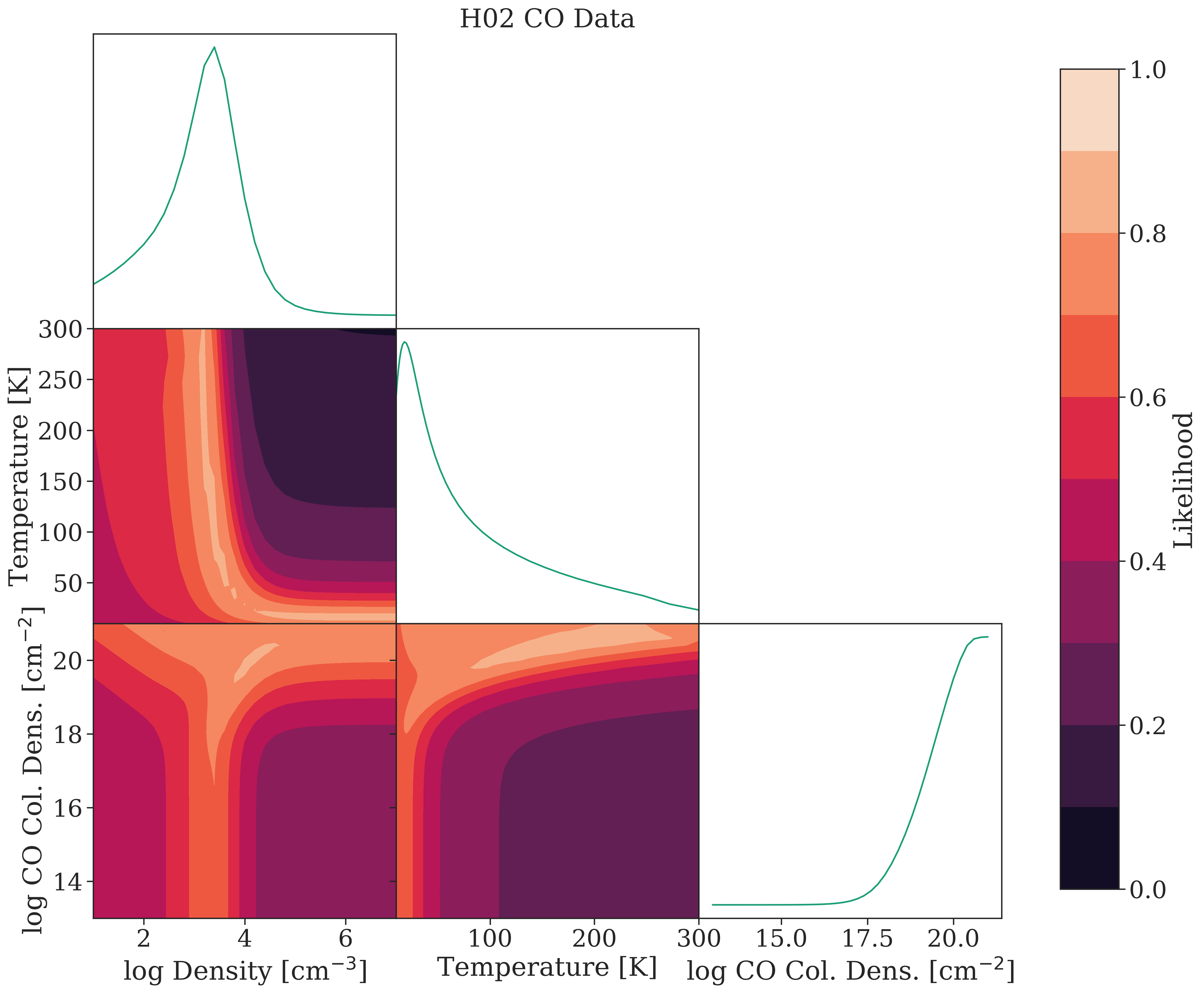}
\includegraphics[width=0.5\textwidth]{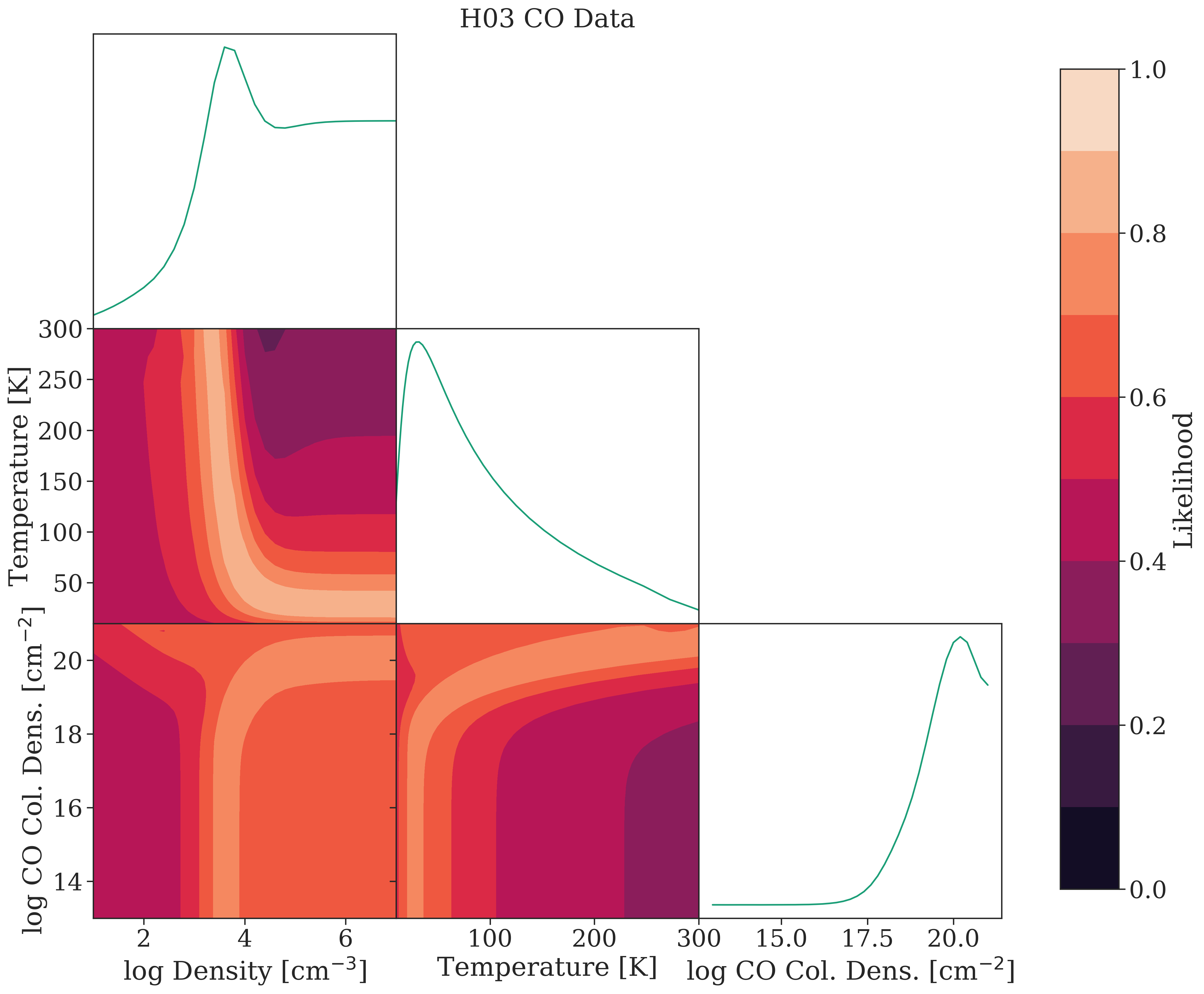}
\includegraphics[width=0.5\textwidth]{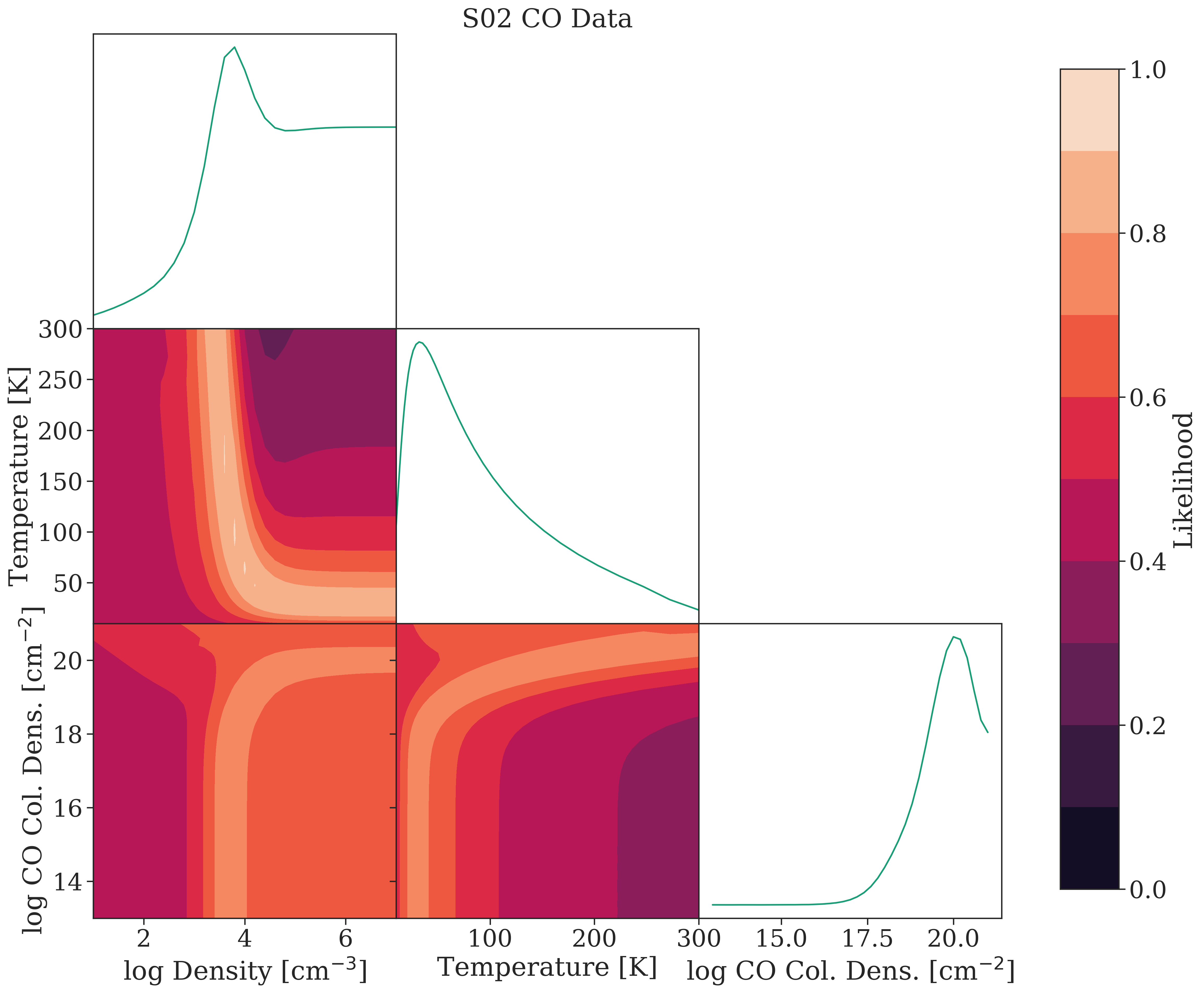}
\caption{Corner plot of likelihoods of temperature, density, and CO column density comparing RADEX models to our observations of CO (1--0), (2--1), and (3--2) for each galaxy. Line plots show the likelihood marginalized over the other two parameters, and contour plots show the likelihood marginalized over the third parameter to show the degeneracies present between parameters. For each galaxy, the best-fit models favor low densities ($\log n/[\rm{cm}^{-3}] \sim 3.4-3.8$) and temperatures $T\sim 15-30$ K, similar to early type galaxies and lower than typical LIRGs. }
\label{fig:radex_co}
\end{figure*}

The inferred density and temperature values are highly degenerate unless the full rise and turnover of the CO SLED can be sampled \citep[e.g.,][]{Carilli2013, Kamenetzky2018}. We visualize these degeneracies by plotting the likelihoods for the temperature, density, and column density of each source in Figure \ref{fig:radex_co}. Densities of $\log n \sim 3.4-3.8$ cm$^{-3}$ and temperatures of $T \sim 15-30$ K are favored, although clear degeneracies can be seen between the cases of low density with relatively unconstrained temperature and higher density with low temperature. These values are similar to molecular gas densities and temperatures in early type galaxies \citep{Bayet2013}, which have densities of $\log n \sim 3-4$ cm$^{-3}$ and temperatures of $T \sim 10-70$ K. The post-starburst galaxies have both temperatures and densities less than typical LIRGs. LIRGs have a wide range of densities of $\log n \sim 2.5-6.5$ cm$^{-3}$ and temperatures of $T \sim 30-120$ K \citep[e.g.,][]{Greve2009,Papadopoulos2012}.

\begin{figure*}
\includegraphics[width=0.5\textwidth]{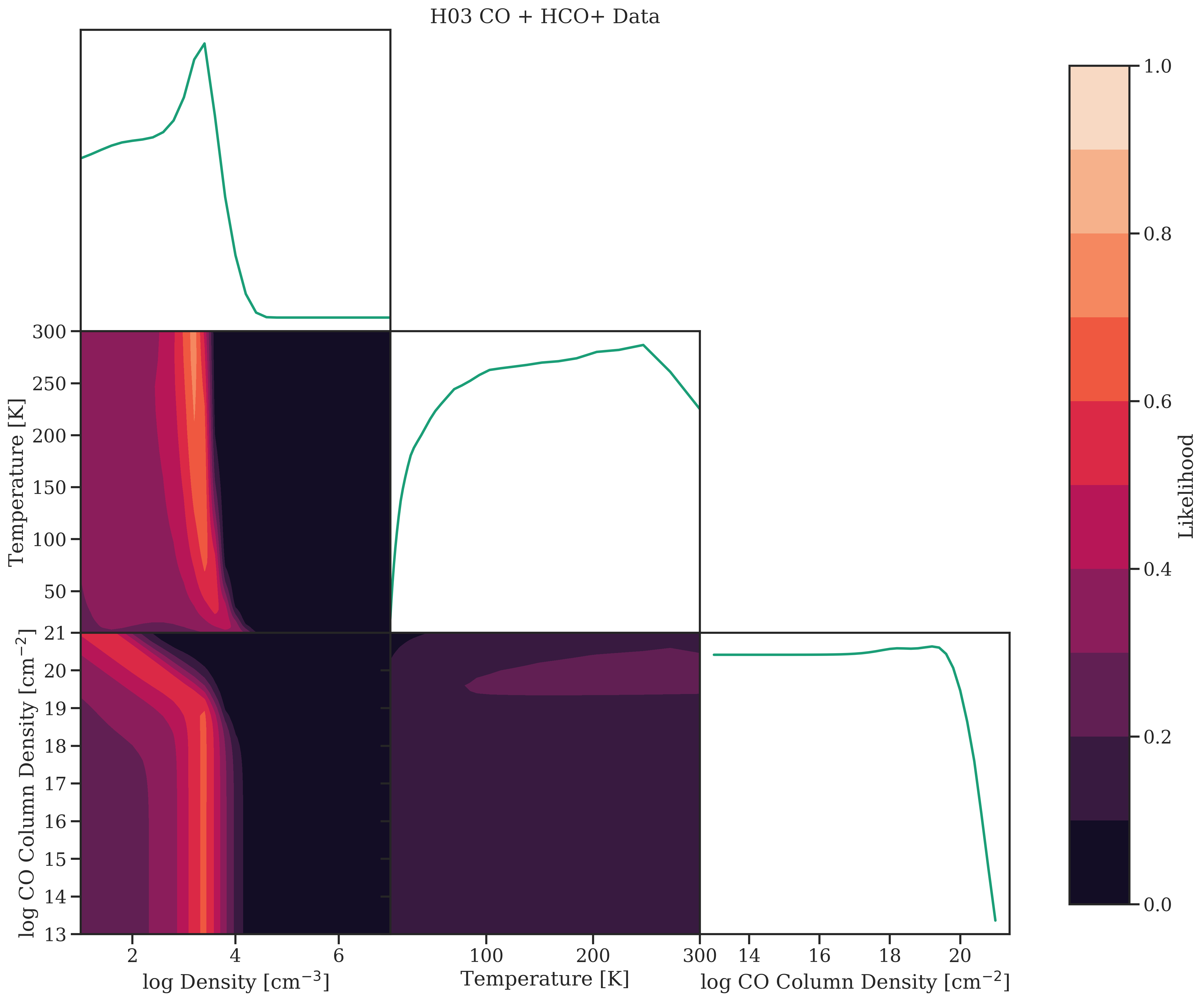}
\includegraphics[width=0.5\textwidth]{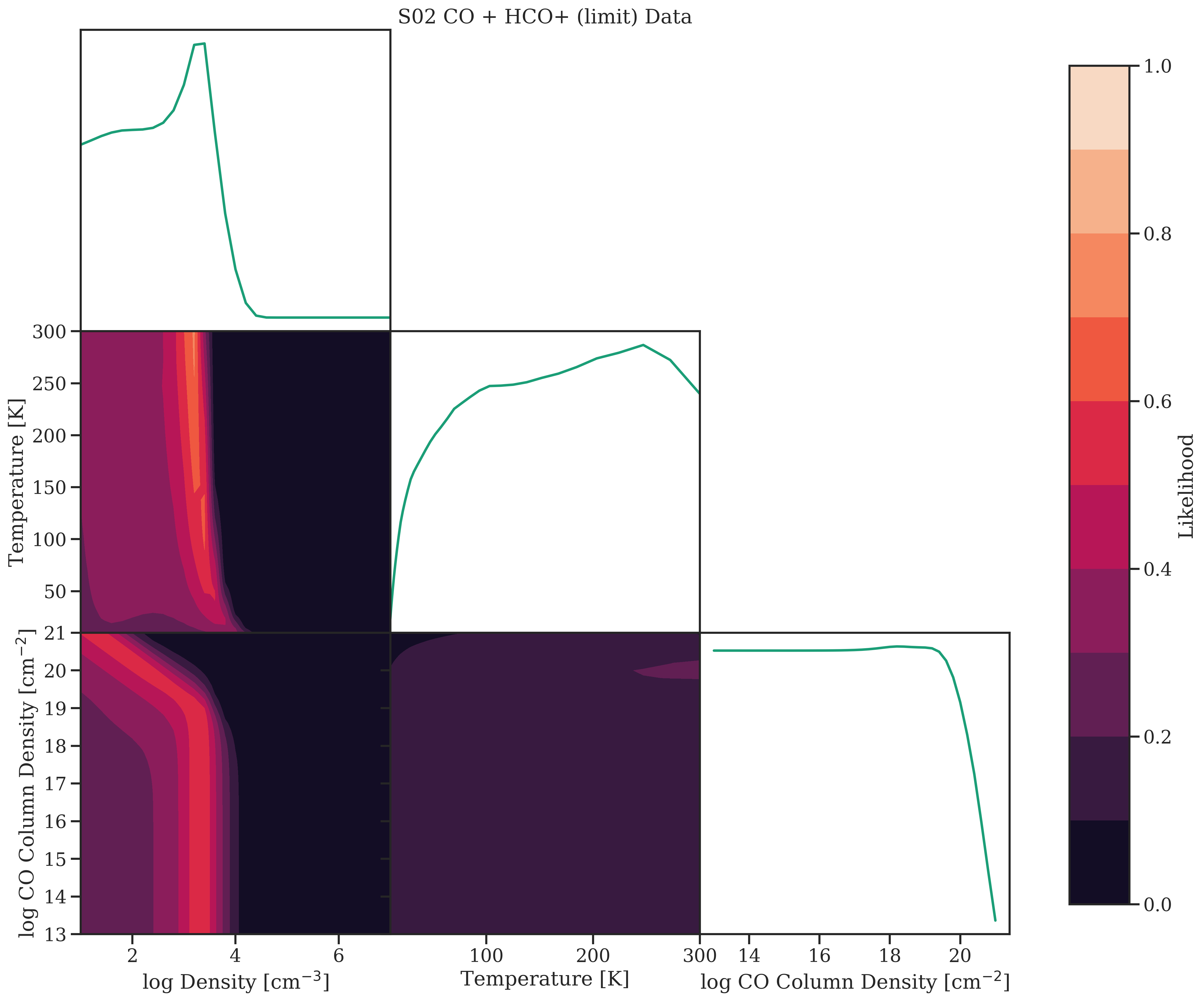}
\includegraphics[width=0.5\textwidth]{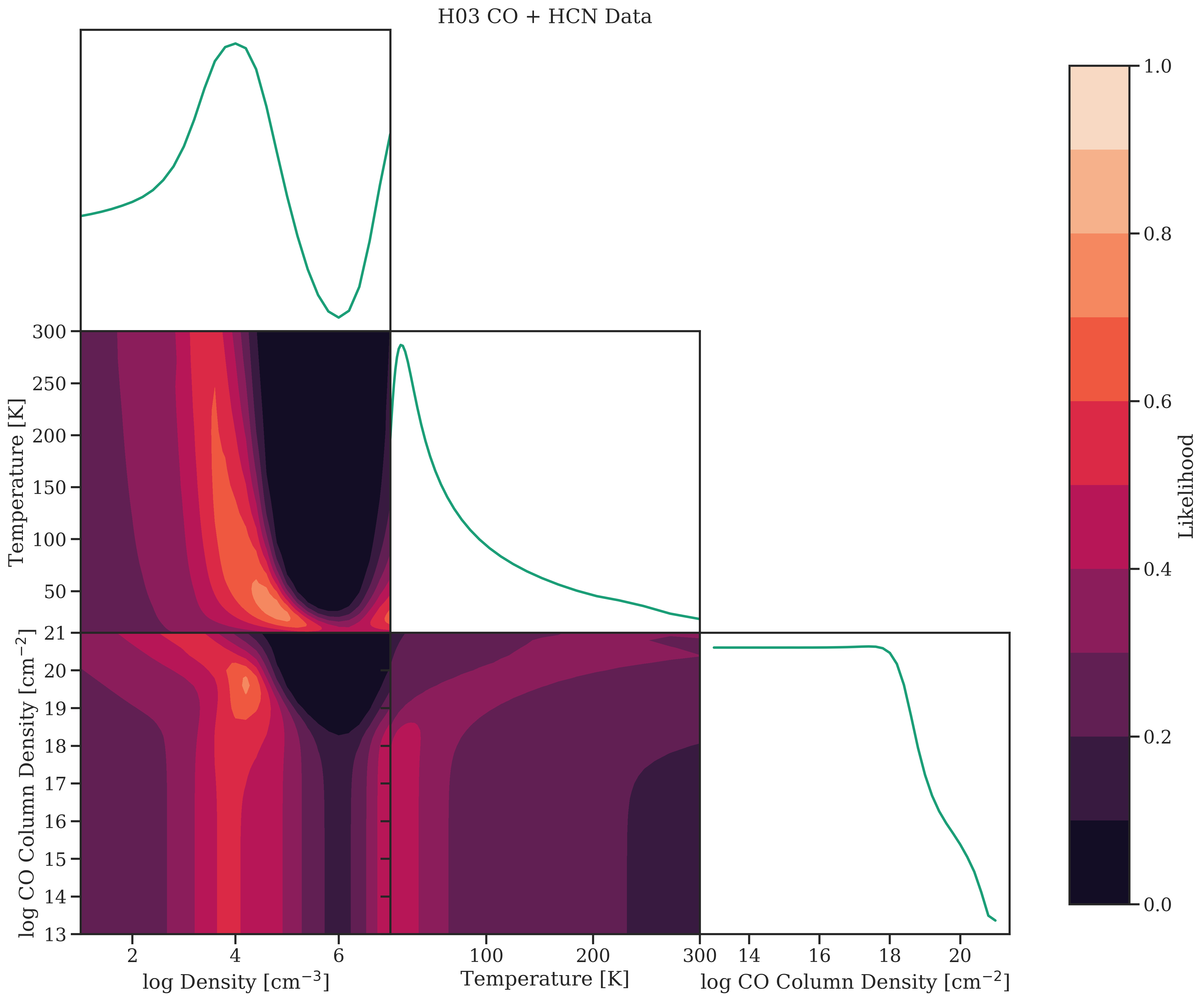}
\includegraphics[width=0.5\textwidth]{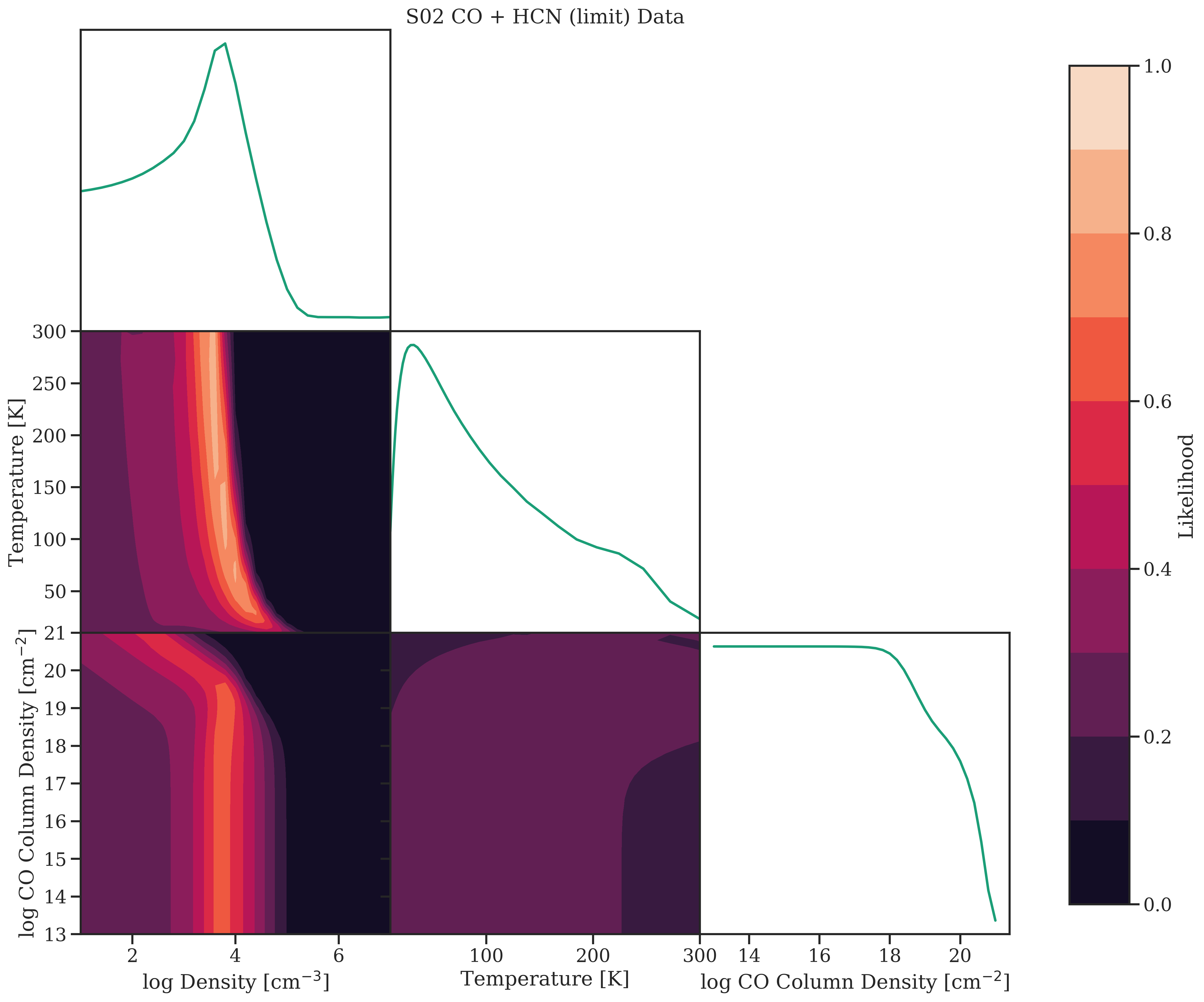}
\caption{Same as Figure \ref{fig:radex_co}, but with the observations of the dense gas tracers (top row: HCO$^+$ (1--0); bottom row: HCN (1--0)) for the two galaxies with both dense gas measurements and CO (3--2) measurements. We assume an abundance ratio of [HCN]/[CO] = [HCO$^+$]/[CO] = $10^{-4}$, and assume the dense gas flux ratios are just below the detection threshold for the non-detections of galaxy S02. The low ratios of dense molecular gas to total molecular gas inferred from the [HCO$^+$]/[CO] ratios is also reflected here in the steep dropoff in likelihood in densities $\log n/[\rm{cm}^{-3}] > 4$. However, we do not have enough data to constrain the likely abundance ratios, and thus these additional dense gas measurements do not provide useful constraints on the temperature or CO column density.
}
\label{fig:radex_co_dens}
\end{figure*}

We perform additional RADEX modeling of the dense gas tracers HCN (1--0) and HCO$^+$ (1--0) in addition to the CO lines. This analysis is complicated by the need to assume an abundance ratio of either [HCO$^+$/CO] or [HCN/CO]. For both dense gas tracers, we assume an abundance ratio with respect to CO of $10^{-4}$ \citep[consistent with the range of measurements by][]{Krips2008,Aalto2012}\footnote{This assumption only affects the results shown in Figure \ref{fig:radex_co_dens}; the results in Figure \ref{fig:radex_co} are independent, as are the results described in subsequent sections}. Due to the uncertainties in what is affecting the HCO$^+$/HCN ratio (\S\ref{sec:densegasstate}), we consider each dense gas tracer individually. In Figure \ref{fig:radex_co_dens}, we show the parameter likelihood corner plots considering CO (1--0), CO (2--1), CO (3--2) and either HCO$^+$ (1--0) or HCN (1--0) for the two post-starburst galaxies (H03, S02) with both datasets. 
For S02, neither dense gas tracer is detected, so we assume the dense gas flux ratios are just below the detection threshold. When the HCO$^+$ constraints are included, the best-fit densities are similar, with sharp cutoffs in likelihood above $\log n \sim 4$, the effective excitation density of the dense gas tracers \citep{Shirley2015}. 

The cold molecular gas temperatures indicated by the low CO excitation do not preclude the existence of an additional component of warm or hot gas, as the low J CO lines cannot predict the full SLED in the event of a secondary component of high temperature molecular gas. In our RADEX analysis, we have assumed the molecular gas is composed of a single temperature component. However, studies of star-forming and starburst galaxies \citep[e.g.,][]{Valentino2020} have found evidence for multiple components using higher $J$ CO lines than available here. A secondary component of high excitation molecular gas was also found for NGC 1266 by \citet{Pellegrini2013}, better fit by shock models than PDR models. The post-starburst galaxies considered here may also have multiple components of molecular gas, which would require additional CO lines to uncover, and may have different filling fractions. Indeed, we expect that such a component exists, as traced by mid-IR H$_2$ rotational lines \citep{Smercina2018}. The extremely high H$_2/$TIR ratios observed by \cite{Smercina2018} are indicative of shocks heating a portion of the molecular gas to high temperatures, with a ``high-soft" radiation field affecting the dust. By modeling the mass of H$_2$ using the mid-IR warm H$_2$ lines \citep{Togi2016} and extrapolating down in temperature to the low temperature regime probed by the $J\le3$ CO lines, \citet{Smercina2018} find gas masses typically within a factor of $2-4\times$ the mass of cold gas inferred from CO (1--0). The higher temperature gas traced by the mid-IR lines  comprises a relatively small fraction of the overall mass. By combining these observations, we conclude that the bulk of the gas by mass remains at temperatures lower than typical LIRGs, while a fraction of the molecular gas is highly excited.

\begin{figure}
\includegraphics[width=0.5\textwidth]{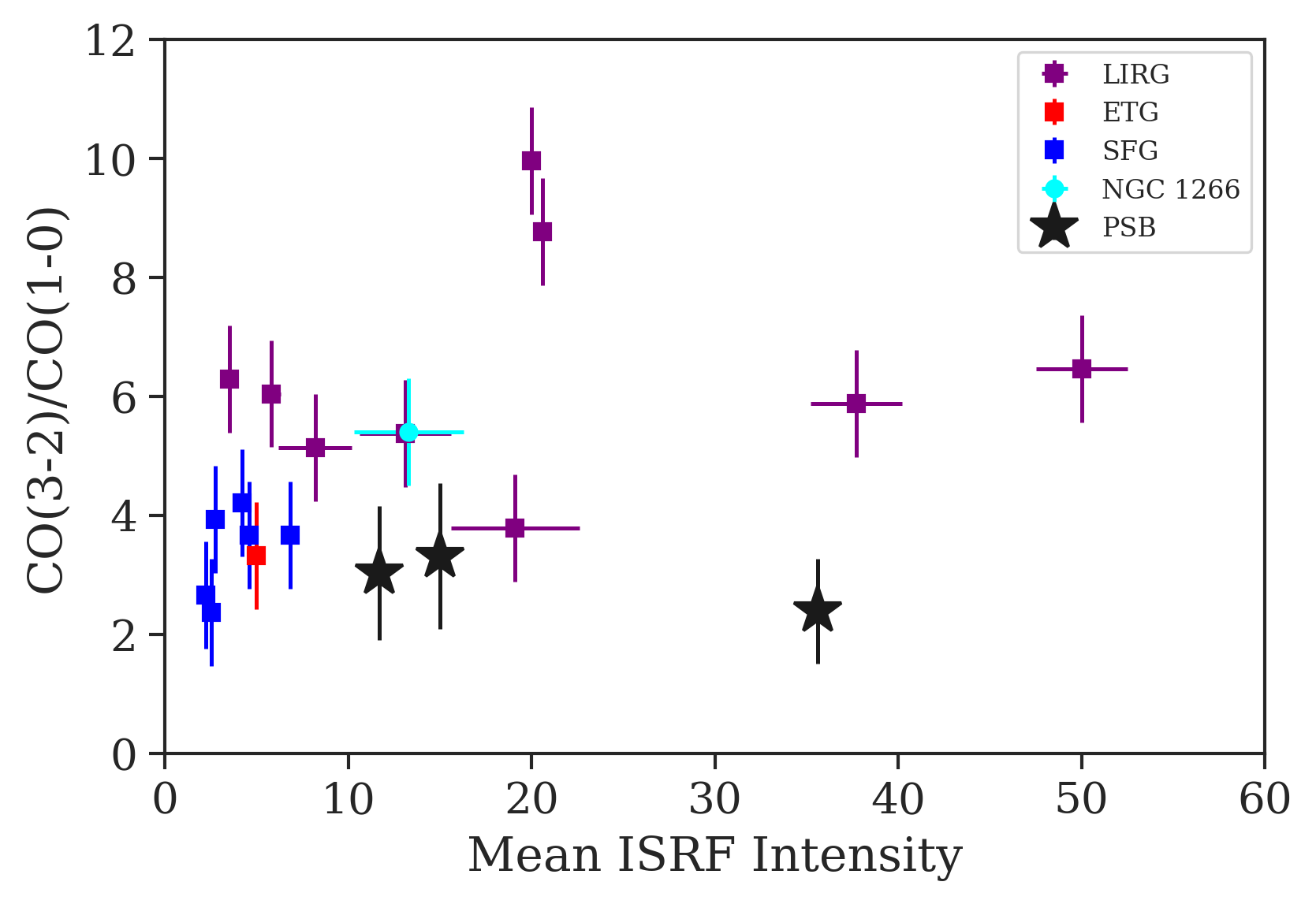}
\caption{CO (3--2) / CO (1--0) intensity ratio (sensitive to gas temperature) vs. mean interstellar radiation field intensity \meanu calculated from dust SED fitting (sensitive to dust temperature). Post-starburst galaxy \meanu is measured from the best-fit parameters in \citet{Smercina2018} using the relation in \citet{Draine2007b}. LIRG \meanu measurements are from \citet{Liu2021} and star-forming and early type galaxy measurements are from \citet{Draine2007b}. CO measurements are as above. While post-starburst galaxies can have high \meanu values similar to LIRGs, these dust temperatures are not well-coupled to the colder molecular gas.}
\label{fig:r31_u}
\end{figure}

We compare the CO (3--2) / CO (1--0) intensity ratio as a tracer of the molecular gas temperature to the mean interstellar radiation field (ISRF) intensity \meanu calculated from dust SEDs in Figure \ref{fig:r31_u}. The \meanu is closely coupled to the dust temperature. \citet{Smercina2018} used \citet{Draine2007a} models to fit the infrared SEDs of post-starburst galaxies, including the galaxies considered here. We use the best-fit values of the minimum ISRF intensity $U_{min}$ and power law index $\gamma$ from \citet{Smercina2018} and the conversion to \meanu from \citet{Draine2007b} to calculate \meanu $= 35.6, 15, 11.7$ for H02, H03, and S02, respectively. Cross matching with the CO SLED data above, we obtain \meanu measurements for LIRGs from \citet{Liu2021} and for the star forming and early type galaxies from \citet{Draine2007b}. For normal star forming and star bursting galaxies, the CO excitation is correlated with \meanu. This is consistent with previous work, e.g., \citet{Liu2021} find the CO R52 ratio to be well correlated with \meanu. While the post-starburst galaxies have high \meanu values similar to the LIRGs, their low CO excitation is more consistent with star forming and early type galaxies. The post-starburst galaxies lie in an unusual region of this parameter space, having high \meanu values similar to LIRGs, but low CO excitation, indicating the dust and gas temperatures are not well-coupled. The decoupled dust and gas temperatures stand in contrast to the coupled time evolution observed during the post-starburst phase \citep{Smercina2018, Li2019}.

\subsection{Spatially-resolved and velocity-resolved observations}

\begin{figure}
\includegraphics[width=0.5\textwidth]{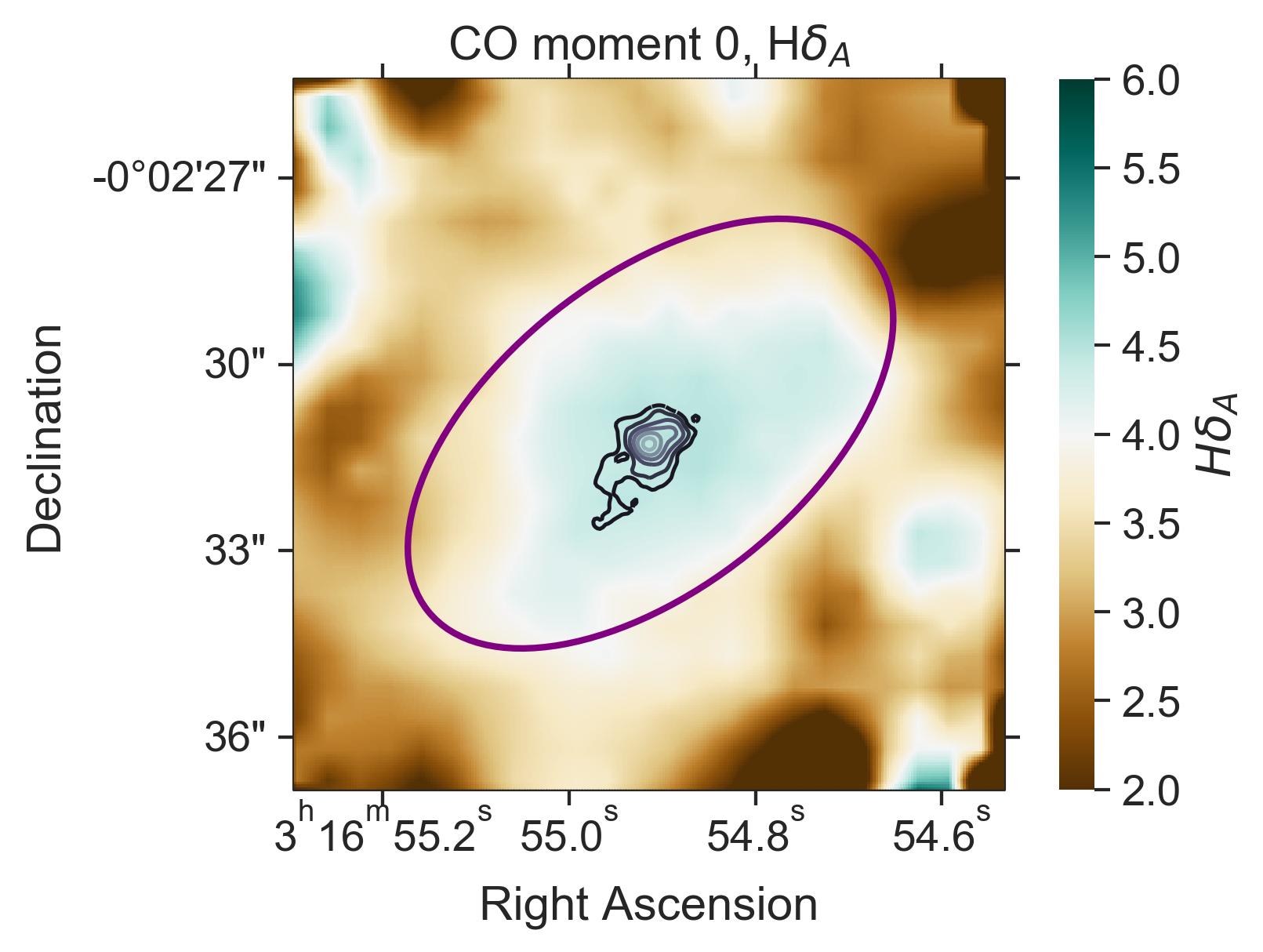}
\caption{ALMA CO (3--2) observations of S02 (contours) with complementary MaNGA \citep{manga} observations (MaNGA plateID 8080-3072) \citep{Westfall2019, marvin}. The half light ellipse is shown in purple. The H$\delta_A$ index traces the young A-star dominated stellar population and indicates the post-starburst regions of the galaxy. The molecular gas region is smaller than the post-starburst region (teal; H$\delta_A \gtrsim 4$). }
\label{fig:s02_hdelta_comp}
\end{figure}

We construct zeroth-, first-, and second- order moment maps for the three post-starburst galaxies with CO (3--2) observations, presented in Appendix \ref{sec:appendix}. The CO (3--2) emission for each of the three post-starburst galaxies observed has limited extent relative to the optical emission, consistent with the CO (2--1) results from \citet{Smercina2022}, which assume a 2D Gaussian profile. The half-light radius of the CO (3--2) emission for each galaxy is shown in Table \ref{tab:sizes}. The sizes range from 0.38--0.57 arcsec, or 367--400 pc. On average, the CO (3--2) sizes are 6.3$\times$ smaller than the $r$-band sizes from the SDSS imaging. Two galaxies (H02 and H03) have both CO (3--2) and CO (2--1) observations; these galaxies have sizes consistent between the two tracers, although differences would be difficult to determine with these observations, as $>50$\% of the flux is contained within a central unresolved beamsize in both tracers for both galaxies. The velocity and velocity dispersion maps are broadly consistent as well. The galaxy S02 has archival MaNGA \citep{manga} IFU data. In Figure \ref{fig:s02_hdelta_comp}, we compare the CO (3--2) emission to the post-starburst region with high H$\delta$ absorption. Even with the coarser $\sim2.5$\arcsec\ resolution of MaNGA, the post-starburst region with strong H$\delta$ absorption is resolved, and extends over most of the half light ellipse of the galaxy. The molecular gas is smaller than the extent of the young stellar population traced by the H$\delta_A$ index.

\begin{figure*}
\includegraphics[width=\textwidth]{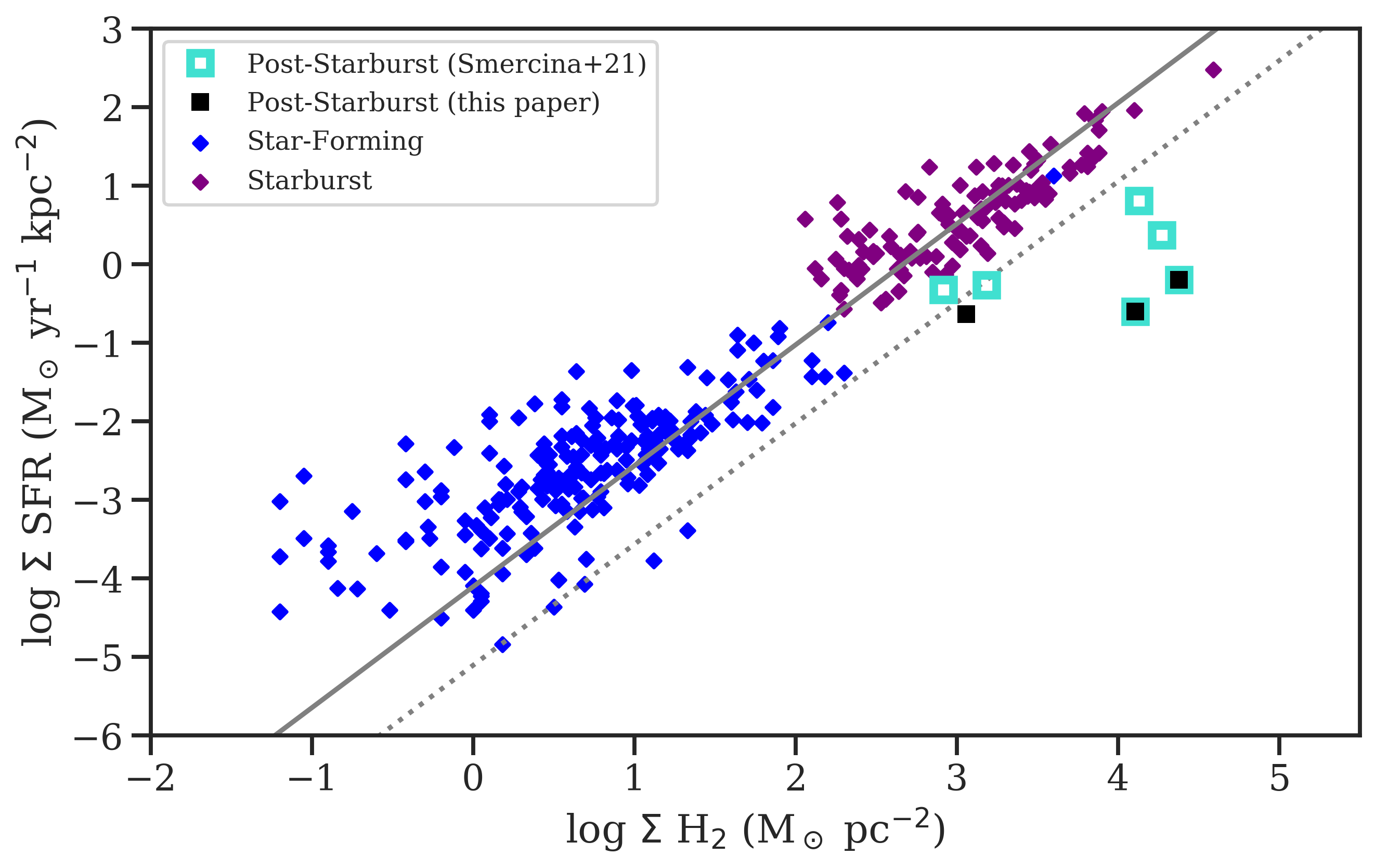}
\caption{Molecular gas surface density vs. SFR surface density for post-starburst galaxies from this work with CO (3--2) sizes and from \citet{Smercina2022} with CO (2--1) sizes, as well as comparison samples of star-forming galaxies from \citet{delosReyes2019} and starbursting galaxies from \citet{Kennicutt2021}. The best-fit relation from \citet{Kennicutt2021} for the total gas density vs. star formation density is plotted in grey, with a dotted line indicating a factor of $10\times$ below the relation. The post-starburst galaxies have very high molecular gas surface densities, yet they lie below the comparison galaxies, with low star formation rate surface densities for their molecular gas surface densities. The post-starburst galaxies have SFR surface densities $5.5\times$ below the median for the comparison samples ($17\times$ lower than the starburst sample alone). We find qualitatively similar results when using other SFR tracers, see Appendix \ref{sec:appendix_ir} and Figure \ref{fig:ks_ir}.}
\label{fig:ks}
\end{figure*}

The small CO (3--2) and CO (2--1) sizes imply very high surface densities of molecular gas and residual star formation. Because of the consistency in sizes between the CO (3--2) measurements here and the CO (2--1) sizes measured by \citet{Smercina2022}, we assume these sizes in calculating the molecular gas surface density $\Sigma_{\rm H_2}$ from the CO (1--0) luminosities from \citet{French2015}. Using the half-light sizes, we calculate the densities as:
\begin{equation}
    \Sigma_{\rm H_2} = 0.5 \frac{M_{H_2}}{\pi R_{1/2}^2}
\end{equation}

\begin{equation}
    \Sigma_{\rm SFR} = 0.5 \frac{SFR}{\pi R_{1/2}^2}.
\end{equation}
We compare our observations to the samples of star-forming galaxies from \citet{delosReyes2019} and starbursting galaxies from \citet{Kennicutt2021} in Figure \ref{fig:ks}. This set of comparison galaxies is low redshift and covers a similar stellar mass range to the post-starburst galaxies presented here. While the comparison galaxies are on average at lower distance, the difference in average redshift is not enough to introduce cosmological effects. We use a CO-to-H$_2$ conversion factor of 4 \unitxco for all samples. The post-starburst galaxies have very high molecular gas surface densities, yet they lie below the comparison galaxies, with low star formation rate surface densities for their molecular gas surface densities. If we define the scatter in the star formation rate surface densities relative to the best-fit relation from \citet{Kennicutt2021} as $\sigma$, the median post-starburst galaxy is 2.5$\sigma$ below the median for the comparison samples and 3.3$\sigma$ below the median for the starburst sample alone. The effects of star formation tracer on these results is explored further in \S \ref{sec:sfr} and Appendix \ref{sec:appendix_ir}; we observe qualitatively similar results using infrared tracers instead of the H$\alpha$ SFR tracers used for the post-starburst galaxies here, even though the SFR derived from the TIR luminosity is biased high for the post-starbursts (see further discussion in \S \ref{sec:sfr}). When using the TIR luminosity to trace SFR, we find the median post-starburst galaxy to be 2.1$\sigma$ below the median for the comparison samples and 3.3$\sigma$ below the median for the starburst sample alone; when using the Neon-based SFRs, we find the median post-starburst galaxy to be 1.9$\sigma$ below the median for the comparison samples and 2.4$\sigma$ below the median for the starburst sample alone, using the scatter in the comparison sample to be $\sigma$ as described above. 
Six of the seven galaxies presented in Figure \ref{fig:ks} are the sample from \citet{Smercina2022}; the addition of the measurement for S02 does not change the interpretation of this result. The values of $\Sigma_{\rm H_2}$ and $\Sigma_{\rm SFR}$ differ slightly from those presented by \citet{Smercina2022} due to the inclusion of the [CII] tracer and a different method for calculating the surface densities. but the qualitative conclusions are consistent. We use the half-light sizes, as these can be more robustly determined than outer sizes like the 95 percentile size used by \citet{delosReyes2019} and \citet{Kennicutt2021}, though we verify that our qualitative conclusions do not change with the use of 95 percentile radii, regardless of SFR tracer. Because the relation between $\Sigma_{\rm H_2}$ and $\Sigma_{\rm SFR}$ is close to linear, changes to the size will not significantly affect the inferred offset from this relation.

Evidence of rotation is seen in the velocity fields of all three sources, though the peak velocity dispersion values are higher than the peak velocities. These behaviors are consistent with the range in CO (2--1) velocity and velocity dispersion maps seen by \citet{Smercina2022}. The high CO(2--1) velocity dispersion and small size measured by \citet{Smercina2022} for H02 and H03, which led them to infer a high turbulent pressure, is also observed here. In the case of one galaxy, S02, we see evidence for a blueshifted outflow, which we explore further in the next section.

\begin{table}[]
    \centering
    \begin{tabular}{llll}
    \hline
    Galaxy & CO (3--2) size &  CO (3--2) size & $r$-band R50$^a$\\
     & (arcsec) & (pc) & (arcsec) \\
    \hline
    H02 & 0.38 & 400 & 1.71 \\
    H03 & 0.41 & 355 & 3.17 \\
    S02 & 0.57 & 267 & 3.72 \\
    \hline
    \end{tabular}
    \caption{CO (3--2) Galaxy Half-light Sizes. $^a$ From the SDSS.}
    \label{tab:sizes}
\end{table}

\subsection{Blueshifted Outflow}

\begin{figure*}
\includegraphics[width=\textwidth]{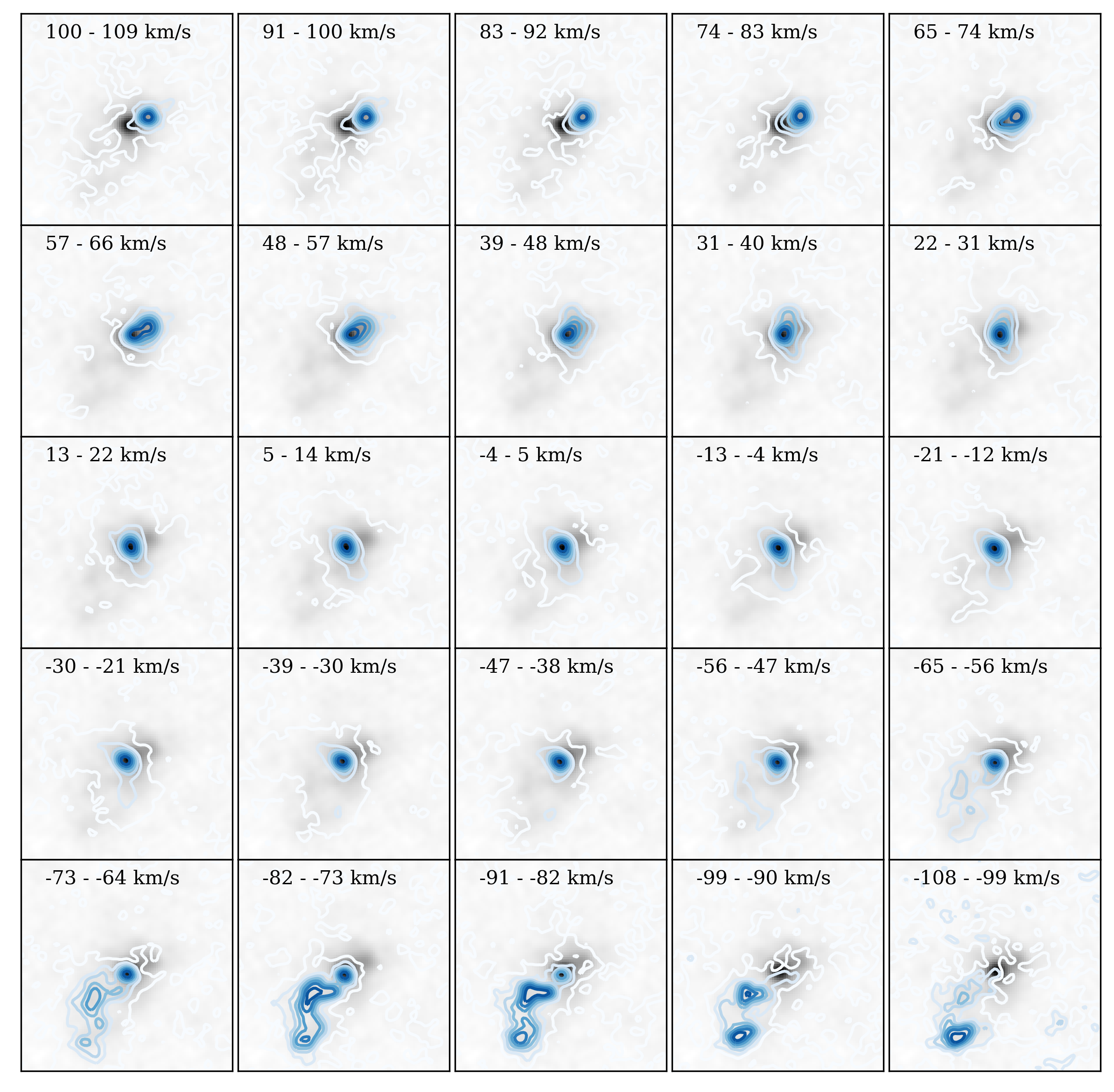}
\caption{CO (3-2) channel maps for S02, grouped by velocity bin (blue contours). Each image is $4\times4$\arcsec. In greyscale, the full moment zero map is shown for reference. A blueshifted outflow is seen to the lower left of the galaxy at velocities $\sim$ -64 -- -108 \kms, in the bottom row of this Figure. This component is roughly in line with the rotational axis of the gas in the rest of the galaxy (and the stellar velocity observed from MaNGA observations). The velocity range shown here is symmetric, and yet only a blueshifted outflow is seen without a redshifted counterpart.}
\label{fig:s02_grid}
\end{figure*}

\begin{figure}
\includegraphics[width=0.5\textwidth]{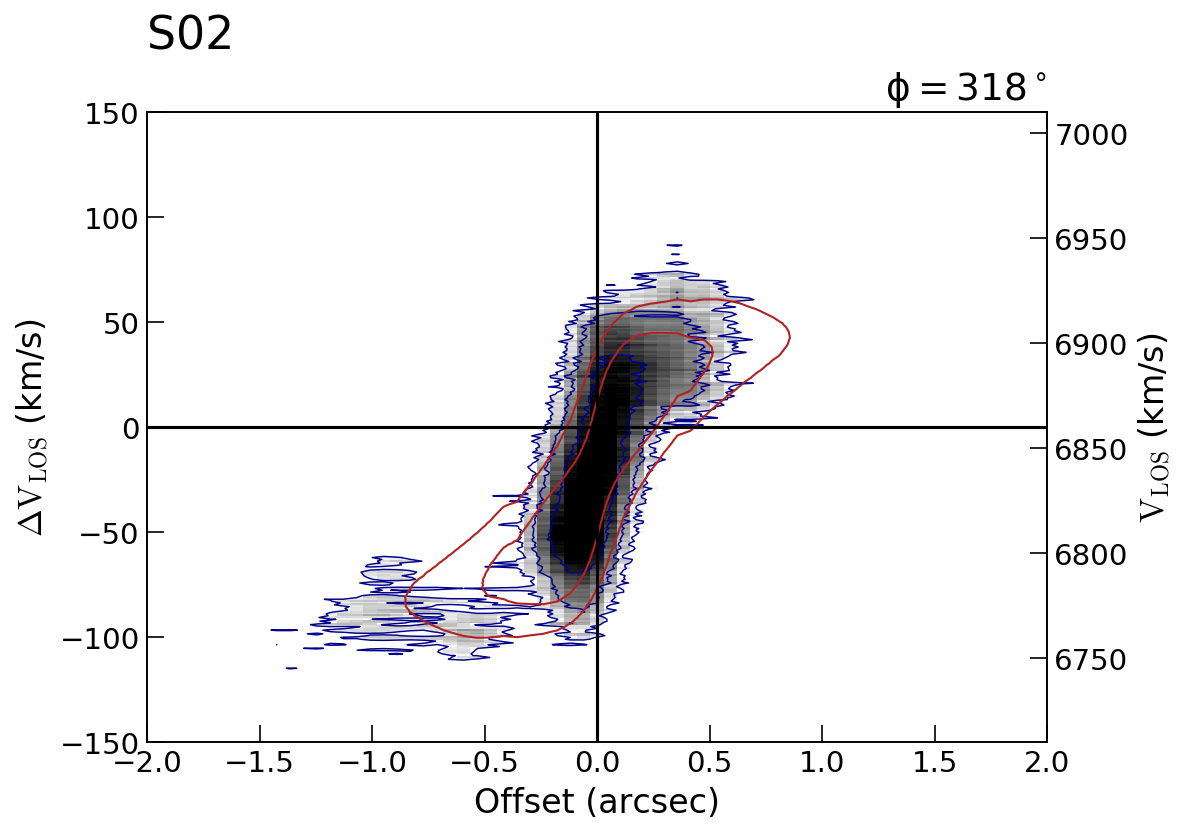}
\caption{Position-Velocity diagram along major axis (from MaNGA stellar velocity field), with best-fit model from Barolo \citep{barolo} overplotted in red contours. The blueshifted feature observed above in the moment maps is distinct from the rest of the molecular gas in the galaxy, suggesting it is an outflow rather than a continuation of a rotating disk.
}
\label{fig:s02_pv}
\end{figure}

\begin{figure*}
\includegraphics[width=0.5\textwidth]{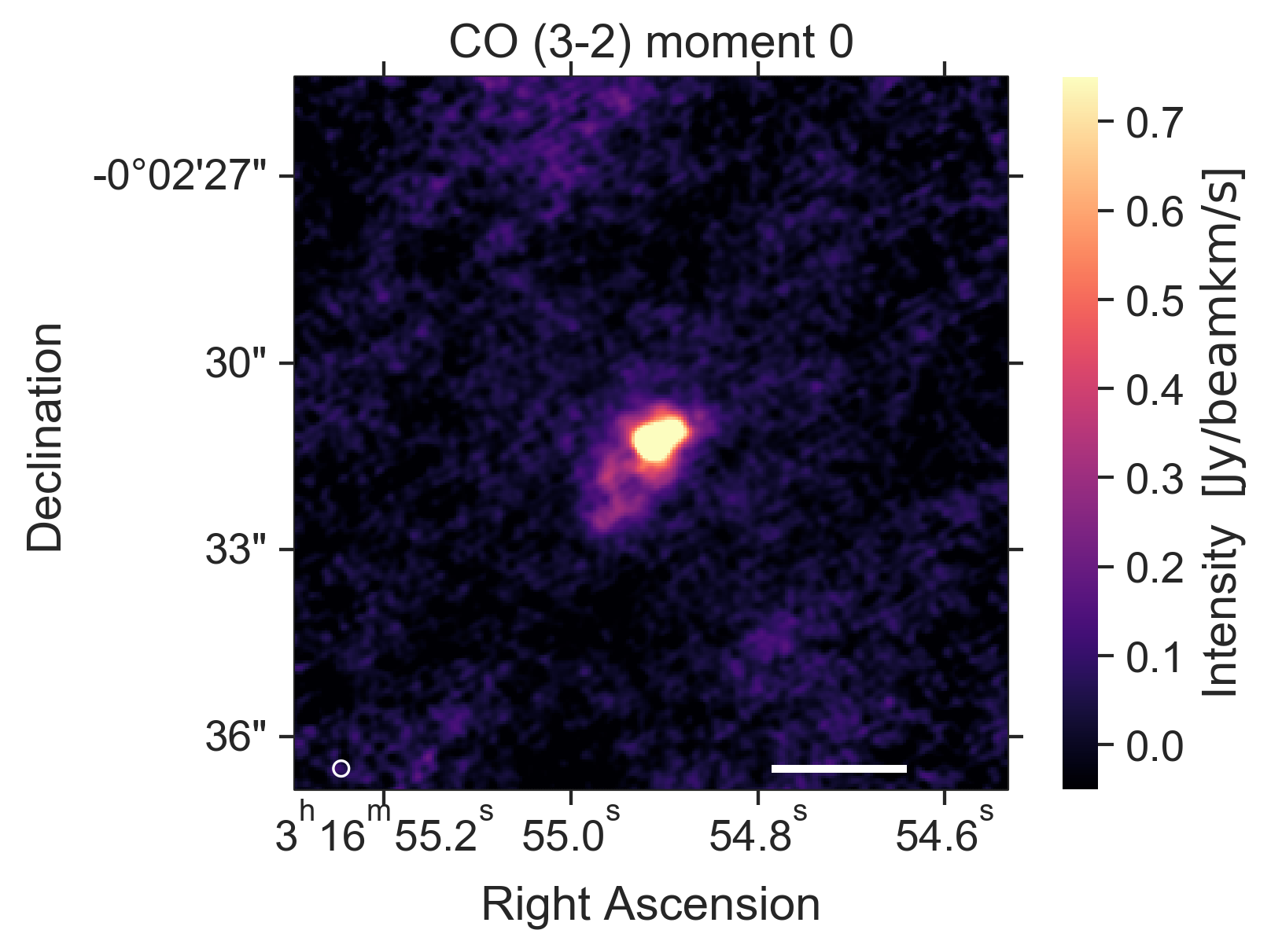}
\includegraphics[width=0.5\textwidth]{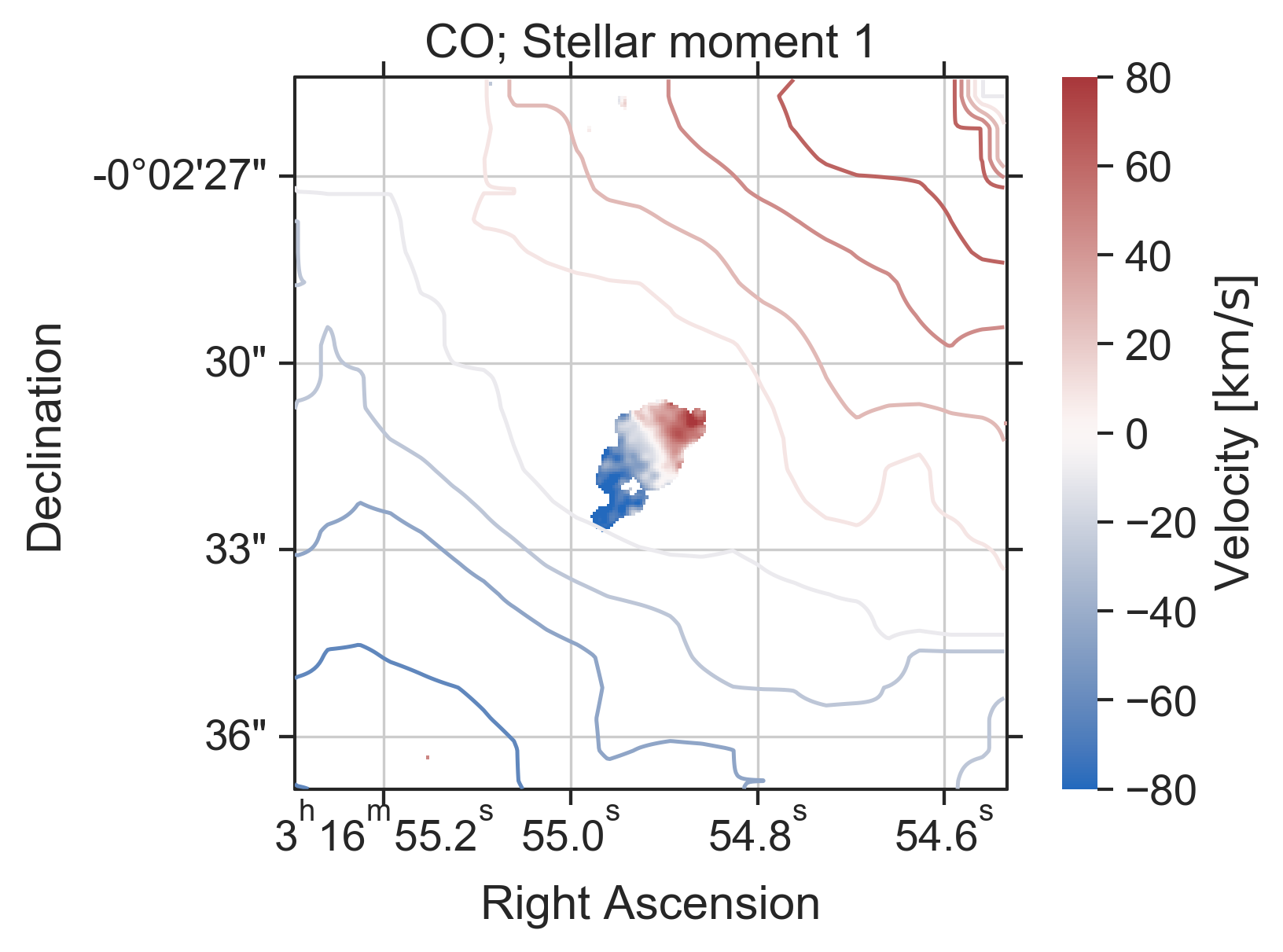}
\includegraphics[width=0.5\textwidth]{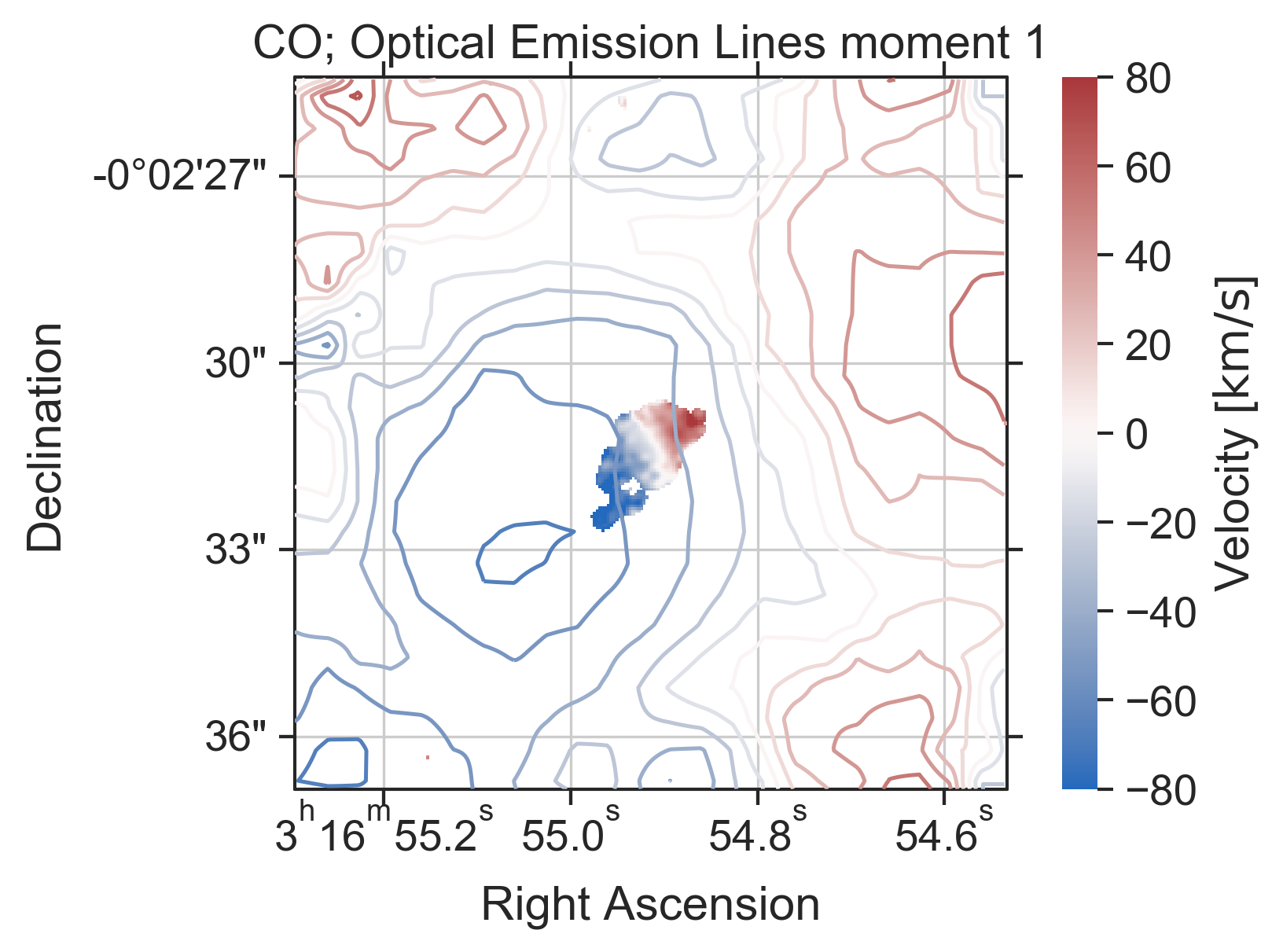}
\includegraphics[width=0.4\textwidth]{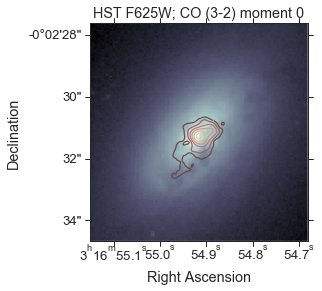}
\caption{ALMA CO (3--2) observations of S02 with complementary MaNGA \citep{manga} observations (MaNGA plateID 8080-3072) accessed via Marvin \citep{marvin}. Top left: CO (3--2) moment 0 map, with beam size shown in bottom left. The galaxy has a faint trail of gas extending to the lower left. Top right: CO (3--2) moment 1 map, with stellar velocity contours from MaNGA overlaid in the same color scale. The stellar velocity is aligned with the bulk of the molecular gas velocity. The faint component to the lower left is blueshifted with respect to the surrounding stellar velocity field, indicative of an outflowing component. Bottom left: CO (3--2) moment 1 map, with optical emission line velocity contours from MaNGA overlaid in the same color scale. The ionized gas is disturbed with respect to the stellar velocity. The blueshifted CO component is aligned with the blueshifted ionized gas region, indicating a multiphase outflow.  Bottom right: Comparison of CO (3--2) moment 0 map with \textit{HST} F625W image. The blueshifted component to the lower left is aligned with a dust lane in the \textit{HST} image.
}
\label{fig:s02}
\end{figure*}

S02 has a blueshifted component to the south east of the CO (3--2) centroid. We explore the nature of this component further in Figures \ref{fig:s02_grid}, \ref{fig:s02_pv}, and \ref{fig:s02}. A channel map of the CO (3--2) observations is shown in Figure \ref{fig:s02_grid}, overlaid on the moment 0 map. The blueshifted component at velocities -64 -- -108 km s$^{-1}$ is roughly aligned with the rotational axis of the galaxy, but extends to larger radius (1\arcsec, 460 pc)) and does not have a symmetric redshifted component on the northwest side, as we would expect if this component were caused by rotation instead of an outflow.

We fit a series of tilted 3D ring models to S02 using Barolo \citep{barolo} to model the kinematics of the molecular gas. The position-velocity diagram of the data and best-fit model are shown in Figure \ref{fig:s02_pv}. The best-fit kinematic major axis of the molecular gas is consistent with that of the stellar velocity. In the position-velocity diagram, the excess flux in the most blueshifted component at $\sim100$ \kms can be seen, inconsistent with the model. This component can also be seen in the asymmetric double peak of the integrated spectrum shown in Figure  \ref{fig:co32_spec}.

We compare the CO (3--2) observations for S02 with archival \textit{HST} imaging (PI: Zabludoff, ID 11643) and MaNGA \citep{manga} IFU data in Figure \ref{fig:s02}. The stellar velocity field from MaNGA is aligned with the bulk of the molecular gas velocity. The southeastern component is blueshifted with respect to the surrounding stellar velocity field, indicative of an outflowing component. The ionized gas traced by the optical emission lines\footnote{The velocities for all optical emission lines are fixed to one another during the fitting process \citep{Westfall2019}.} is disturbed in morphology with respect to the stellar velocity, with the ionized gas blueshifted by $\sim40$ \kms relative to the stellar velocity field. The ionized gas velocity can be measured with good signal to noise, with a median SNR$=4$ over the entire MaNGA field of view, and SNR$=15-20$ over the blueshifted component. The blueshifted CO component is aligned with the blueshifted ionized gas region, indicating a multiphase (both ionized and molecular gas) outflow. When we compare the location of this component to the \textit{HST} image, it is aligned with a dust lane. It is unlikely that the dust lane is the source of the abnormal ionized gas velocity field due to extinction effects, as the blueshifted ionized gas component extends several arcseconds outside of the dust lane. Instead, the dust may be tracing the outflow similar to the case seen in the post-starburst galaxy IC 860 by \citet{Luo2022}. \citet{Luo2022} observe a neutral gas outflow traced by NaD that aligns with a dust feature, with tentative evidence for a molecular gas outflow, consistent with the correlation of dust extinction and neutral gas outflows observed for AGN \citep{Veilleux2005}. 

Thus, while the blueshifted component is aligned with the stellar velocity field and bulk rotation of the molecular gas, it is more likely to be an outflow than an extension of the molecular gas rotation due to (1) the spatially-distinct morphology of the outflow (see in Figure \ref{fig:s02_grid}), (2) the asymmetric nature and lack of a counterpart on the northwest side, as evidenced by the excess flux above the best-fit Barolo model (see Figure \ref{fig:s02_pv}), and (3) the velocity excess over the surrounding stellar velocity field (see Figure \ref{fig:s02}). The blueshifted component is unlikely to be a parcel of infalling gas from the recent merger, as the smooth stellar velocity field suggests any merger components have already coalesced. In gas-rich early type galaxies that have experienced a recent minor merger, position-velocity diagrams do not display as significant of asymmetric components as seen here \citep{vandevoort2018}, meaning a merger where the stellar field has coalesced more quickly than the molecular gas is unlikely.

We consider the bulk properties of this component, assuming a distance from the center of the galaxy of 1\arcsec (460 pc) and a velocity of 100 km s$^{-1}$, based on the blueshifted component's position in the channel map and position-velocity diagram. If this component is an outflow, its characteristic timescale $\tau$ is $4.5$ Myr. The south east component consists of $\sim15$\% of the total flux. Scaling the CO (1--0) inferred molecular gas mass from \citet{French2015} by 15\%, the expected mass in this outflowing component is 7.3$\times10^7$ \Msun, and the average outflow rate over the timescale of the outflow is 16 \sfrunit. 
This is consistent with the (large) range of typical molecular gas depletion rates inferred for the combined post-starburst population by \citet{French2018}. The mass traced by the molecular outflow will dominate over the mass in the ionized gas component, as ionized gas outflow rates are typically low $\sim10^{-2}$ \sfrunit \citep{Baron2019}.

In order to determine the likely fate of this outflowing gas, we compare the outflow velocity to the escape velocity at its radius of 1\arcsec (460 pc). We use the stellar mass from the SDSS \citep{Strauss2002,Brinchmann2004, Tremonti2004} and the {\it HST} F625W observations to estimate the stellar mass profile of the galaxy. The fraction of stellar mass within the central 1\arcsec\ is 0.17, which we use to scale the total mass from SDSS. The estimated stellar mass within this radius is  M$_\star = 2.04\times10^9$ M$_\odot$. The escape velocity is thus 195 \kms. The outflow velocity of $\sim100$ km s$^{-1}$ is less than the escape velocity, but given inclination effects, if the angle of the outflow from our line of sight is $>60^\circ$, may exceed the escape velocity from the center of the galaxy.

The inferred outflow kinetic power using $P = \frac{1}{2} M_{\rm outflow} v^2 /\tau $ is $8\times10^{40}$ erg s$^{-1}$. The molecular outflow is comparable to those in star-forming galaxies and LINERs, and has less kinetic power than those in luminous AGN \citep{Cicone2014}. Another source of energy may be tidal disruption events (TDEs), which occur at higher rates in post-starburst galaxies than normal galaxies \citep{French2016,French2020}. As in \citet{Smercina2022}, we consider the energy input of $\sim10^{51}-10^{53}$ erg per TDE \citep{Mockler2021} and a typical TDE rate of $10^{-3}$ per year per post-starburst galaxy \citep{French2016}, resulting in a total energy source of $\sim3\times10^{40-42}$ erg s$^{-1}$. The feasibility for energy driving this outflow to come from intermittent AGN/LINER activity or from TDEs will depend heavily on the coupling of energy from the AGN/LINER or TDE to the molecular gas in these galaxies.

\begin{figure*}
\centering
\includegraphics[width=0.8\textwidth]{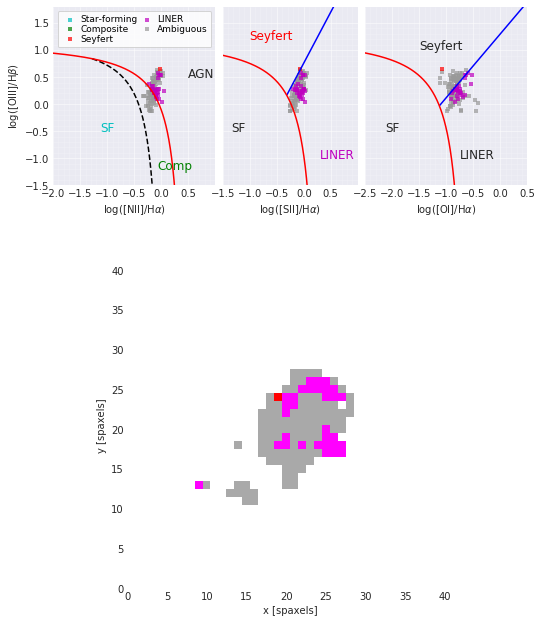}
\caption{Resolved BPT \citep{Baldwin1981} diagrams (top row) for S02, shown with diagnostic lines from \citet{Kewley2001} and \citet{Kauffmann2003}. Each spaxel is classified using its emission line ratios. The bottom plot shows the spatial distribution of the spaxels colored by classification. Most of the galaxy has low emission line fluxes such that the classification is ambiguous, but the spaxels that can be classified are in the LINER part of the BPT diagram (with one Seyfert spaxel). This LINER-like signature extends outside of the nucleus over $\sim5$ arcsec (10 spaxels), significantly more than the 2.5 arcsec FWHM of MaNGA's spatial resolution \citep{manga_drp}. It is thus more likely to be caused by post-AGB stars \citep{Yan2012} or shocks \citep{Rich2015} than current AGN activity. This galaxy does not have current AGN activity. However, the flickering timescale for AGN is shorter than the characteristic timescale of the outflow, so the outflow could have been launched by a previous episode of AGN activity.
}
\label{fig:s02_bpt}
\end{figure*}

The MaNGA data for this source also allow us to investigate the nature of the LINER-like emission seen in the SDSS spectrum. Using the MaNGA data to construct a resolved BPT \citep{Baldwin1981, Kewley2001, Kauffmann2003} diagram, most of the galaxy has low emission line fluxes such that the classification is ambiguous, but the spaxels that can be classified are in the LINER part of the BPT diagram (Figure \ref{fig:s02_bpt}). This LINER-like signature extends outside of the nucleus over $\sim5$ arcsec (10 spaxels), significantly more than the 2.5 arcsec FWHM of MaNGA's spatial resolution \citep{manga_drp}. It is thus more likely to be caused by post-AGB stars \citep{Sarzi2010, Yan2012} or shocks \citep{Rich2015} than low luminosity AGN activity. However, if the ``ring" like structure is a real feature, it could be an echo of previous nuclear activity. Given the light travel time from the nucleus to the edge of this feature, the echo would trace nuclear activity on a timescale of $\tau\sim3000$ years ago. Further data would be required to determine if this ring like structure is real, due to the low signal to noise of these weak emission lines, especially for H$\beta$ and in the center of the galaxy.

While this galaxy does not show evidence for ongoing luminous AGN activity, the timescale for AGN to vary is much smaller than the characteristic timescale of this outflow. AGN are observed to change dramatically, turning on and off on timescales $\sim10^4-10^5$ years \citep{Keel2012,Sartori2018,Shen2021}. The molecular gas outflow we observe would have been launched 4.5 Myr ago, when the galaxy could have had now-faded AGN activity.

\subsection{Evolution During the Post-Starburst Phase}
\label{sec:age}

The evolution of the molecular gas state during the post-starburst phase can provide more information than comparing the average properties to other classes of galaxies. We found in \citet{French2018} that the CO-traced molecular gas fraction declined during the post-starburst phase, after 90\% of the starburst is complete, with a timescale of $\sim200$ Myr. A similar decline has been seen in the CO-traced molecular gas in higher redshift post-starburst galaxies \citep{Bezanson2021, Suess2021}. The low SFRs during this period of decline cannot be responsible for depleting this gas via consumption or stellar feedback, suggesting that another mechanism like AGN feedback is operating late into the post-starburst phase. The dust fraction is also observed to decline during the post-starburst phase \citep{Smercina2018, Li2019}, with a timescale consistent with that of the CO-traced molecular gas \citep{Li2019}. \citet{Li2019} considered the evolution of the star formation efficiency (traced by SFR/($\alpha_{CO}$ \lpco), finding a rapid decline in SFE during the first 200 Myr of the post-starburst phase, which levels off to a shallower decline in after 200 Myr post-burst. This two-phase evolution in SFE implies a two-phase evolution in SFR and a faster initial timescale of SFR evolution compared to the CO-traced molecular gas. 

Such a two-phase evolution in SFR with post-burst age was observed by \citet{Rowlands2015} in both the H$\alpha$ and FIR-based SFRs. A short $\sim30$ Myr decline was followed by a plateau of $\sim 400$ Myr. The overall timescale for the SFR to drop was $\sim200-300$ Myr. 

\begin{figure*} 
\includegraphics[width=\textwidth]{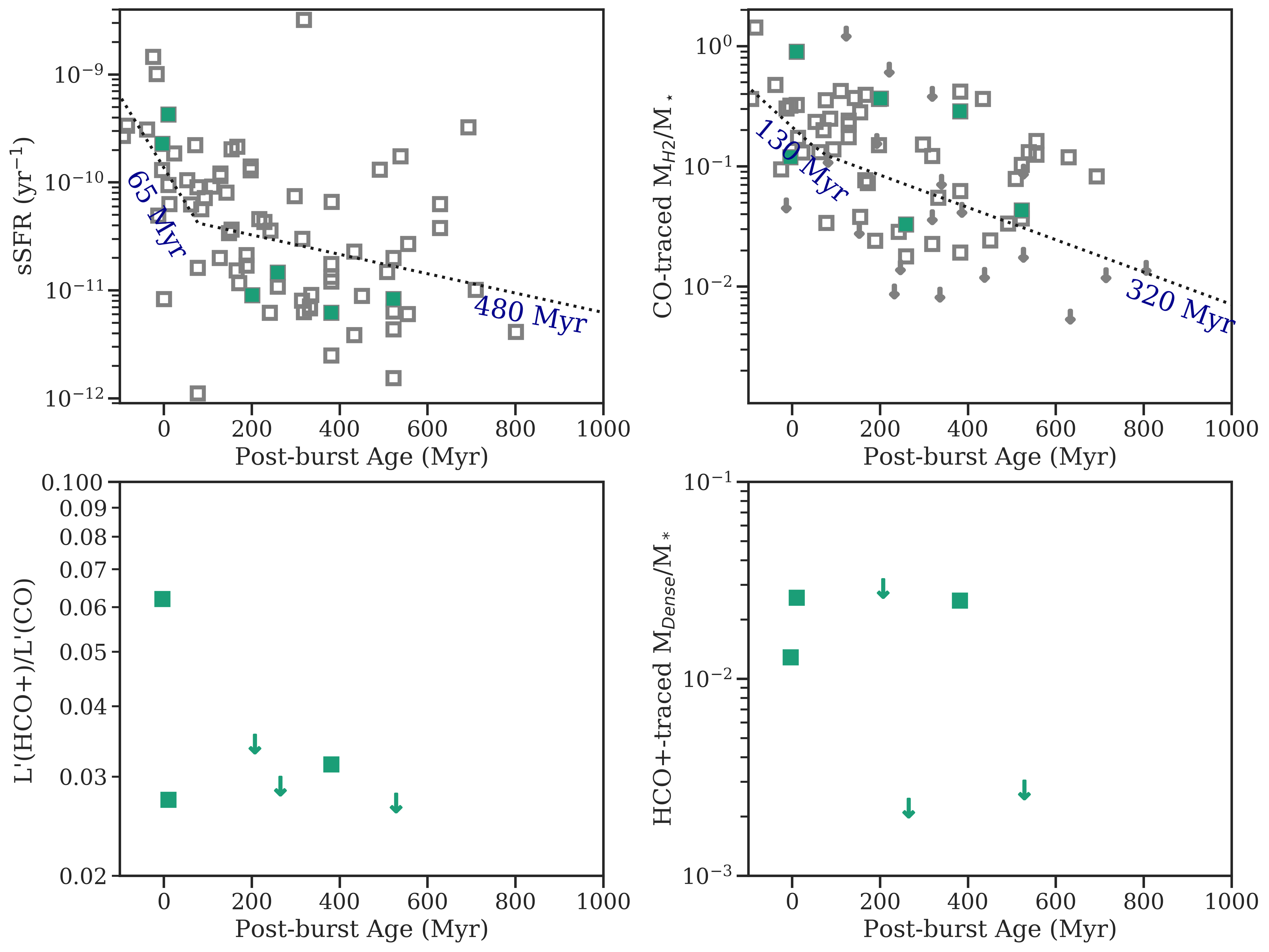}
\caption{Trends of specific SFR, molecular gas fraction (traced by CO (1-0)), and dense gas fraction (traced by HCO$^+$/CO (1-0) vs. post-starburst age \citep{French2018}. The ALMA targets considered in this study are highlighted in green.  (Top left): SFR (traced by H$\alpha$) for galaxies with CO observations from \citet{French2015}, \citet{Rowlands2015}, and \citet{Alatalo2016b}. We fit these data using a two-component exponential decline (dotted line), finding a significant difference in the early vs. late time slopes (timescales indicated on figure). (Top right): CO-traced molecular gas to stellar mass fraction (assuming $\alpha$CO = 4) for galaxies from \citet{French2015,Rowlands2015,Alatalo2016b}. Using the same two-component model, we observe no significant difference in the early vs. late timescales in the CO-traced gas fraction. The time evolution of the CO-traced gas fraction is consistent with a single component model, unlike the sSFR trend. (Bottom left): Dense gas fraction traced by HCO$^+$/CO for the six PSBs considered here. Dense gas ratio (HCO$^+$/CO) suppression occurs throughout PSB phase and begins early ($\sim$10-200 Myr after burst ends). Observations of a larger sample will be required to determine whether the dense molecular gas has a rapid early decline similar to that seen in the sSFR, or a slower decline throughout the post-starburst phase more like the evolution seen in the CO-traced gas. (Bottom right): The dense gas mass fraction normalized by stellar mass does not appear to decline as steeply as the HCO$+$/CO ratio, as the dense gas mass fraction of H03 remains high.}
\label{fig:age}
\end{figure*}

We illustrate the evolution of the specific SFR (sSFR) and various tracers of the molecular gas in Figure \ref{fig:age}. We plot the sSFR traced by H$\alpha$ for the three post-starburst samples from \citet{French2015, Rowlands2015, Alatalo2016b}. Post-burst ages are from \citet{French2018} and measure the time elapsed since the end of the recent starburst (the age since 90\% of the stars were formed\footnote{The post-burst age will be shorter than the time since the beginning of the burst and results in some cases of small negative ages in galaxies for which the best-fit SFH has a still-declining burst}). The sSFR trend qualitatively appears to follow that expected from \citet{Rowlands2015} and \citet{Li2019}, with a rapid initial decline that levels off at later times. \citet{Rowlands2015} observed a third phase of more rapid decline after 400 Myr in the H$\alpha$ SFRs, which we do not observe here. We see no evidence of a significant correlation between age and sSFR for ages $>400$ Myr using a Spearman or Pearson correlation test. This may be driven by differing treatments of the SFR in galaxies in the AGN portion of the BPT diagram in the limiting cases of very low SFR. 

We explore the evolution during the post-starburst phase quantitatively by fitting the sSFR and CO-traced molecular gas fraction trends with two-component linear fits to $x - ln(y)$, with the two lines required to meet in the middle. This model has four free parameters: the early slope, the late slope, the break point, and a y-offset. We assume a constant uncertainty in sSFR of 0.1 dex and take into account the measured molecular gas fraction uncertainties and upper limits. For this analysis, we do not include the uncertainty on age. For the sSFR - age comparison, the best fit early slope is 65 Myr, the best fit late slope is 480 Myr, and the pivot point is 77 Myr. The two-slope fit is significantly preferred over a single-slope fit using either the reduced $\chi^2$ ($\chi^2/\nu_{1 \ \rm{slope}} = 208$ vs. $\chi^2/\nu_{2 \  \rm{slope}} = 19$) or Bayesian Information Criterion (BIC) ($\rm{BIC}_{1 \  \rm{slope}} =1.5\times10^4$ vs. $\rm{BIC}_{2  \ \rm{slope}} =1326$) tests. The best fit early slope is comparable to the typical duration of the recent bursts \citep{French2018} and may be a continuation of the decline in SFR as the burst ended. Given the uncertainties on the derived parameters, the early and late slopes are significantly different. In contrast, fitting the same two-component function to the CO-traced gas fraction - age comparison, there is no significant difference between the early and late slope. We see no evidence to support a two-phase evolution in the decline of the CO-traced gas with time. These lines are overplotted on Figure \ref{fig:age}.

The dense molecular gas to total molecular gas fraction, traced by the \lhco/\lpco ratio, is proportional to the fraction of the molecular gas reservoir in the denser states probed by HCO$^+$. In several cases, the post-starburst galaxies show evidence that the \lhcn luminosity is enhanced relative to the \lhco, similar to many AGN and ULIRGs (see further discussion in \S\ref{sec:densegasstate}), so we consider \lhco here as a more accurate tracer of the dense gas mass. Most (5/6) of the post-starburst galaxies have low (\lhco/\lpco $<0.04$) fractions, except for R02, which is the youngest post-starburst galaxy in our ALMA-targeted sample. This may indicate a rapid ($<<100$ Myr) decline in the dense gas fraction at the start of the starburst, but a larger sample will be required to determine whether this trend is significant, as it is driven by the observations for a single galaxy. If we instead consider the dense gas to stellar mass fraction, using an $\alpha_{HCO^+}=10$ \unitxco, \ \footnote{The conversion factor $\alpha_{dense} = M_{dense}/L_{dense}$ will scale as $\alpha_{dense} \approx 2.1 \frac{\sqrt{\langle n \rangle}}{T_b}$, where $n$ is the density of H$_2$ molecules and $T_b$ the brightness temperature \citep{Papadopoulos2007}. \citet{Gao2004a} use this argument to estimate the conversion factor for HCN given the expected conditions, resulting in a value of $\alpha_{HCN}=10$ \unitxco. The dense gas luminosity to mass conversion factor should be similar for both HCO$^+$ and HCN (1--0), as they trace similar gas conditions, so we use this same conversion factor for HCO$^+$. The result shown in Figure \ref{fig:age} will only be affected by differences in $\alpha_{dense}$ if there is significant variation from source to source. } this rapid decline is not observed. These differing trends may be caused by scatter in the CO-traced gas fraction amongst the sample. Observations of a larger sample will be required to determine whether the dense molecular gas has a rapid early decline similar to that seen in the sSFR, or a slower decline throughout the post-starburst phase more like the evolution seen in the CO-traced gas.

\subsection{Star Formation in the Dense Gas}
\label{sec:sfinthedensegas}

\begin{figure*}
\includegraphics[width=\textwidth]{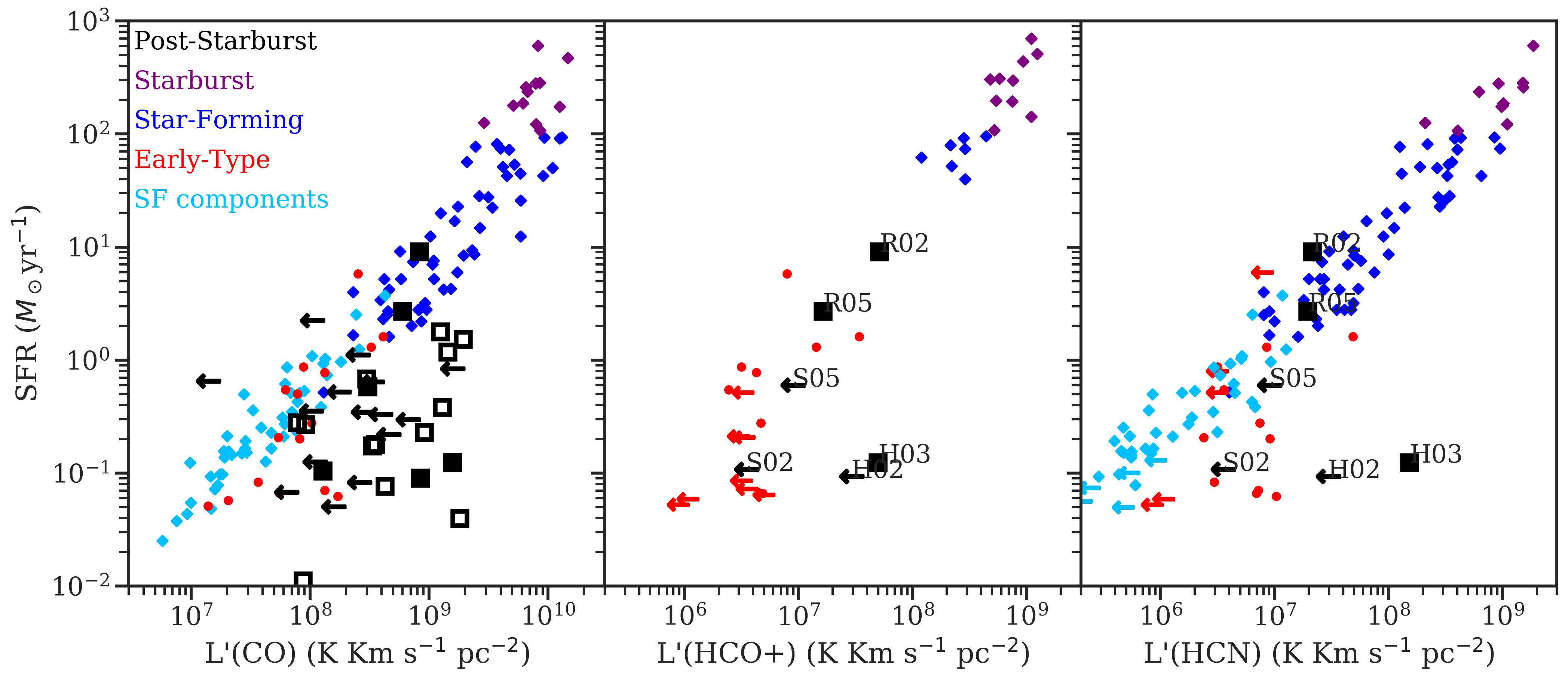}
\caption{(Left:) \lpco vs. SFR. Early type galaxies \citep{Crocker2012}, starburst and star-forming galaxies \citep{Gao2004}, and star forming galaxy components \citep{Usero2015a} are correlated with low scatter, but post-starburst galaxies \citep{French2015} have low SFRs for their CO luminosities. Black squares indicate post-starburst detections and arrows indicate $3\sigma$ upper limits. Filled squares indicate the ALMA targets considered here (including two galaxies from the \citet{Rowlands2015} sample). SFRs for the post-starburst galaxies are measured using H$\alpha$ emission (see \S\ref{sec:sdss}).  (Middle:) \lpco vs. SFR for the same samples of galaxies. (Right: )\lhcn vs. SFR for the same samples of galaxies. Uncertainties in post-starburst \lhcn and \lhco luminosities are smaller than the plotted symbols. The post-starburst galaxies are more consistent with the comparison samples in their dense gas - star formation relations, with the exception of H03, for which different SFR tracers predict a wide range of SFRs (see full discussion in \S\ref{sec:sfr}. The post-starburst galaxies have dense molecular gas properties consistent with either early type galaxies or lying between the star-forming and early type samples.
}
\label{fig:gas_sfr}
\end{figure*}

Previous work on the dense gas in post-starburst galaxies were motivated by the high CO luminosities relative to the low SFRs. In Figure \ref{fig:gas_sfr}, we plot \lpco vs. SFR for post-starburst galaxies and comparison samples of other galaxy types. The comparison star-forming and starbursting galaxies from \citet{Gao2004}, sub-regions of nearby star-forming galaxies from \citet{Usero2015a} and early type galaxies from \citet{Crocker2012} show SFRs highly correlated with \lpco luminosities, despite the varying ranges of stellar masses and redshifts. We consider the H$\alpha$-traced SFRs for the post-starburst sample here; the impact of SFR tracer on our results is explored further in S\ref{sec:sfr} and in Appendix \ref{sec:appendix_ir}. The post-starburst galaxies are systematically offset to higher \lpco relative to other quiescent galaxies \citep{French2015}. The two highest SFR post-starbursts shown are R02 and R05, from the \citet{Rowlands2015} sample, which uses a different selection method (see \S\ref{sec:sample}). These are also the youngest post-starburst galaxies considered here and have \lpco values consistent with the SFRs. The location of these young post-starbursts on the \lpco--SFR relation is consistent with the evolution of the star formation efficiency ($\propto SFR/$\lpco) seen by \citet{Li2019}. 

\begin{figure*}
\includegraphics[width=\textwidth]{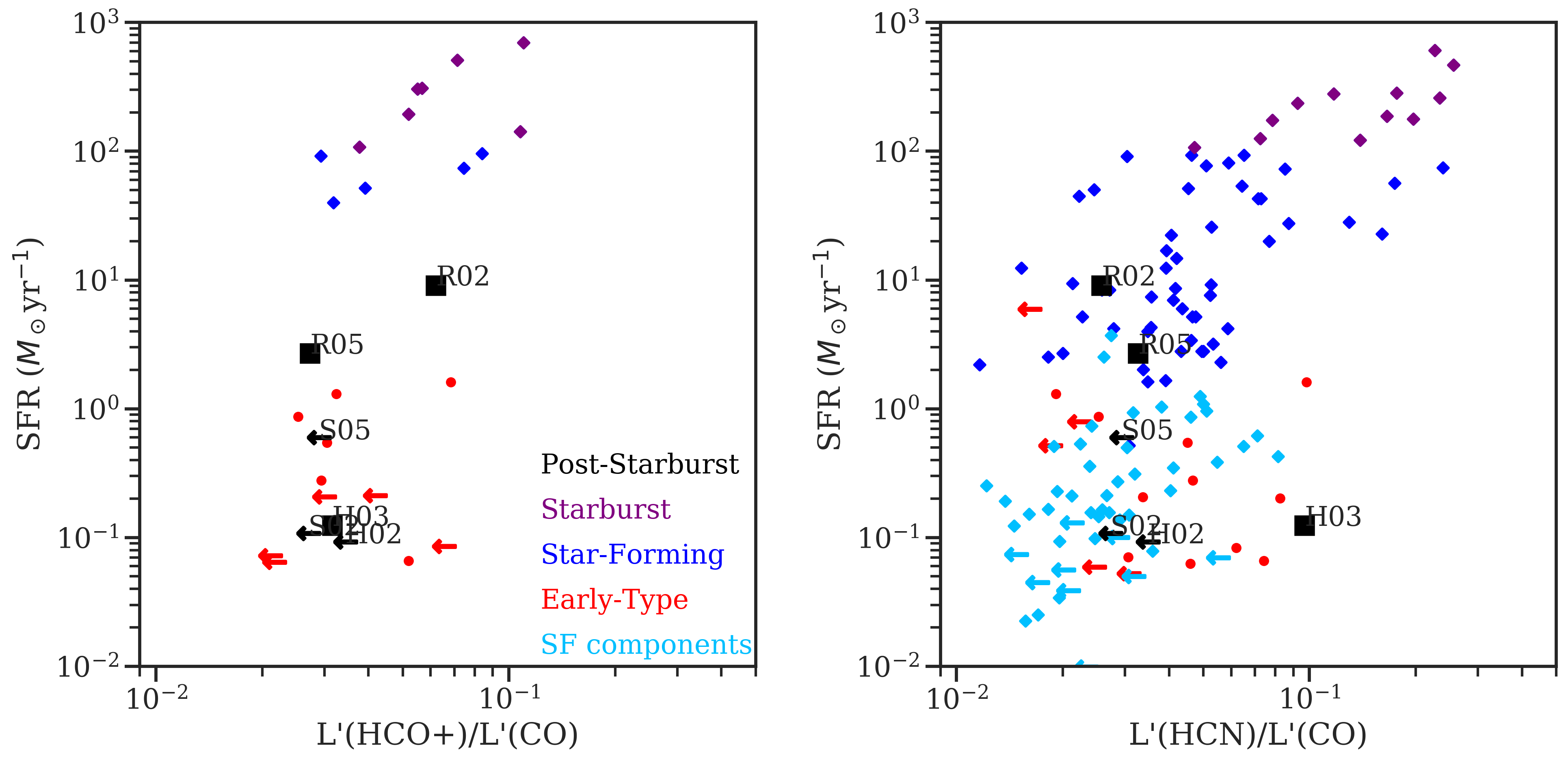}
\caption{(Left:) Dense gas luminosity ratio  \lhco/\lpco for post-starburst galaxies and comparison samples. Plot colors are the same as Figure \ref{fig:gas_sfr}. Post-starburst galaxies have low ratios of dense molecular gas to total molecular gas, more similar to early type galaxies and the low end of the star-forming galaxies than the starburst galaxies, although this may evolve rapidly with time during the early post-starburst phase (Figure \ref{fig:age}). (Right:) Dense gas luminosity ratio \lhcn/\lpco for the same samples.  The post-starburst galaxy with high HCN/CO (H03) may have HCN luminosity increased via mechanical heating or cosmic ray heating, as it has a very high HCN/HCO$+$ ratio (see discussion in \S\ref{sec:densegasstate}), and the HCN/CO ratio may overestimate the dense molecular gas fraction.}
\label{fig:dense_gas_co_ratios}
\end{figure*}

Our previous study of the dense gas in two post-starburst galaxies \citep[H02, S05][]{French2018b} found low limits on \lhcn and \lhco, consistent with their low SFRs. Here, we primarily consider the \lhco instead of \lhcn to trace the dense gas, as the HCN is likely overestimating the dense gas mass (see further discussion in \S\ref{sec:densegasstate}), although both are shown in Figures \ref{fig:gas_sfr} and \ref{fig:dense_gas_co_ratios}. For the four new post-starburst galaxies targeted in this study, we find that the two young post-starburst galaxies from the \citet{Rowlands2015} parent sample (R02 and R05) also have dense gas luminosities consistent with their SFRs (Figure \ref{fig:gas_sfr}). For the two older post-starburst galaxies from the \citet{French2015} parent sample, S02 is not detected in either HCN$+$ or HCN, at levels either consistent with its low SFR or slightly above. H03 remains offset from the \lhco--SFR relation, but the SFR of this galaxy is highly uncertain (see discussion in \S\ref{sec:sfr}).

We consider the ratio of dense molecular gas to total molecular gas traced by \lhco/\lpco in Figure \ref{fig:dense_gas_co_ratios}. 5/6 of the post-starburst galaxies have low values of \lhco/\lpco, lower than at least 63\% of the comparison galaxies, except for the youngest post-starburst galaxy in the sample, R02. As discussed in \S\ref{sec:age}, this may be an evolutionary effect with post-burst age. In \citet{French2018b}, we observed low HCN/CO ratios from two of the HCN limits plotted here. With the addition of the four new galaxies, as well as the addition of the comparison sample of star-forming galaxy components from \citet{Usero2015a}, we see that 5/6 of the post-starbursts have low HCN/CO ratios relative to the comparison samples, though there are some star-forming galaxy components with lower HCN/CO ratios. 5/6 of the post-starburst galaxies have HCN/CO ratios lower than at least 70\% of the comparison samples, with the exception of H03. The post-starburst galaxy with high HCN/CO (H03) may have HCN luminosity increased via mechanical heating or cosmic ray heating, as it has a very high HCN/HCO$+$ ratio (see discussion in \S\ref{sec:densegasstate}). Given the low HCO$+$ luminosity of this galaxy, it is likely that the dense gas fraction is low. We also compare the HCO$+$/CO and HCN/CO ratios for the post-starbursts and comparison samples using a Mann-Whitney U test \citep{MannWhitney1947}. We can exclude the possibility that the post-starbursts and comparison samples are drawn from the same distribution in HCO$+$/CO at the $2\sigma$ level, but cannot exclude the possibility that the HCN/CO distributions are the same. The HCN/CO distributions may be more difficult to distinguish due to the high HCN/CO ratio observed for H03.

\subsection{Dense Gas State}
\label{sec:densegasstate}

\begin{figure}
\includegraphics[width=0.5\textwidth]{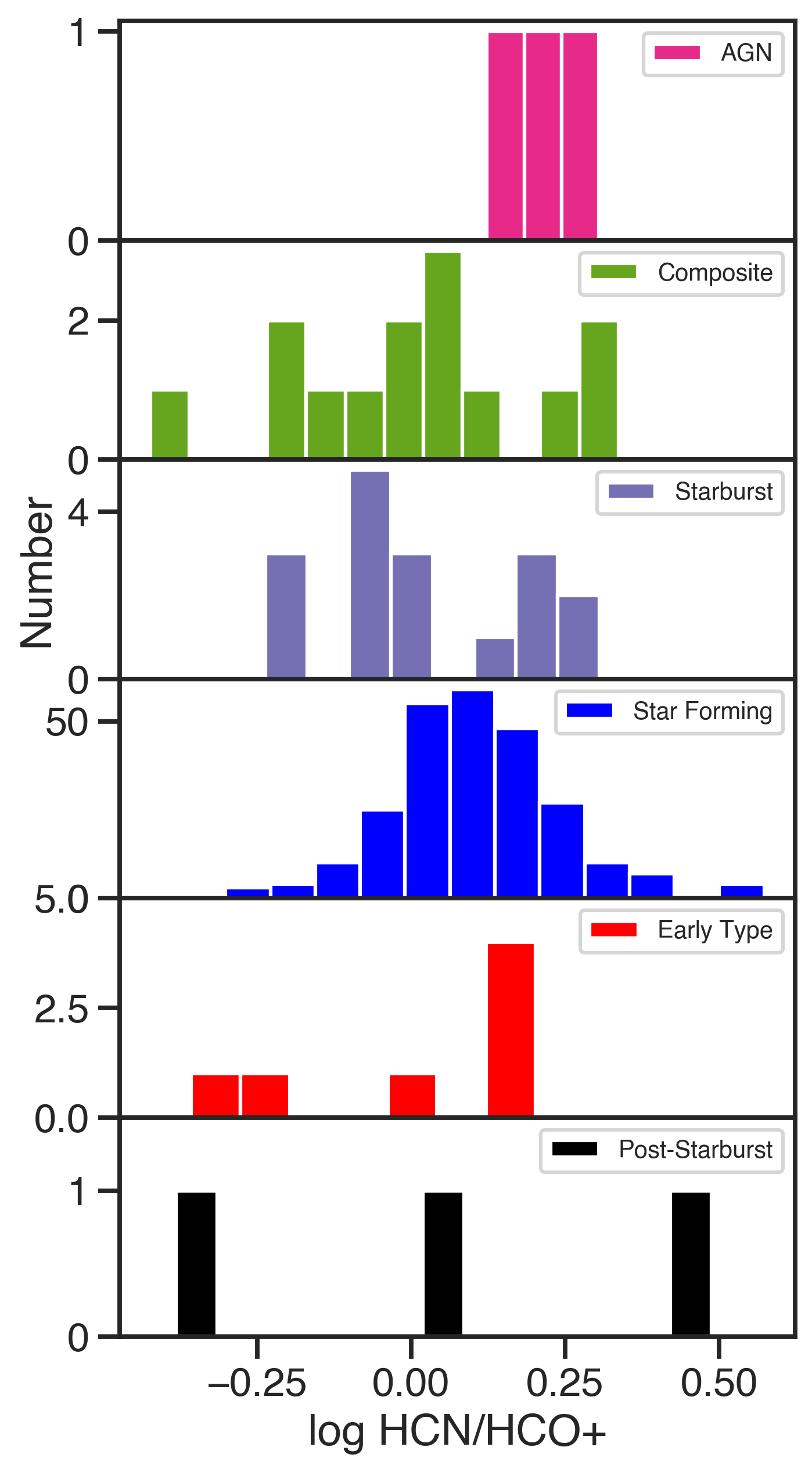}
\caption{Histograms of HCN/HCO$^+$ ratios for different galaxy types. LIRGs from \citet{Privon2015} are divided in to AGN, composite, and starbursting based on their PAH 6.2$\mu$ EWs \citep{Stierwalt2013}. Components of massive star-forming galaxies are from \citet{Jimenez2019}. Early type galaxies are from \citet{Crocker2012}. Only galaxies with significant detections in both lines are plotted here. \citet{Privon2015} found that while the HCN/HCO$^+$ ratio is enhanced in AGN, likely due to mechanical heating, other galaxies may have high HCN/HCO$^+$ ratios without signs of current AGN activity. This may be due to the short duration of AGN signatures compared to the effect on the dense gas, or other processes affecting the dense gas state. One post-starburst galaxy (H03) has a HCN/HCO$^+$ ratio as high as the AGN from \citet{Privon2015}, which may indicate buried or past AGN activity. One post-starburst (R02) has a lower HCN/HCO$^+$ ratio than almost all other comparison galaxies with both lines detected. The HCN/HCO$+$ ratios of the post-starburst galaxies vary considerably, with scatter across the entire sample of comparison galaxies. This scatter is much larger than the uncertainty on the line ratio measurements.
}
\label{fig:hcn_hco_histos}
\end{figure}

Ratios between the dense gas tracers HCN, HCO$+$, and HNC (1--0) are sensitive to various mechanisms that affect the reliability of these lines as tracers of the dense gas mass. AGN and some ULIRGs have high HCN/HCO$+$ ratios \citep{Kohno2001, Imanishi2004, Imanishi2007, Privon2015}, which have been attributed to IR pumping, XDR (X-ray dominated region)-dominated chemistry, mechanical heating, or cosmic ray heating \citep{Aalto2007, Loenen2008, Meijerink2011, Bayet2011, Privon2015, Privon2020}.

We compare the HCN/HCO$+$ ratios of the three post-starburst galaxies with both line measurements with other types of galaxies in Figure \ref{fig:hcn_hco_histos}. The HCN/HCO$+$ ratios of the post-starburst galaxies vary considerably, with scatter across the entire sample of comparison galaxies. This scatter is much larger than the uncertainty on the line ratio measurements. One of the PSBs (H03) has a very high HCN/HCO$+$ ratio (low HCO$^+$/HCN), even compared to AGN and ULIRGs.

\begin{figure*}
\includegraphics[width=\textwidth]{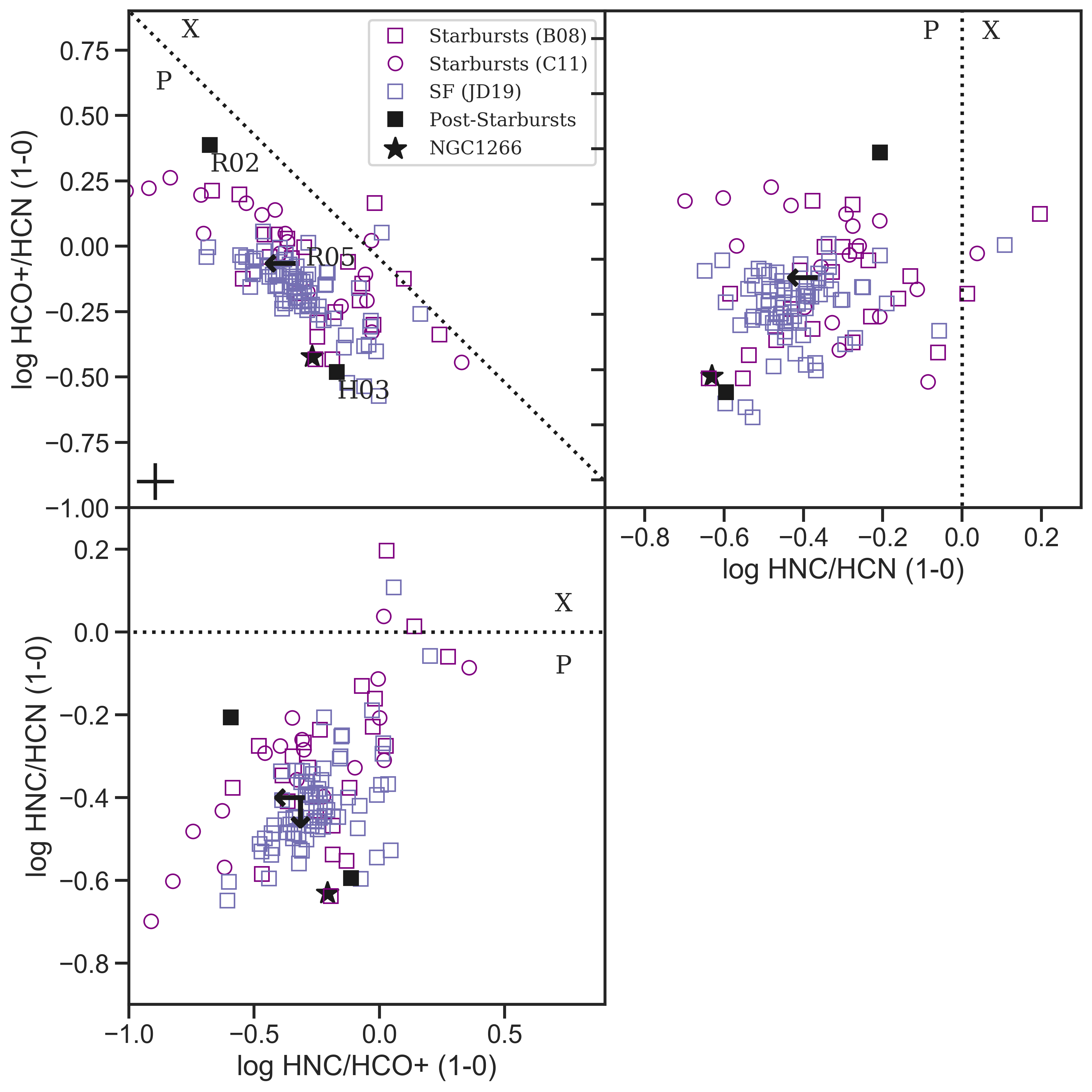}
\caption{Ratios of dense gas tracers HCN (1-0), HCO$^+$ (1-0), and HNC (1-0) \citep[adapted from][]{Baan2008, Loenen2008}. Post-starburst galaxies from this study are shown with comparison starburst samples \citep{Baan2008, Costagliola2011}, comparison components of massive star-forming galaxies \citep{Jimenez2019}, and the shocked post-starburst galaxy NGC 1266 (K. Alatalo, private communication). A characteristic errorbar is shown in the top left panel. The HNC/HCN ratio is sensitive to the ionization state of the ISM, and can distinguish XDR from PDR dominated regions (dotted lines). The PSBs, like most of the starbursts, are consistent with PDR-dominated ionization. HCN/HCO$^+$ ratio varies as much within the PSB sample as within ULIRGs. The low HCO$^+$/HCN ratios of the PSBs are consistent with mechanical heating as modelled by \citet{Loenen2008}. 
}
\label{fig:hcn_hnc_hco}
\end{figure*}

The HNC/HCN ratio is sensitive to the ionization state of the ISM, and can distinguish XDRs from PDRs (photon dominated regions). We plot the comparison of the HCN, HCO$+$, and HNC (1--0) ratios in Figure \ref{fig:hcn_hnc_hco}. The PSBs have similar HNC/HCN values as most of the comparison starbursts and star-forming galaxies, and are consistent with PDR-dominated ionization. 

\section{Discussion}

\subsection{Star Formation Rates}
\label{sec:sfr}

The measurement of current SFRs in post-starburst galaxies is complicated by a number of factors (see discussion in \citealt{French2021} \S5). For this work, our goal is to measure the current SFR (over timescales $<10$ Myr), with low contamination from other sources of excitation, and with accurate correction applied for dust extinction. Contamination from the young stellar populations, AGN, and shocks would cause SFRs to be over-estimated, while high dust obscuration beyond that probed by Balmer decrement corrections would cause the SFRs to be under-estimated. We consider here the comparison of the H$\alpha$-based SFRs used above (described in \S\ref{sec:sdss}) to other SFR tracers to assess the possible biases in these measurements.

\begin{figure*}
\includegraphics[width=0.5\textwidth]{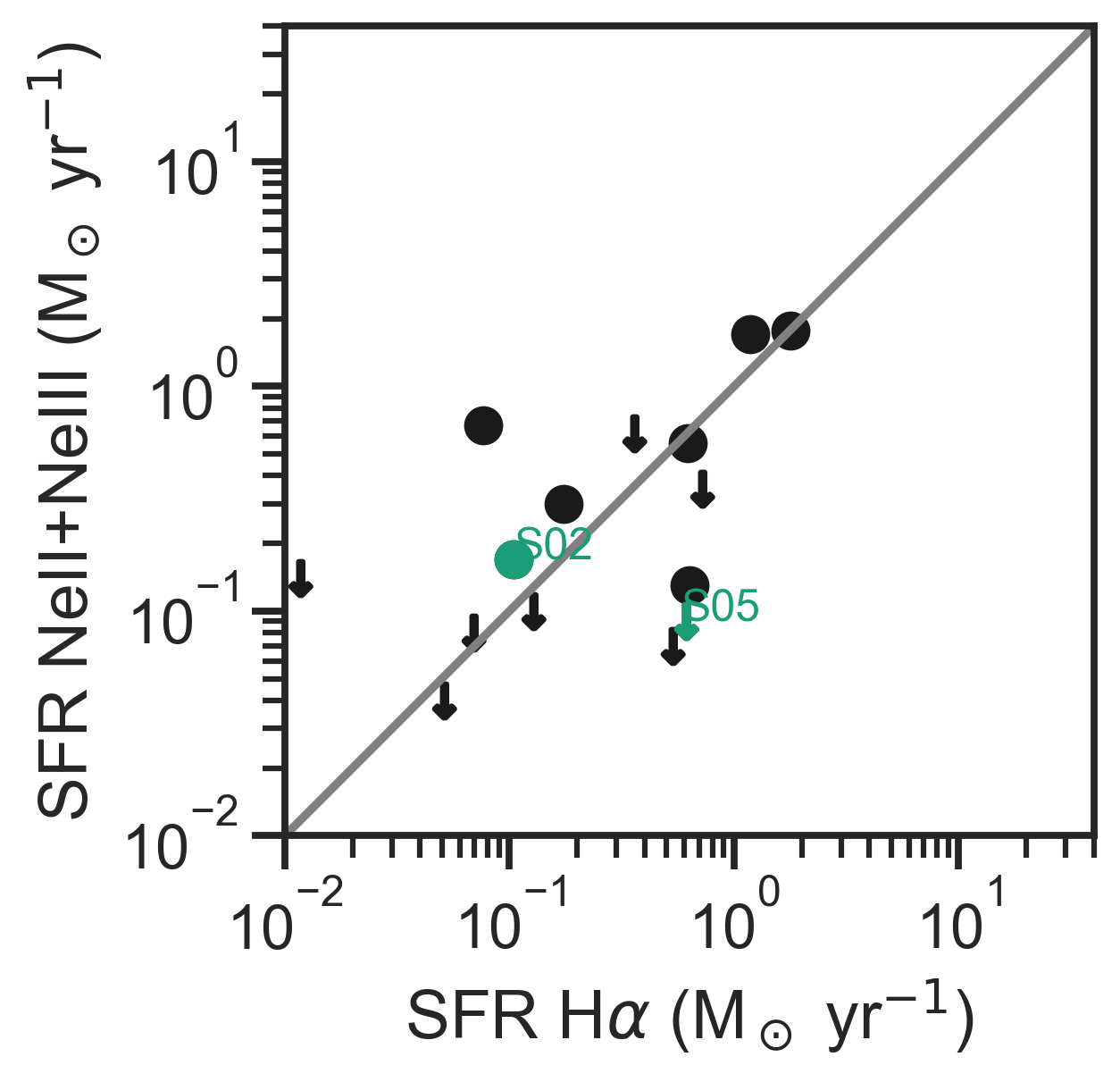}
\includegraphics[width=0.5\textwidth]{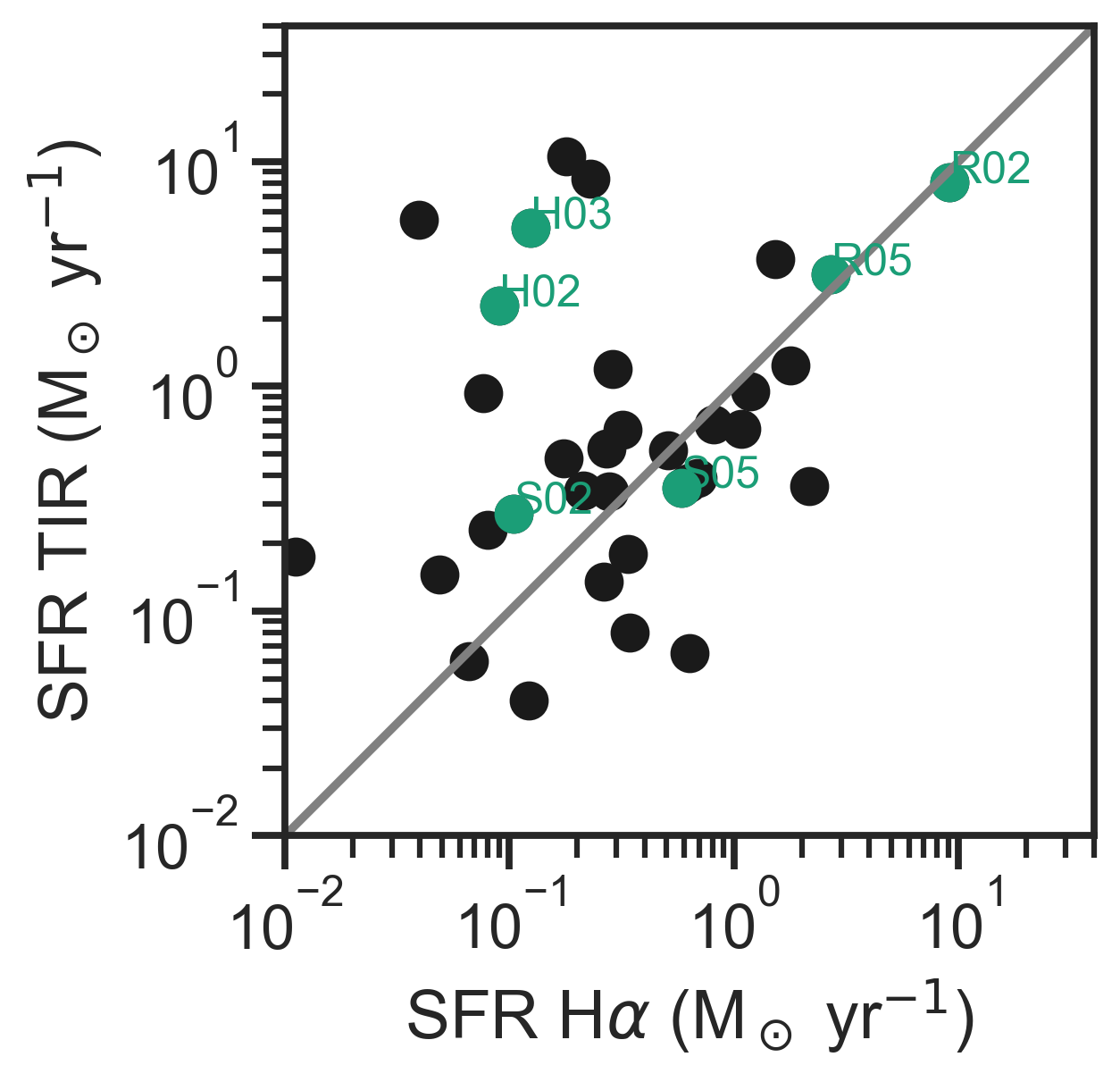}
\includegraphics[width=0.5\textwidth]{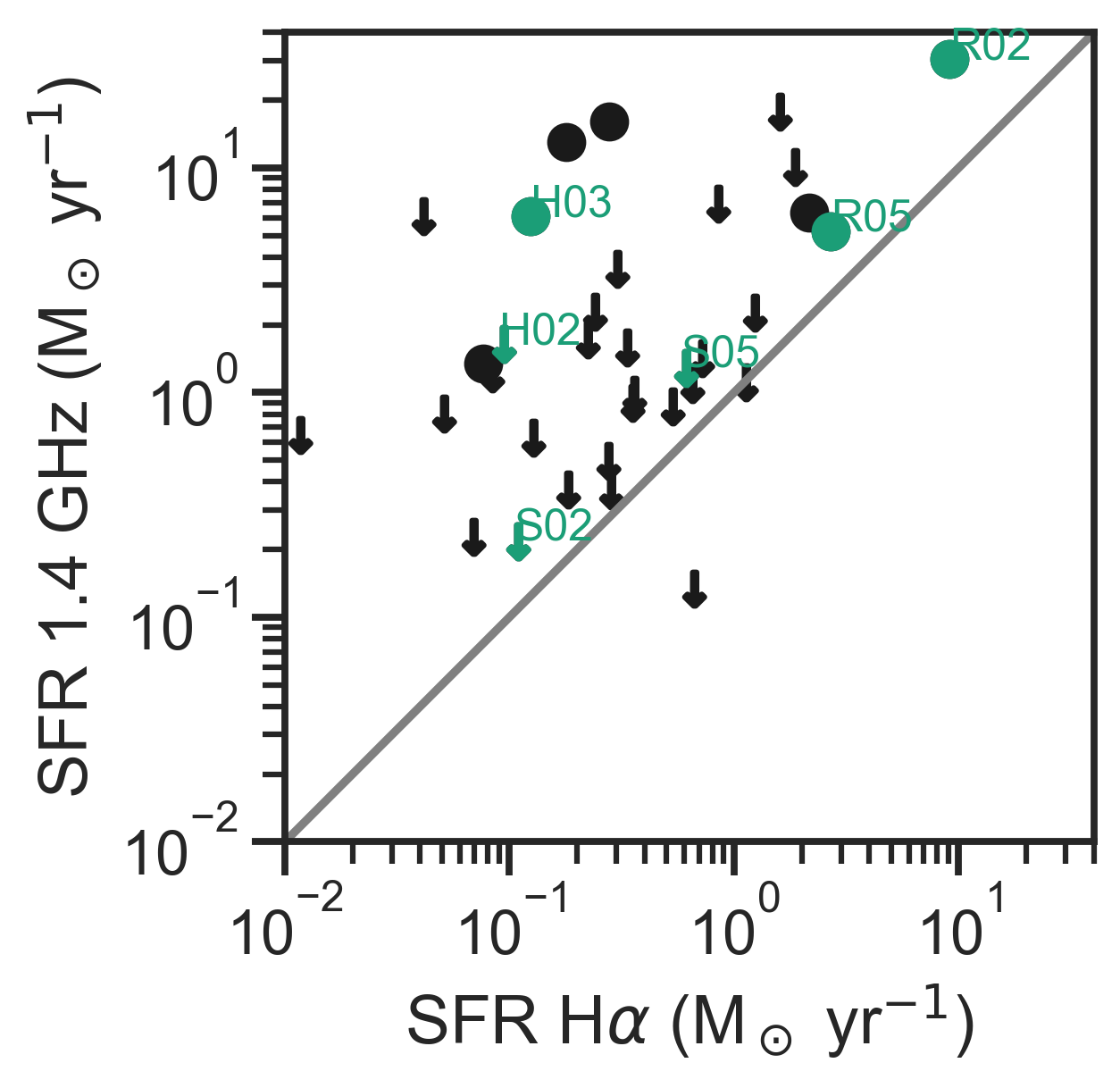}
\includegraphics[width=0.5\textwidth]{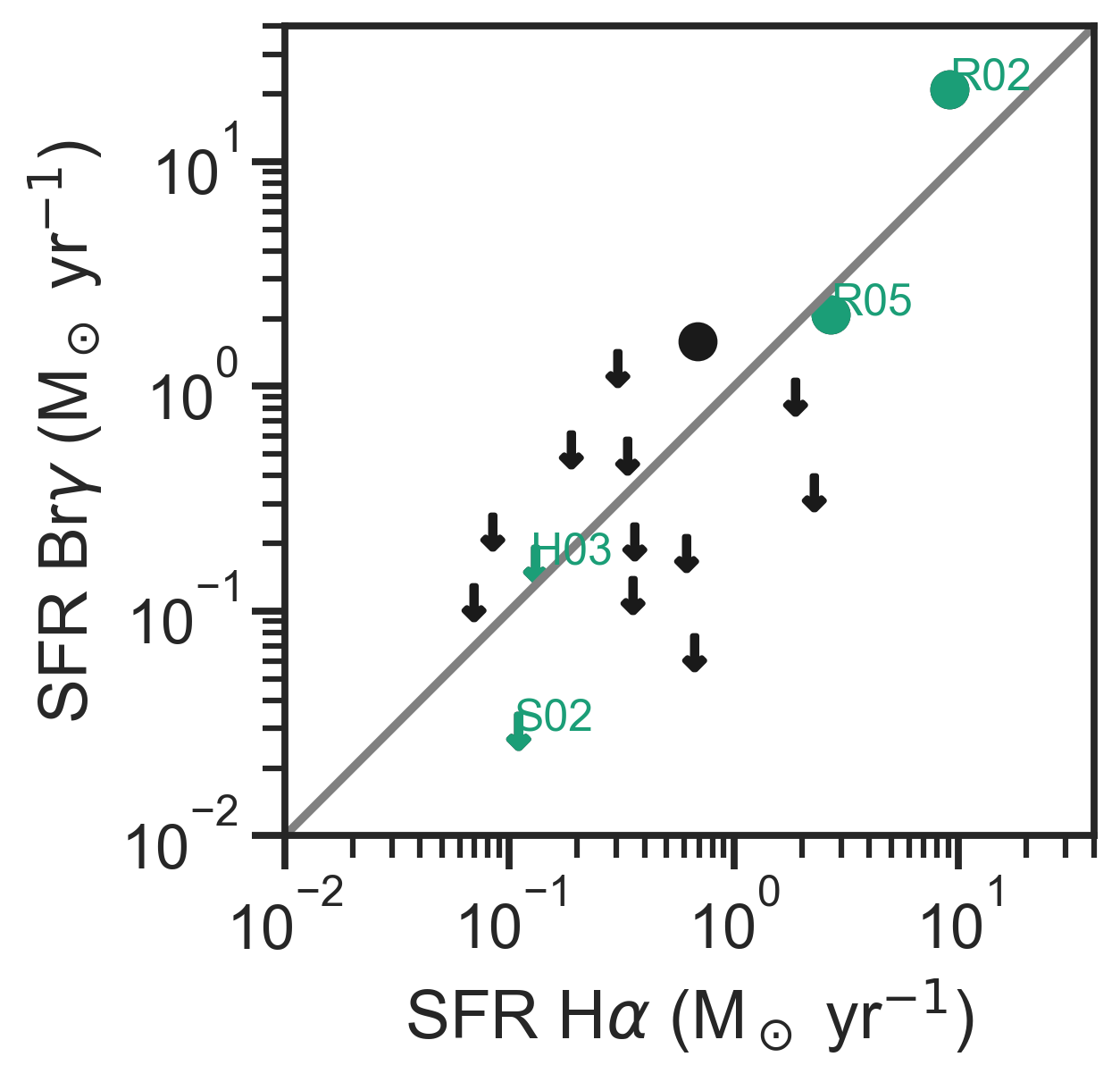}
\caption{Comparison of SFR tracers (discussed in \S\ref{sec:sfr}) with the H$\alpha$-based SFRs used elsewhere in this study. Post-starburst galaxies from \citet{French2015, Rowlands2015, Smercina2018} are shown in black, and the ALMA targets are highlighted in green. All limits are $3\sigma$ upper limits. The H$\alpha$ SFRs are consistent with short-duration infrared tracers from Neon and Br$\gamma$, indicating that the use of H$\alpha$ is not biased low due to dust obscuration. \citet{Smercina2018} observed a systematic offset between TIR and Neon SFRs, such that the TIR-based SFRs were on average $2\times$ higher than the Neon SFRs, likely due to dust heating by young stars. We apply this correction to the TIR SFRs used here. The 1.4 GHz flux is likely to be contaminated by any AGN or LINER activity in the post-starbust sample \citep[e.g.,][]{Moric2010}, especially for H03, given the large offset between the 1.4 GHz and Br$\gamma$ indicators. 
}
\label{fig:sfr_comp}
\end{figure*}

The combination of the [NeII] 12.8$\mu$m and [NeIII] 15.6$\mu$m lines traces HII regions with relatively low bias compared to other SFR tracers \citep{Ho2007,Whitcomb2020}. Dust extinction scales as $A_\lambda/A_V \sim \lambda^{-2}$ in the near- and mid-IR; using the extinction law measured by \citep{Wang2019}, the extinction at the wavelengths of [NeII] and [NeIII] is $500-800\times$ lower than $A_V$. Thus, the Neon-based SFRs will not be subject to underestimation due to dust obscuration, even in galaxies with ULIRG-like central dust obscuration. \citet{Smercina2018} measured Neon-based SFRs for a sample of 15 post-starburst galaxies, two of which (S02 and S05) are studied in this work. We compare the Neon-based SFRs to H$\alpha$-based SFRs in Figure \ref{fig:sfr_comp}. For the sample of seven post-starburst galaxies with Neon detections, we observe no systematic bias between the H$\alpha$- and Neon-based SFRs. Of the two galaxies considered here, the measurements of S02 agree well, and the Neon-based SFR for S05 is lower than the H$\alpha$ SFR. This indicates that neither galaxy has significant obscured star formation. In section \ref{sec:appendix_ir}, we consider the impact of using Neon-based SFRs on the Kennicutt-Schmidt relation, finding the post-starburst galaxies to lie offset from the relation formed by star forming and starburst galaxies, consistent with the results from using H$\alpha$-based SFR tracers.

\begin{figure*}[t]
\centering
\includegraphics[width=0.7\textwidth]{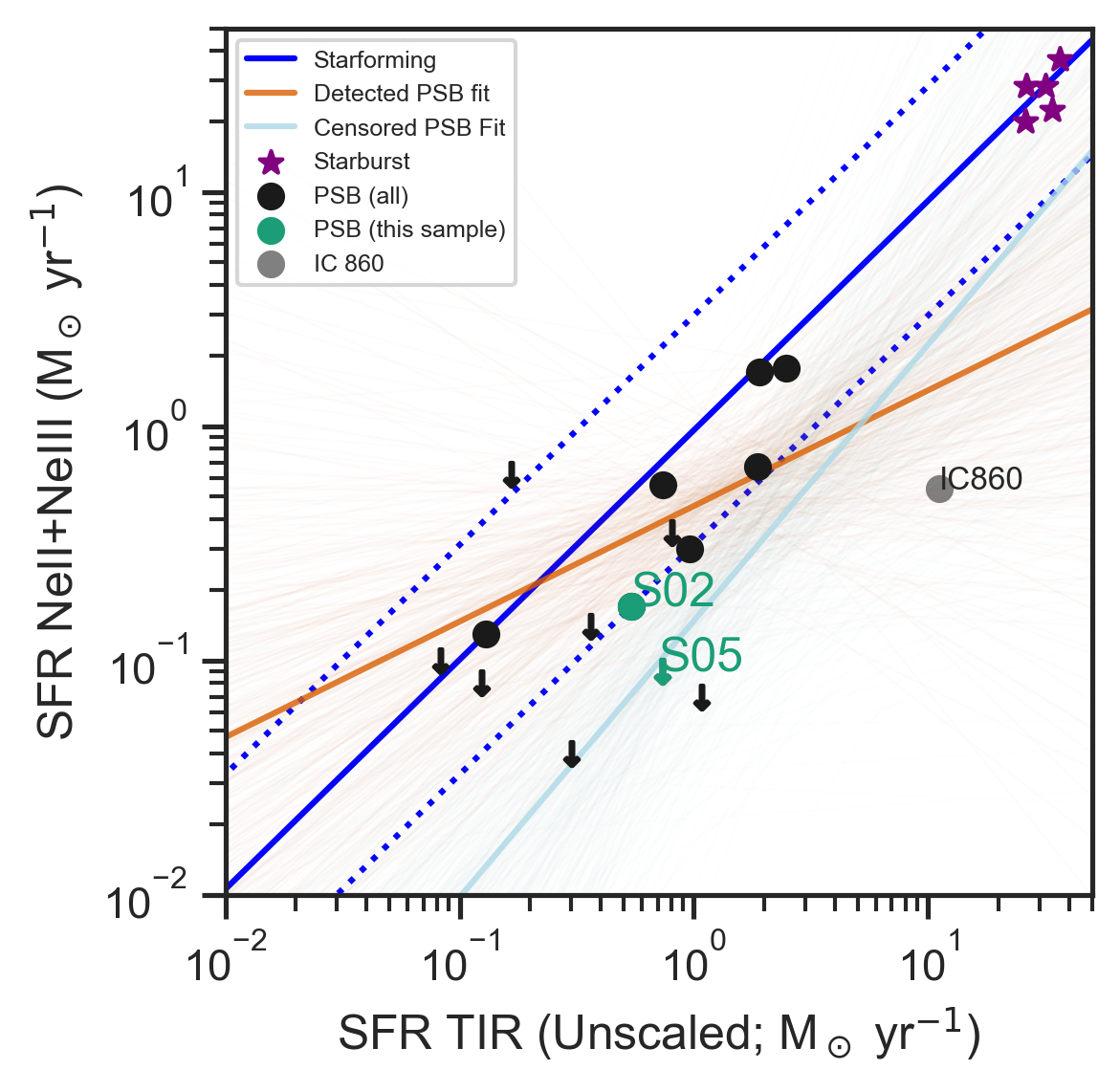}
\caption{Comparison of TIR-based SFRs and Neon-based SFRs for post-starburst galaxies observed with \textit{Spitzer} by \citet{Smercina2018} and \citet{Luo2022} (IC860). For comparison, we also plot the best-fit line (blue) and scatter (blue dashed) for star-forming galaxies from \citet{Ho2007} and the stacks of starburst galaxies with varying AGN properties from \citet{Stone2022} (upper right). The post-starburst galaxies (with the two targets observed with ALMA in this work highlighted in green) are systematically below the \citet{Ho2007} calibration, indicating the TIR-based SFRs are likely overestimating the true SFR. This offset is not seen for starbursting galaxies considered by \citet{Stone2022}. Using {\tt linmix} \citep{Kelly2007}, we fit a linear relation between log(SFR-TIR) and log(SFR-Neon). If we consider only the post-starbursts with Neon detections (orange), the best-fit line indicates a factor $2\times$ offset below the 1:1 line; including information from the $3\sigma$ upper limits (light blue) indicates a factor $8\times$ offset. This comparison suggests the TIR overestimates the SFR that would be measured with the Neon lines by $2-8\times$ in post-starburst galaxies. Because this effect is not present in the starbursting sample, it may be due to the longer duration of TIR as a SFR tracer, resulting in  the TIR-based SFR being contaminated by the recent starburst for the post-starburst galaxies.
}
\label{fig:sfr_comp_irne}
\end{figure*}

The TIR luminosity is also sensitive to obscured star formation, yet the TIR traces star formation on longer timescales, which can be comparable to the time since the recent starburst. \citet{Smercina2018} found that TIR-based SFRs overestimate the SFR traced by Neon for post-starburst galaxies. For the seven post-starburst galaxies with Neon detections and eight galaxies with upper limits, the ratio of SFR-TIR to SFR-Neon is $>2$. \citet{Luo2022} observed a TIR-based SFR 20$\times$ higher than the Neon-based SFR for the post-starburst galaxy IC 860. In Figure \ref{fig:sfr_comp_irne}, we compare the SFRs from TIR and Neon for post-starburst galaxies, stacks of starbursting galaxies with varying AGN properties from \citet{Stone2022}, and the best-fit relation from star-forming galaxies from \citet{Ho2007}. Each of the Neon-detected post-starburst galaxies and all but one of the upper limits are consistent with SFR-TIR $>$ SFR-Neon. In contrast, the starburst galaxy stacks from \citet{Stone2022} scatter evenly around the \citet{Ho2007} relation calibrated on star-forming galaxies. We use {\tt linmix} \citep{Kelly2007} to further quantify the offset observed for the post-starburst sample (including IC 860) by fitting a linear relation to log(SFR-TIR) vs. log(SFR-Neon). If we consider only the post-starburst galaxies with Neon detections, the best-fit line indicates a factor $2\times$ offset below the 1:1 line, consistent with our estimate from the median values. If we  include information from the $3\sigma$ upper limits for Neon non-detections, the best fit line indicates a factor $8\times$ offset for the range of SFR-TIR values\footnote{Excluding IC 860, we observe a $2\times$ offset for the Neon detections and a $7\times$ offset for the censored fit including non-detections.}. This comparison suggests the TIR overestimates the SFR that would be measured with the Neon lines by $2-8\times$ in post-starburst galaxies. Because this effect is not present in the starbursting sample, it may be due to the longer duration of TIR as a SFR tracer being contaminated by the recent starburst for the post-starburst galaxies. For the analysis of SFR-TIR of the post-starburst galaxies, we thus adopt a correction factor to decrease the SFR-TIR by a factor of 2.

We compare the H$\alpha$-based SFRs to TIR-based SFRs from \citet{Smercina2018} for the galaxies from \citet[H02, H03, S02, S05]{French2015} and full SED fit SFRs (including \textit{Herschel} photometry) from \citet{Rowlands2015} for R02 and R05 in Figure \ref{fig:sfr_comp}. The TIR-based SFRs are divided by a factor of two as described above, to account for the likely overestimate due to the longer duration of this tracer.  With this correction, there is no systematic bias between the two SFR indicators, although the scatter is large. We note that for H02 and H03, the large offset indicates there may be obscured star formation. Recently, \citet{Baron2022} found that some samples of post-starburst galaxies have high SFRs $\sim10-100$ \sfrunit when traced by their IR luminosity. The samples most heavily affected are post-starburst samples chosen to optimize for young post-starburst ages or to not select against AGN activity. We note that the SFRs for the post-starburst galaxies considered here have significantly lower SFRs, even using the highest estimates from the TIR luminosity.

We compare the H$\alpha$-based SFRs to 1.4 GHz based SFRs using data from the VLA FIRST survey \citep{Becker1995}. We convert the 1.4 GHz fluxes and flux limits to SFRs following \citet{Nielsen2012} by using the calibrations from \citet{Condon1992, Yun2001}. Of the galaxies considered here, R02, R05, and H03 are detected by FIRST, and the rest of the galaxies are not detected. The 1.4 GHz flux is likely to be contaminated by any AGN or LINER activity in the post-starbust sample \citep[e.g.,][]{Moric2010}, especially for H03, given the large offset between the 1.4 GHz and Br$\gamma$ indicators. 

Another infrared line used to calculate SFRs with minimal effects from dust attenuation is the Br$\gamma$ line \citep[see][]{Pasha2020}. We have conducted a survey of NIR spectroscopy of the parent samples considered here from \citet{French2015, Rowlands2015, Smercina2018} using Magellan/FIRE (Tripathi, French et al. in preparation). Using the calibration from \citet{Kennicutt1998}, we compare the Br$\gamma$ SFRs to the H$\alpha$ SFRs in Figure \ref{fig:sfr_comp}. Br$\gamma$ is detected for R02 and R05. Our data provide useful upper limits on the SFRs for H03 and S02\footnote{Unfortunately, the faint K-band magnitude of H02 prohibited its inclusion in the FIRE sample.}. Despite the high SFR $\sim5$ M$_\odot$ yr$^{-1}$ for H03 that would be inferred from its TIR or 1.4 GHz luminosities, H03 is not detected in Br$\gamma$, at a level consistent with its H$\alpha$ flux. This unusual TIR/Br$\gamma$ ratio is not seen for other galaxies, even those with extremely high extinction like Arp 220 \citep{Pasha2020}. Further observations, especially Near- and Mid- IR spectroscopy, will be required of these sources to determine the origin of the dust heating. While in \citet{French2018b} we assigned an upper limit on the SFR for H02, here we use the H$\alpha$-based SFRs throughout, motivated by the consistency with the Br$\gamma$ observations.

The Neon-based SFRs are the best available SFR tracer for this sample, as the Neon lines can trace obscured star formation even at the levels of ULIRGs, the contamination from other excitation sources is low, and the timescale for this SFR tracer is much shorter than the typical post-starburst ages \citep{Ho2007, Smercina2018}. Unfortunately, mid-IR spectroscopy is required for this measurement, and not available for the entire sample considered here. We use the cases where both Neon and H$\alpha$ SFRs are available to test the H$\alpha$ SFRs for bias (Figure \ref{fig:sfr_comp}) and find the two tracers to be consistent. This consistency indicates that the post-starburst galaxy samples considered here do not typically have obscured star formation missed by the H$\alpha$ tracer corrected using the Balmer decrement. Two galaxies (H02 and H03) have strong mismatches between the TIR SFR and H$\alpha$ SFR, yet do not have mid-IR spectroscopy for which a Neon SFR can be measured. For one of these galaxies (H03), we have a strong limit on the Br$\gamma$ SFR, which is consistent with H$\alpha$. While Br$\gamma$ is less sensitive to dust obscuration than H$\alpha$, ULIRG-like central dust densities could still obscure a nuclear star-forming region. Such a mismatch between a quiescent host galaxy and a nuclear starburst would be unusual, as LIRGs and ULIRGs have star-forming regions visible outside of the central regions with high dust obscuration, but we cannot fully rule out the possibility of an obscured region with SFR$\sim5$ \sfrunit in H02 and H03. 

In order to assess the impact of SFR tracer on our conclusions, we reproduce the key figures in this work using the TIR-based SFRs in Appendix \S\ref{sec:appendix_ir}. The SFRs for four galaxies (S02, S05, R02, R05) are comparable, while the SFRs for two galaxies (H02 and H03) are higher. Considering the TIR SFRs instead of the H$\alpha$ SFRs, our qualitative conclusions do not change. We find that the post-starburst galaxies still lie offset from the Kennicutt-Schmidt relation, and have high \lpco yet consistent \lhco and \lhcn values for their SFRs.

\subsection{Interpretation}

As galaxies evolve through the post-starburst phase, the galaxies with remaining molecular gas remaining experience an unusual transition in their gas properties. The gas is confined to the central $\sim$kpc, more limited in extent than the optical light or even the young stellar populations. Yet the gas state is such that the typical density is low, as traced by both the lack of strong dense gas emission and by the low CO excitation. What then is suppressing this gas from collapsing to denser states? 

The low CO excitation indicates that the bulk of the molecular gas is not being heated. The outflow observed in S02, as well as the outflow observed in CO (2--1) by \citet{Smercina2022} for another post-starburst, provide a clue that relatively low velocity outflows, lower than or close to the escape velocity, may prevent the gas from re-collapsing. The origin of these outflows is still unclear. We observe no strong AGN activity in these galaxies, yet because the timescale for AGN to vary is shorter than the timescale for us to observe these outflows, past AGN activity may have launched these outflows. Alternatively, weak, low-level AGN activity (if the LINER is a low luminosity AGN) may be enough to sustain quiescence during this phase and deplete the gas over 1--2 Gyr. If weak AGN activity is currently affecting these galaxies, it must be at such a low level as to not result in high CO excitation and not drive AGN-like emission line ratio maps. We speculate that if the energy coupling of TDEs to the molecular gas is efficient compared to AGN energy coupling, the high TDE rate during this phase may act to provide the energy source needed to keep this gas from collapsing to denser states and forming stars \citep[see further discussion in][]{Smercina2022}. 

A key test for models of feedback in simulations aiming to re-create the galaxy population, including galaxies with rapidly-ending star formation, will be to predict the detailed evolution of the densest gas on scales of $n\sim10^4$ cm$^{-3}$ and at cold temperatures $T<100$ K.

\section{Conclusions}

Observations of large fractions of molecular gas remaining in galaxies with recently-ended starbursts have raised questions of what mechanisms act to drive galaxies to quiescence. In this study, we present new observations of CO (3--2) observations for three post-starburst galaxies and dense gas tracers for four post-starburst galaxies, combining with literature measurements for a total dense gas sample size of six.

\begin{enumerate}

\item The post-starbursts have low excitation as traced by the CO spectral line energy distribution (SLED) up to CO (3--2), more similar to early-type than starburst galaxies. The low excitation indicates that lower density rather than high temperatures may suppress star formation during the post-starburst phase, as higher temperatures would result in excitation states more like starbursts. Radiative transfer modelling with RADEX supports this picture; the RADEX models favor low densities ($\log n/\rm{cm}^{-3} \sim 3.4-3.8$) and temperatures ($T\sim 15-30$ K), similar to early type galaxies and lower than typical starbursts. The low CO excitation is in contrast with the high ISRF intensity (and thus high dust temperatures) in these galaxies, suggesting the molecular gas temperature is decoupled from the dust. 

\item The post-starburst galaxies have small CO (3--2) sizes ($\sim 250-400$ pc)  compared to their optical sizes. The CO (3--2) sizes are on average 6.3$\times$ smaller than the $r$-band optical sizes from SDSS images. This result is consistent with the findings of \citet{Smercina2022}.

\item Post-starburst galaxies have high molecular gas surface densities for their star formation rate surface densities, resulting in an offset from the Kennicutt-Schmidt relation, consistent with the findings of \citet{Smercina2022}. We find this same result using both optical (H$\alpha$) and infrared SFR tracers (TIR luminosity and  [NeII] 12.8$\mu$m + [NeIII] 15.6$\mu$m), indicating that this offset is not driven by the presence of dust-obscured star formation. 

\item One galaxy (S02) displays a blueshifted molecular gas outflow traced by CO (3--2). This galaxy has complementary \textit{HST} and MaNGA observations, facilitating multiwavelength comparisons. The MaNGA observations show the ionized gas velocity is disturbed relative to the stellar velocity field, with a blueshifted component aligned with the molecular gas outflow, indicative of a possible multiphase outflow. The inferred mass loss rate is consistent with the CO depletion observed statistically in \citet{French2018}. The energy required to drive this outflow is consistent with lower luminosity AGN/LINERs or TDEs. The feasibility of energy from intermittent AGN activity or from TDEs will depend heavily on the coupling to the molecular gas in these galaxies. 
  
\item Low ratios of HCO$^+$/CO, indicating low fractions of dense molecular gas relative to the total molecular gas, are seen throughout the post-starburst phase beginning $\sim$10-200 Myr after burst ends for 5/6 post-starbust galaxies. Most post-starbursts have low HCO$^+$/CO ratios; the exception is the youngest post-starburst in our sample, suggesting early evolution. Rapid evolution in the dense gas would be consistent with the rapid evolution in sSFR during the early post-starburst phase. However, observations of a larger sample will be required to determine whether the dense molecular gas has a rapid early decline similar to that seen in the sSFR, or a slower decline throughout the post-starburst phase, more like the evolution seen in the CO-traced gas.

\item Post-starburst galaxies have low \lhco-traced dense gas luminosities more consistent with their low SFRs than the CO luminosities would indicate, with the exception of the post-starburst galaxy H03 which has a highly uncertain SFR. This is consistent with our previous work \citep{French2018b} and indicates that the low SFRs in the post-starburst phase are due to a lack of dense gas, in contrast to the large masses of total molecular gas traced by CO (1--0) \citep{French2015,Rowlands2015,Alatalo2016b}. Our qualitative conclusions do not depend on the SFR tracer used. 
    
\item The three post-starbursts with measured HCN/HCO$^+$ ratios show a large variation, spanning the entire range shown by AGN, ULIRGs, and star forming galaxies. This may be due to a range of either mechanical heating or cosmic ray heating of the HCN, but the origin is uncertain.

\end{enumerate}

\begin{acknowledgments}

We thank the referee for their thoughtful review, which has improved this paper. We thank Antonio Usero and Adam Leroy for providing star forming galaxy comparison observations. K.D.F. thanks Dalya Baron for valuable discussions on the SFR tracers. A.S.\ was supported by NASA through grant \#GO-14610 from the Space Telescope Science Institute, which is operated by AURA, Inc., under NASA contract NAS 5-26555. D.N. acknowledges support from NSF grant AST-1908137. Y.Y.'s research was supported by the Basic Science Research Program through the National Research Foundation of Korea (NRF) funded by the Ministry of Science, ICT \& Future Planning (NRF-2019R1A2C4069803). 

This paper makes use of the following ALMA data: 2016.1.00881, 2017.1.00930, 2017.1.00925, 2018.1.00948. ALMA is a partnership of ESO (representing its member states), NSF (USA) and NINS (Japan), together with NRC (Canada), MOST and ASIAA (Taiwan), and KASI (Republic of Korea), in cooperation with the Republic of Chile. The Joint ALMA Observatory is operated by ESO, AUI/NRAO and NAOJ. The National Radio Astronomy Observatory is a facility of the National Science Foundation operated under cooperative agreement by Associated Universities, Inc.

Funding for the Sloan Digital Sky 
Survey IV has been provided by the 
Alfred P. Sloan Foundation, the U.S. 
Department of Energy Office of 
Science, and the Participating 
Institutions. 

SDSS-IV acknowledges support and 
resources from the Center for High 
Performance Computing  at the 
University of Utah. The SDSS 
website is www.sdss.org.

SDSS-IV is managed by the 
Astrophysical Research Consortium 
for the Participating Institutions 
of the SDSS Collaboration including 
the Brazilian Participation Group, 
the Carnegie Institution for Science, 
Carnegie Mellon University, Center for 
Astrophysics | Harvard \& 
Smithsonian, the Chilean Participation 
Group, the French Participation Group, 
Instituto de Astrof\'isica de 
Canarias, The Johns Hopkins 
University, Kavli Institute for the 
Physics and Mathematics of the 
Universe (IPMU) / University of 
Tokyo, the Korean Participation Group, 
Lawrence Berkeley National Laboratory, 
Leibniz Institut f\"ur Astrophysik 
Potsdam (AIP),  Max-Planck-Institut 
f\"ur Astronomie (MPIA Heidelberg), 
Max-Planck-Institut f\"ur 
Astrophysik (MPA Garching), 
Max-Planck-Institut f\"ur 
Extraterrestrische Physik (MPE), 
National Astronomical Observatories of 
China, New Mexico State University, 
New York University, University of 
Notre Dame, Observat\'ario 
Nacional / MCTI, The Ohio State 
University, Pennsylvania State 
University, Shanghai 
Astronomical Observatory, United 
Kingdom Participation Group, 
Universidad Nacional Aut\'onoma 
de M\'exico, University of Arizona, 
University of Colorado Boulder, 
University of Oxford, University of 
Portsmouth, University of Utah, 
University of Virginia, University 
of Washington, University of 
Wisconsin, Vanderbilt University, 
and Yale University.

This research is based on observations made with the NASA/ESA Hubble Space Telescope obtained from the Space Telescope Science Institute, which is operated by the Association of Universities for Research in Astronomy, Inc., under NASA contract NAS 5–26555. These observations are associated with program 11643.

\end{acknowledgments}

\software{Astropy \citep{astropy2013, astropy2018}, CASA \citep{casa}, Matplotlib \citep{matplotlib}, NumPy \citep{numpy}}

\bibliographystyle{aasjournal}
\bibliography{refs}

\appendix
\section{Spectra and Moment Maps}
\label{sec:appendix}
\begin{figure*}
\includegraphics[width=0.33\textwidth]{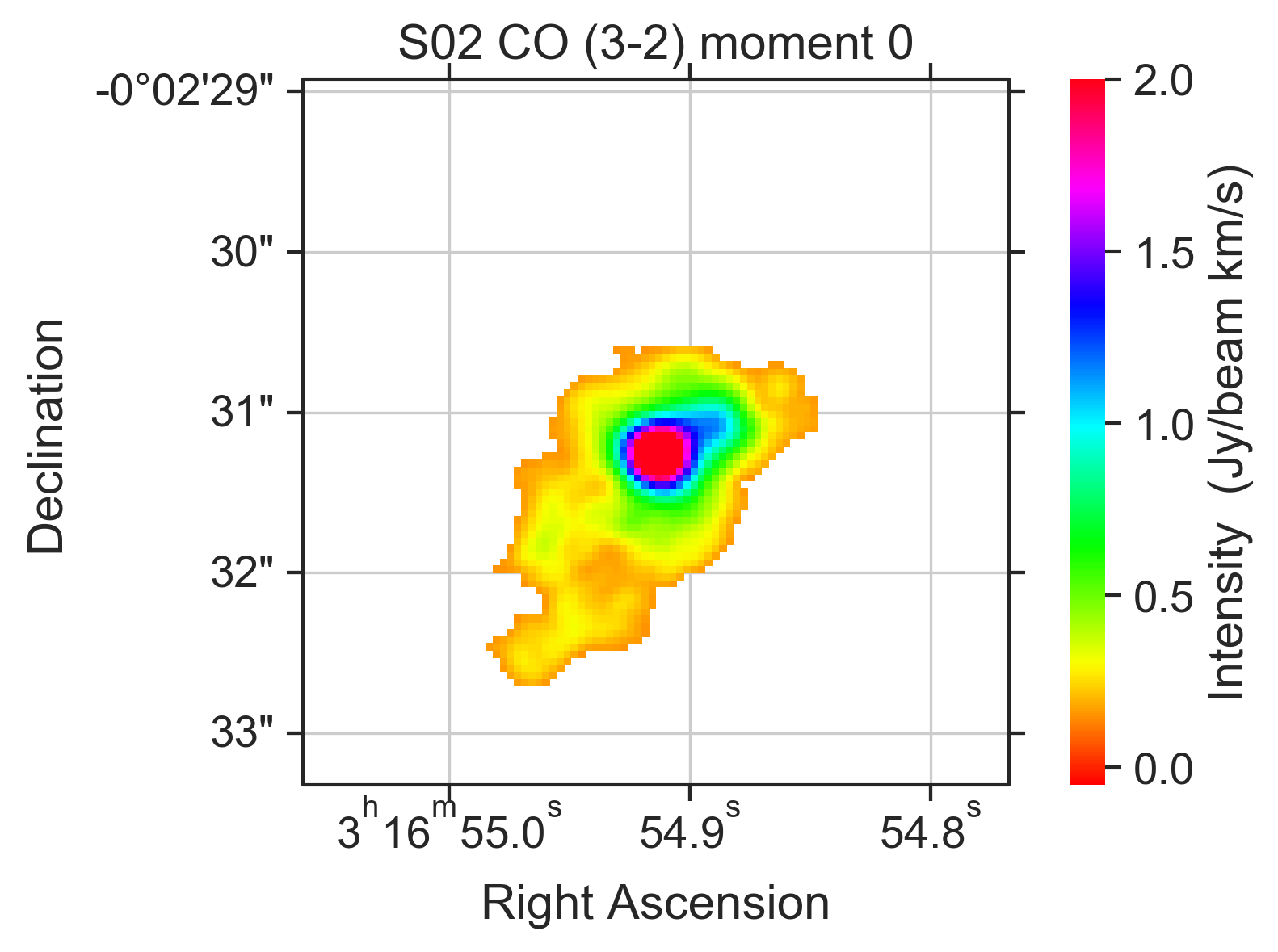}
\includegraphics[width=0.33\textwidth]{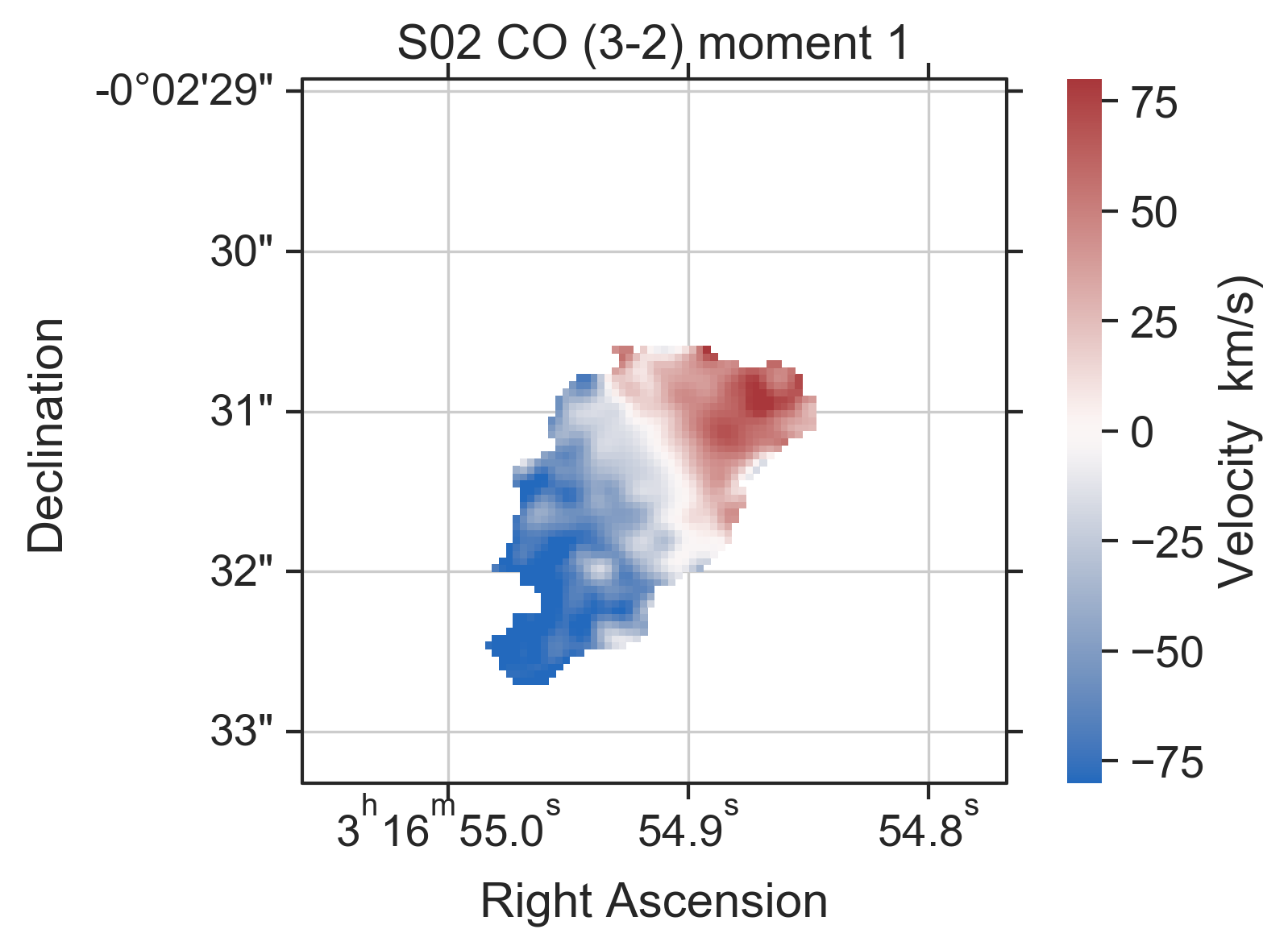}
\includegraphics[width=0.33\textwidth]{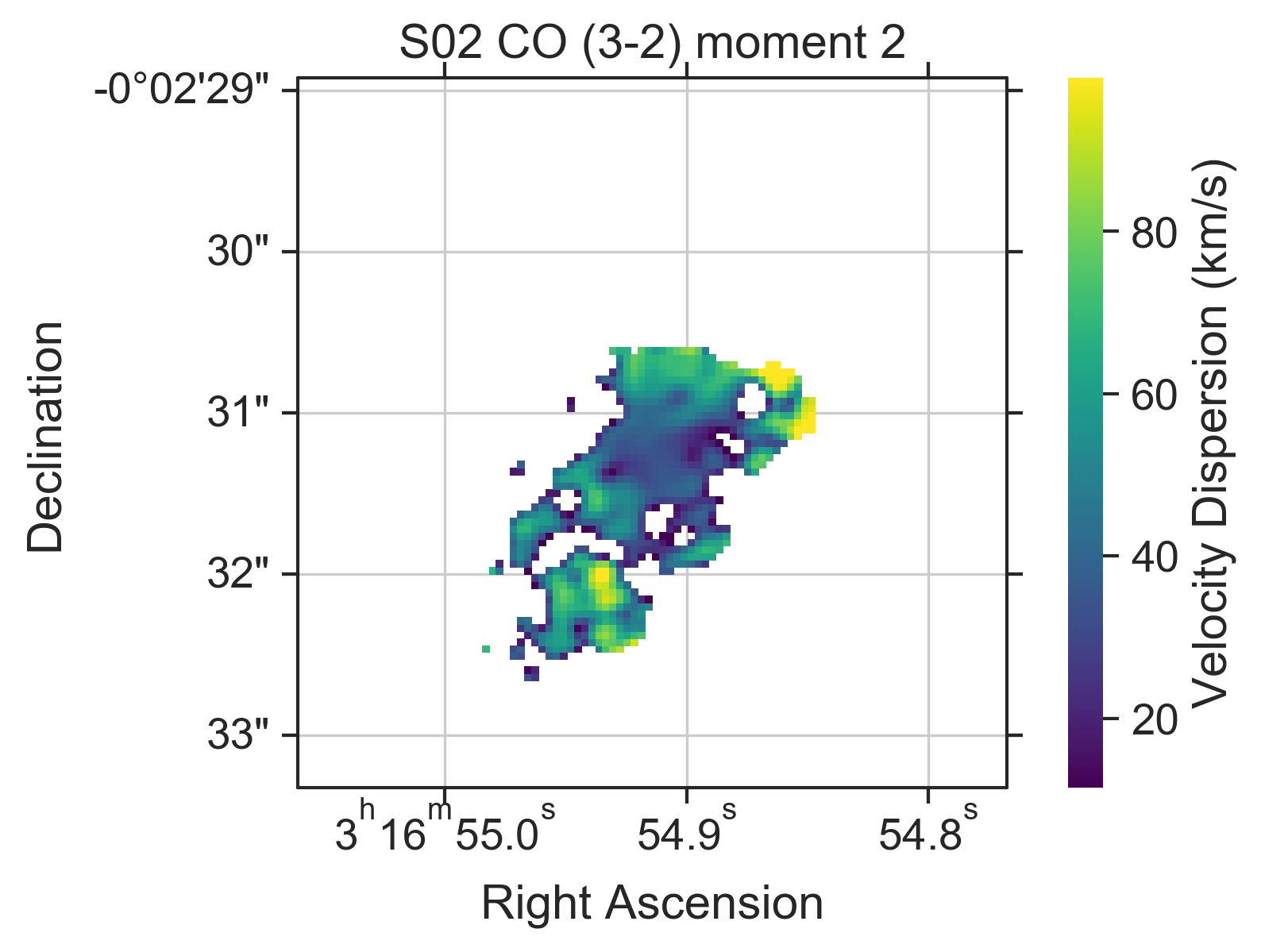}
\includegraphics[width=0.33\textwidth]{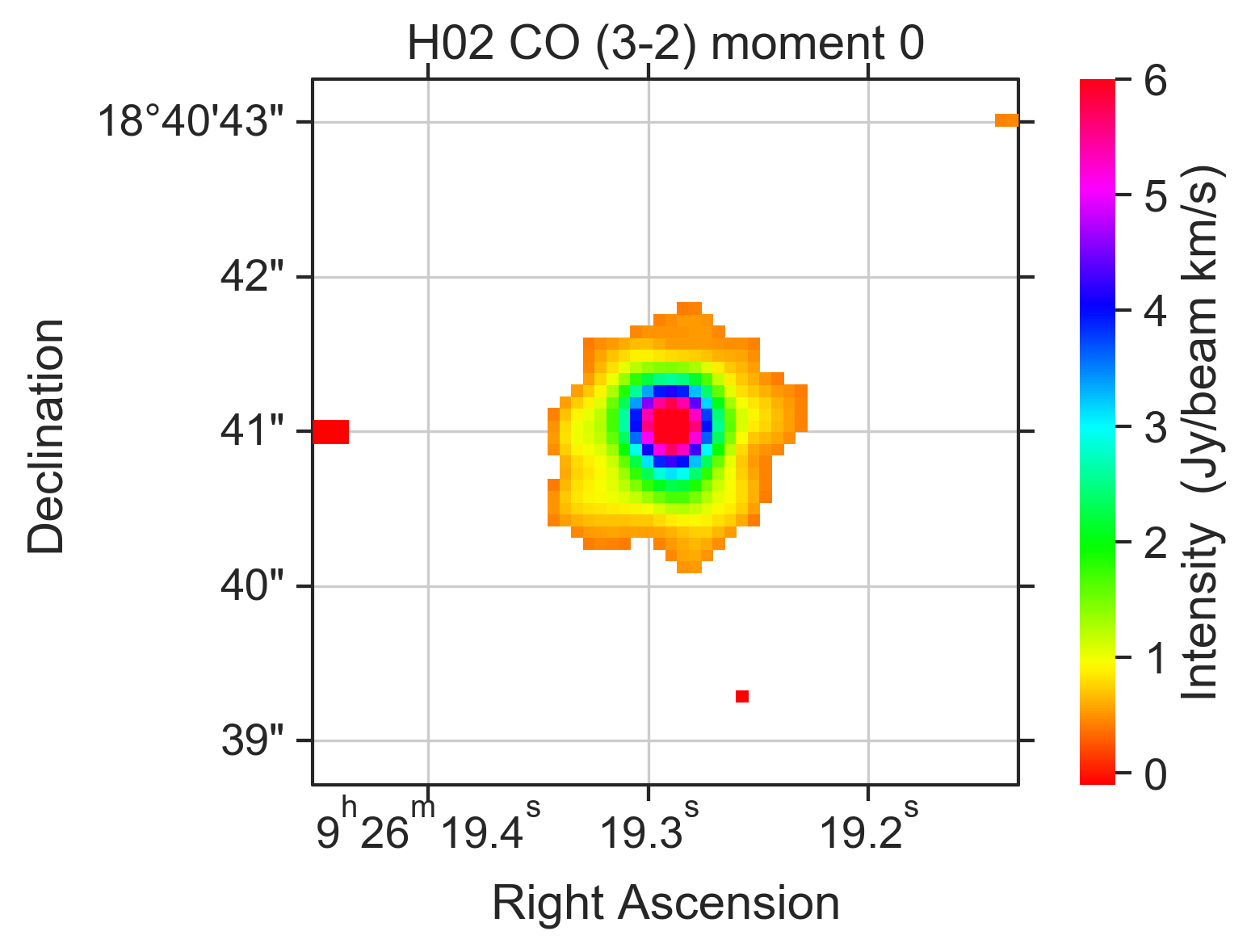}
\includegraphics[width=0.33\textwidth]{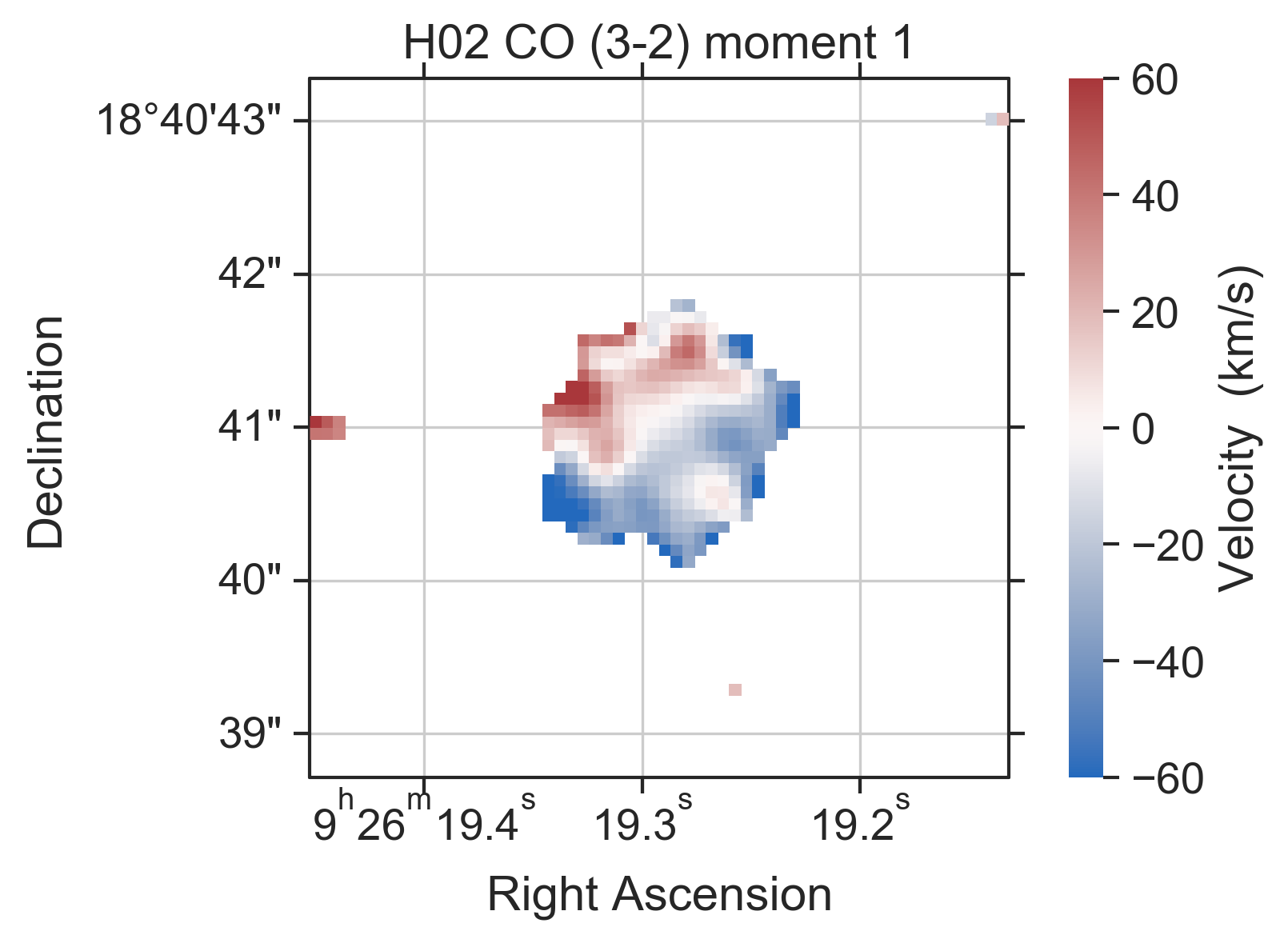}
\includegraphics[width=0.33\textwidth]{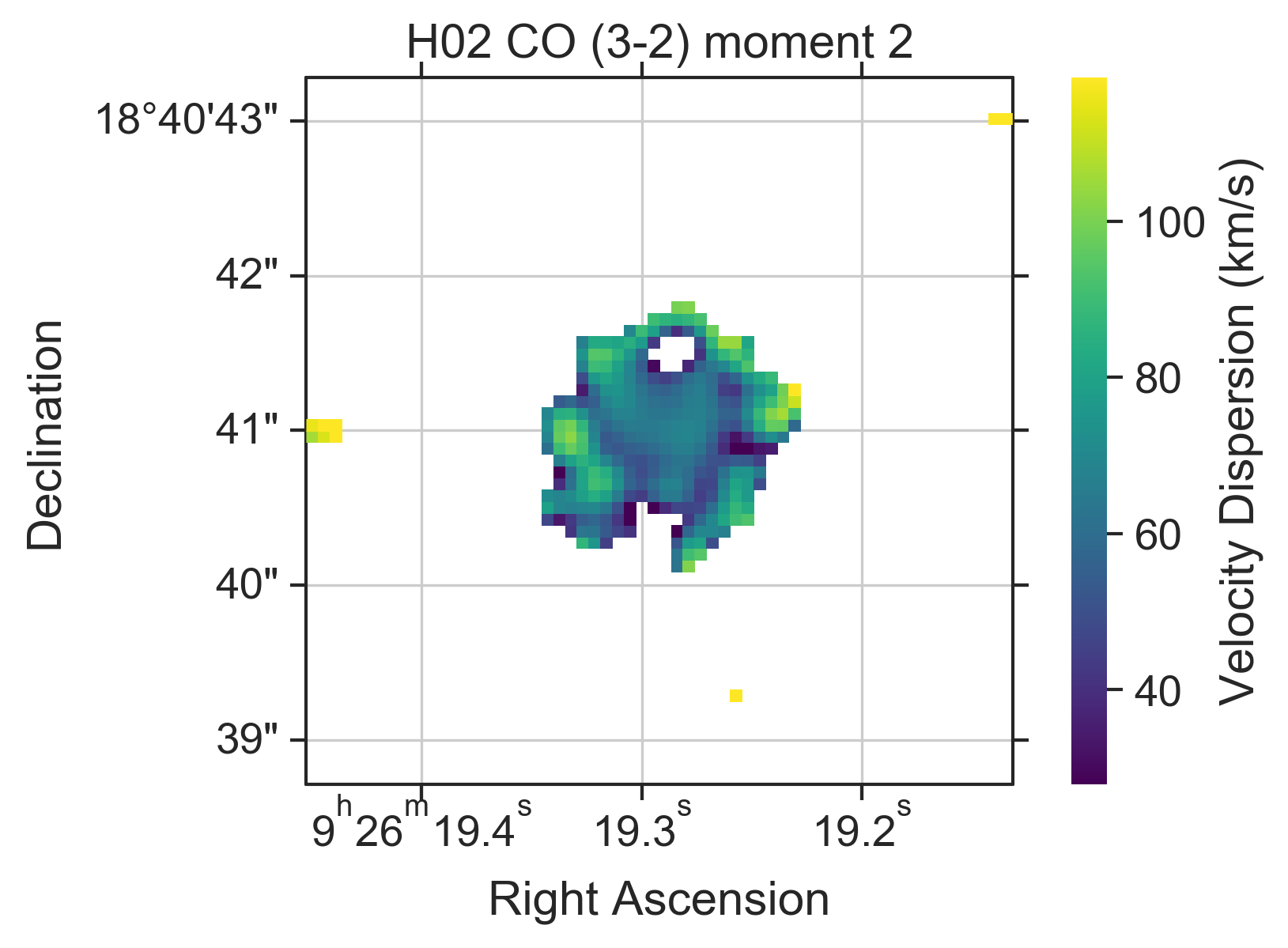}
\includegraphics[width=0.33\textwidth]{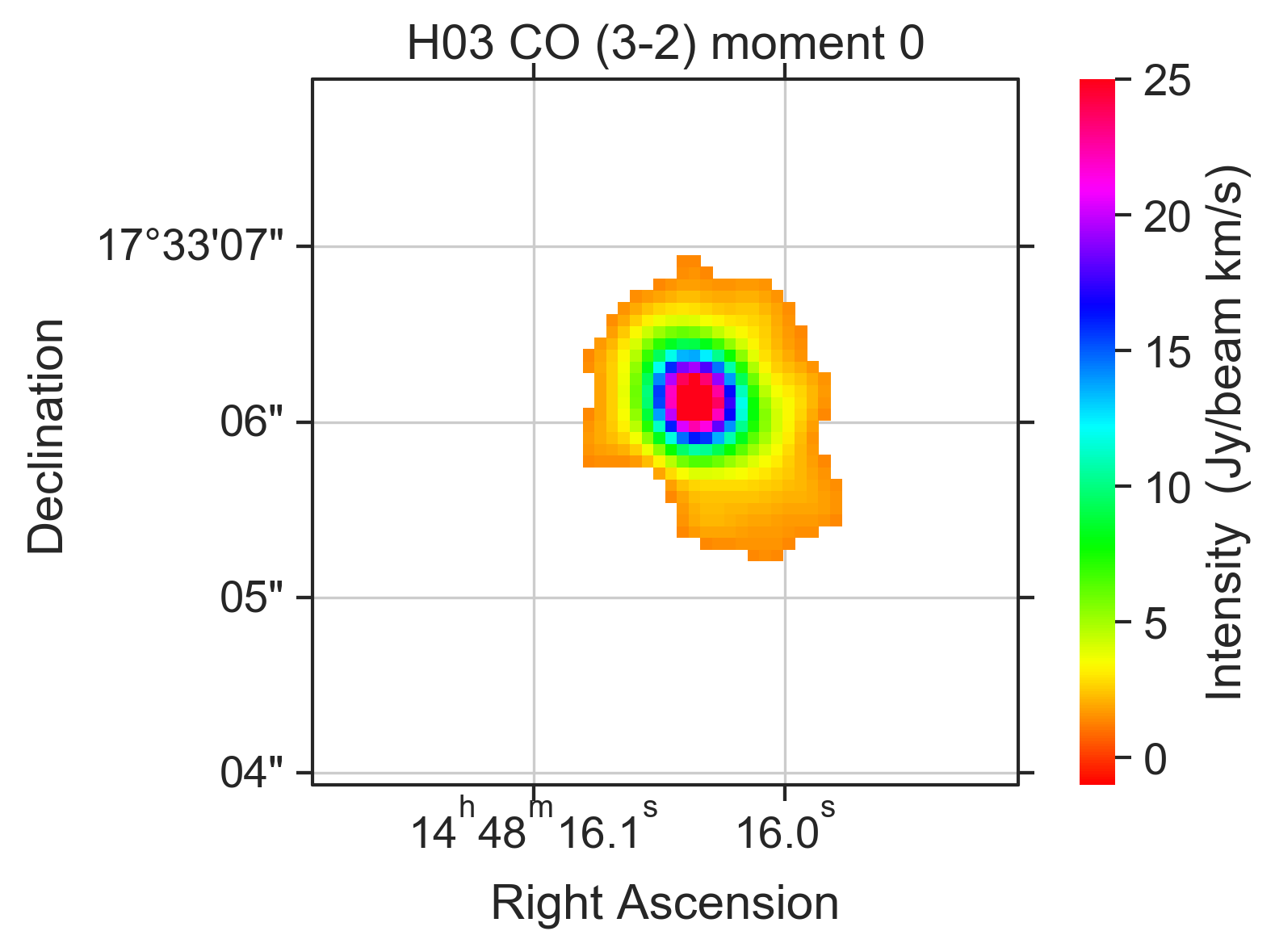}
\includegraphics[width=0.33\textwidth]{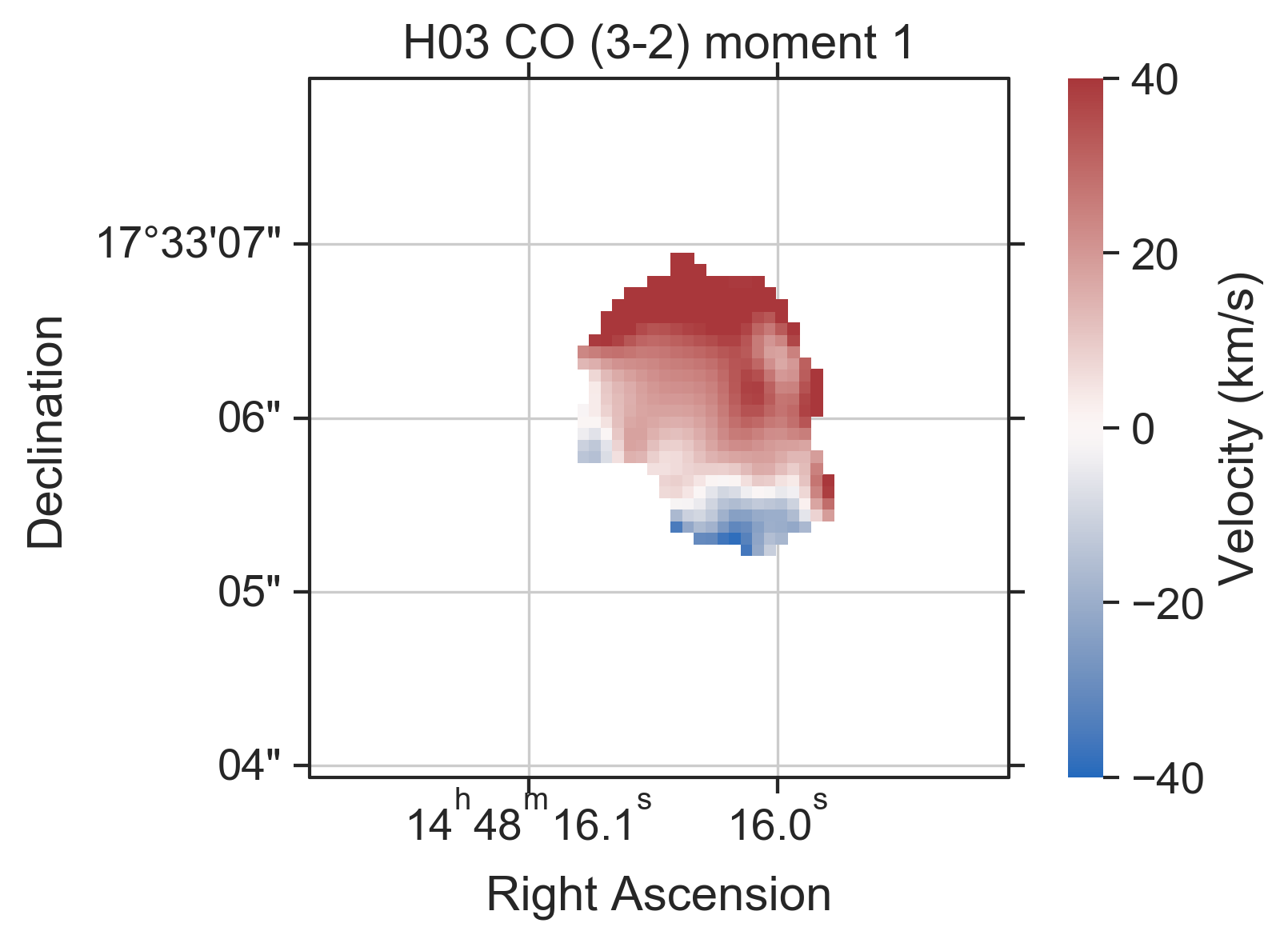}
\includegraphics[width=0.33\textwidth]{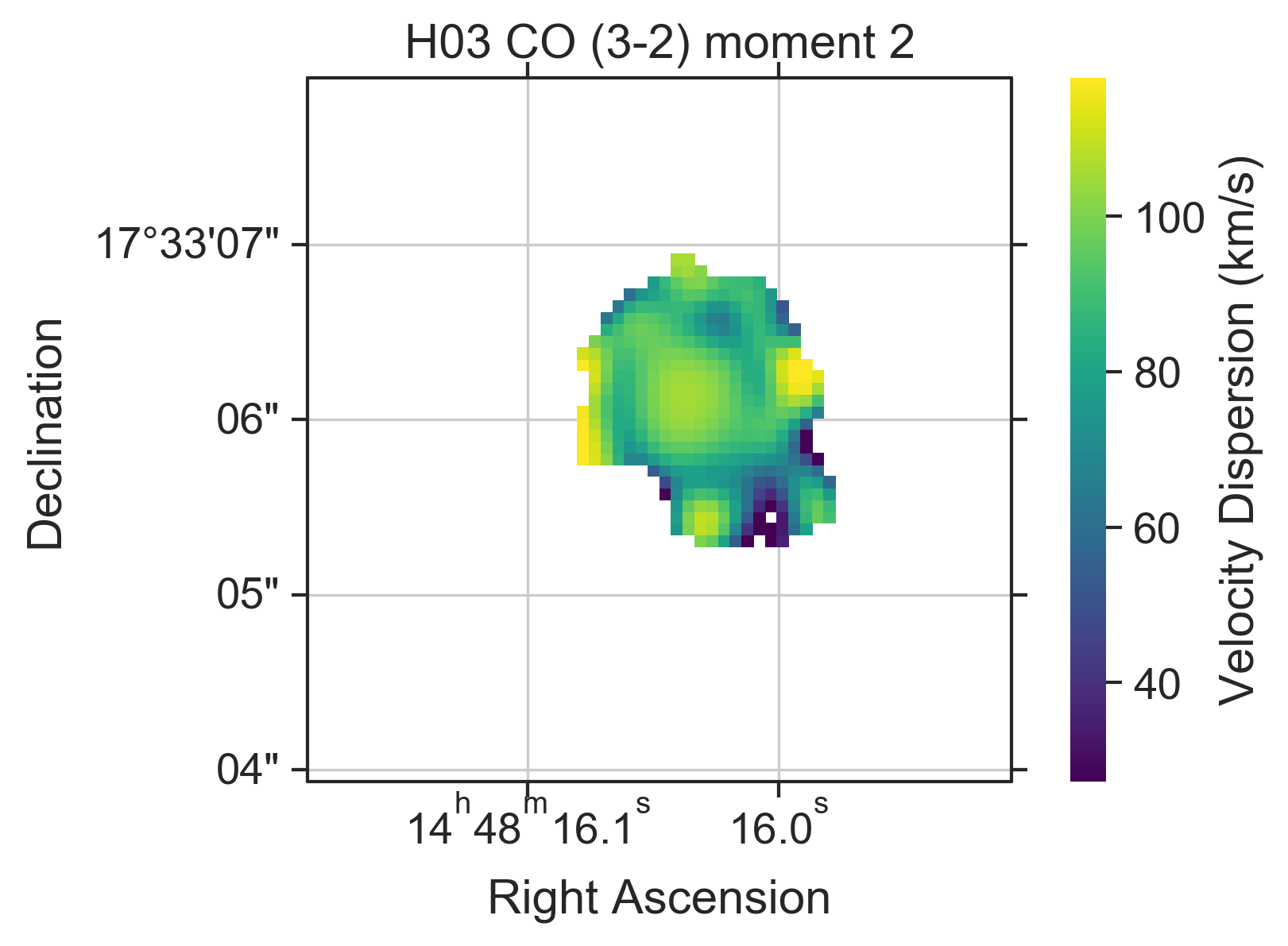}
\caption{CO (3--2) moment maps (left column: moment 0; middle column: moment 1; right column: moment 2). Data with a signal to noise ratio of $<3$ are masked. The molecular gas has limited spatial extent, with CO sizes $<1$ arcsec, significantly smaller than the optical extents of the galaxies ($r$-band R50 values 3.72\arcsec, 1.71\arcsec, and 3.17\arcsec\ for S02, H02, and H03, respectively). S02 has a blueshifted component to the lower left, which we explore further in Figures \ref{fig:s02_grid} -- \ref{fig:s02}. 
}
\label{fig:moments}
\end{figure*}

\begin{figure*}
\includegraphics[width=0.33\textwidth]{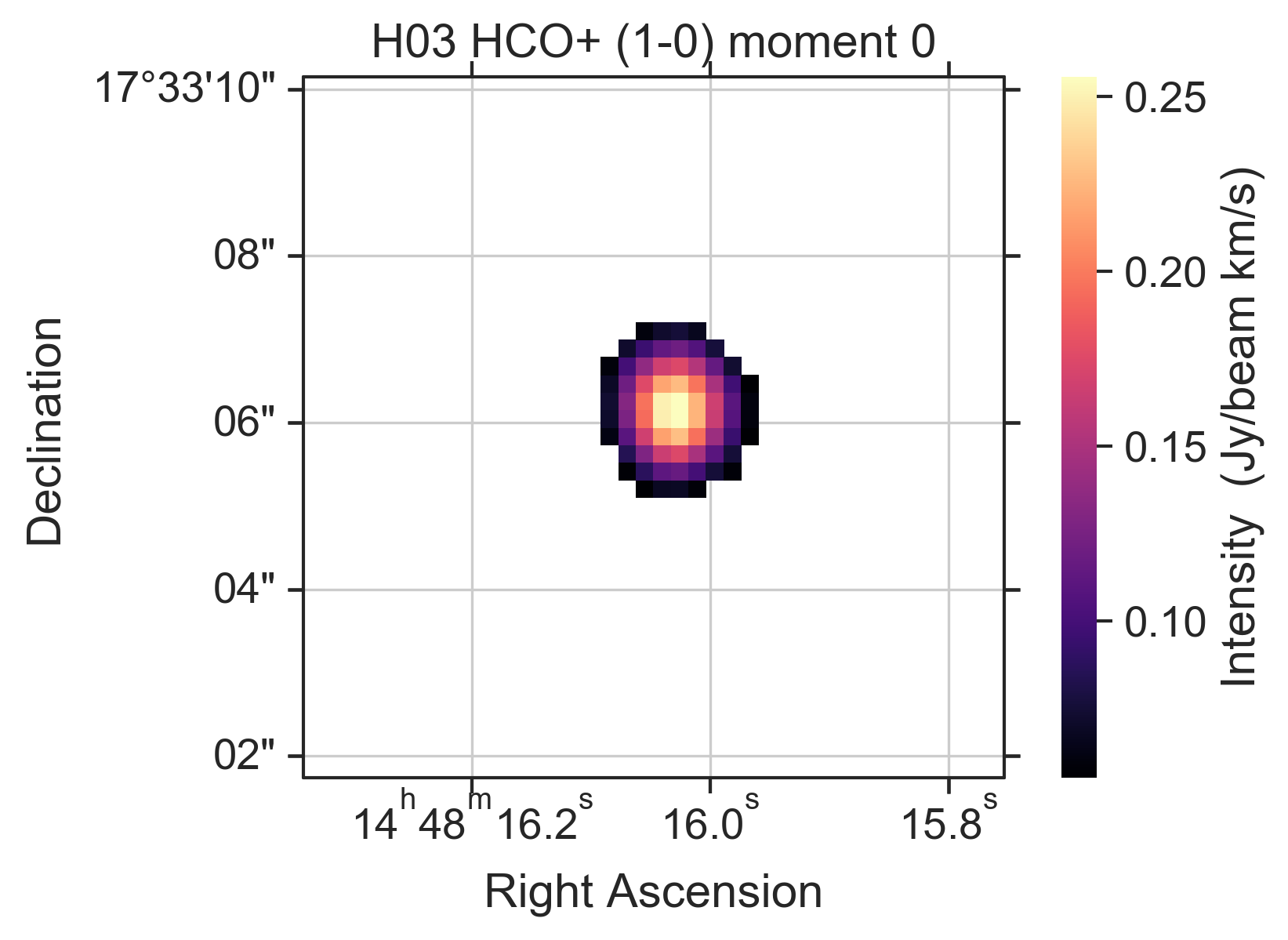}
\includegraphics[width=0.33\textwidth]{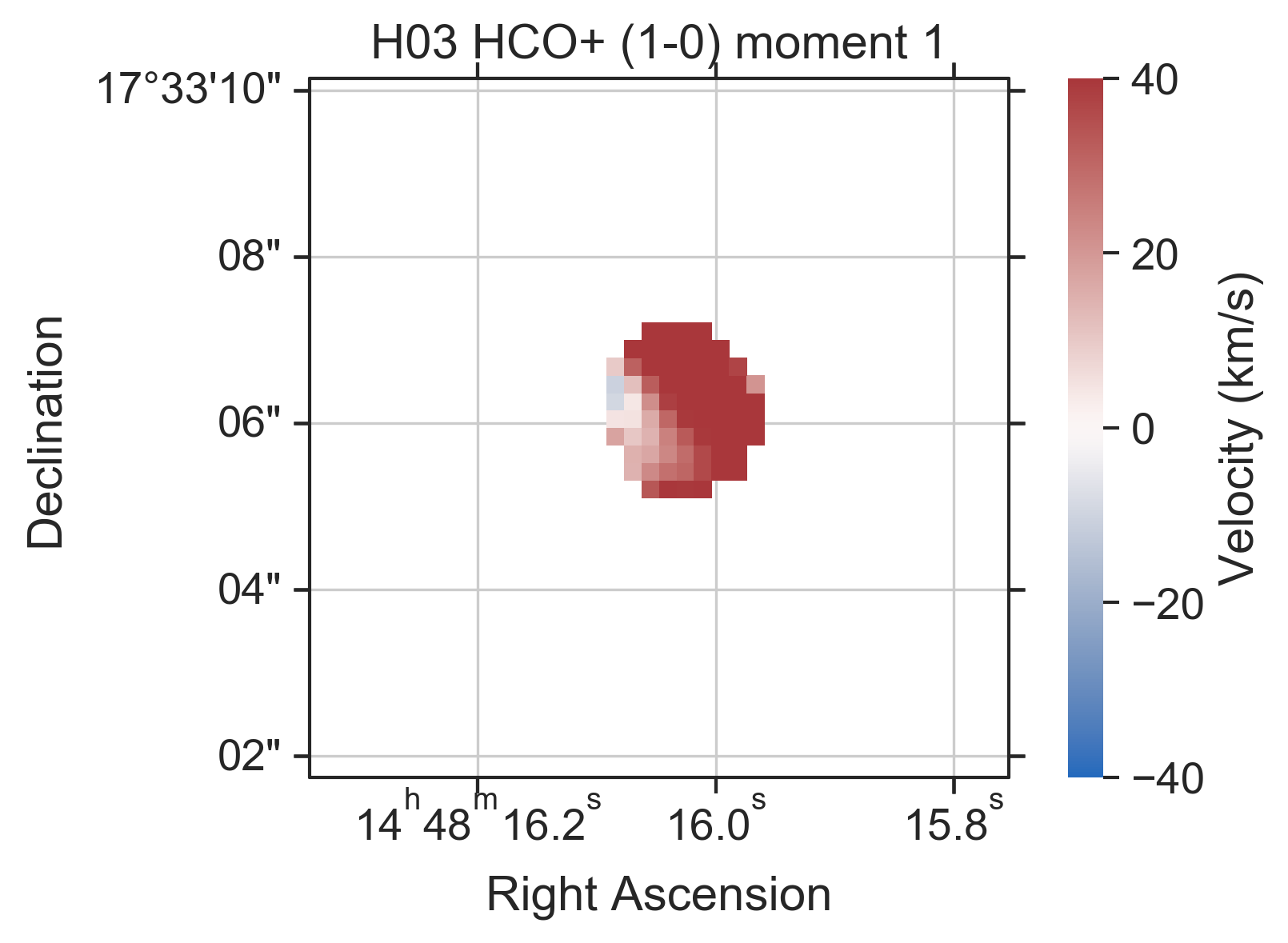}
\includegraphics[width=0.33\textwidth]{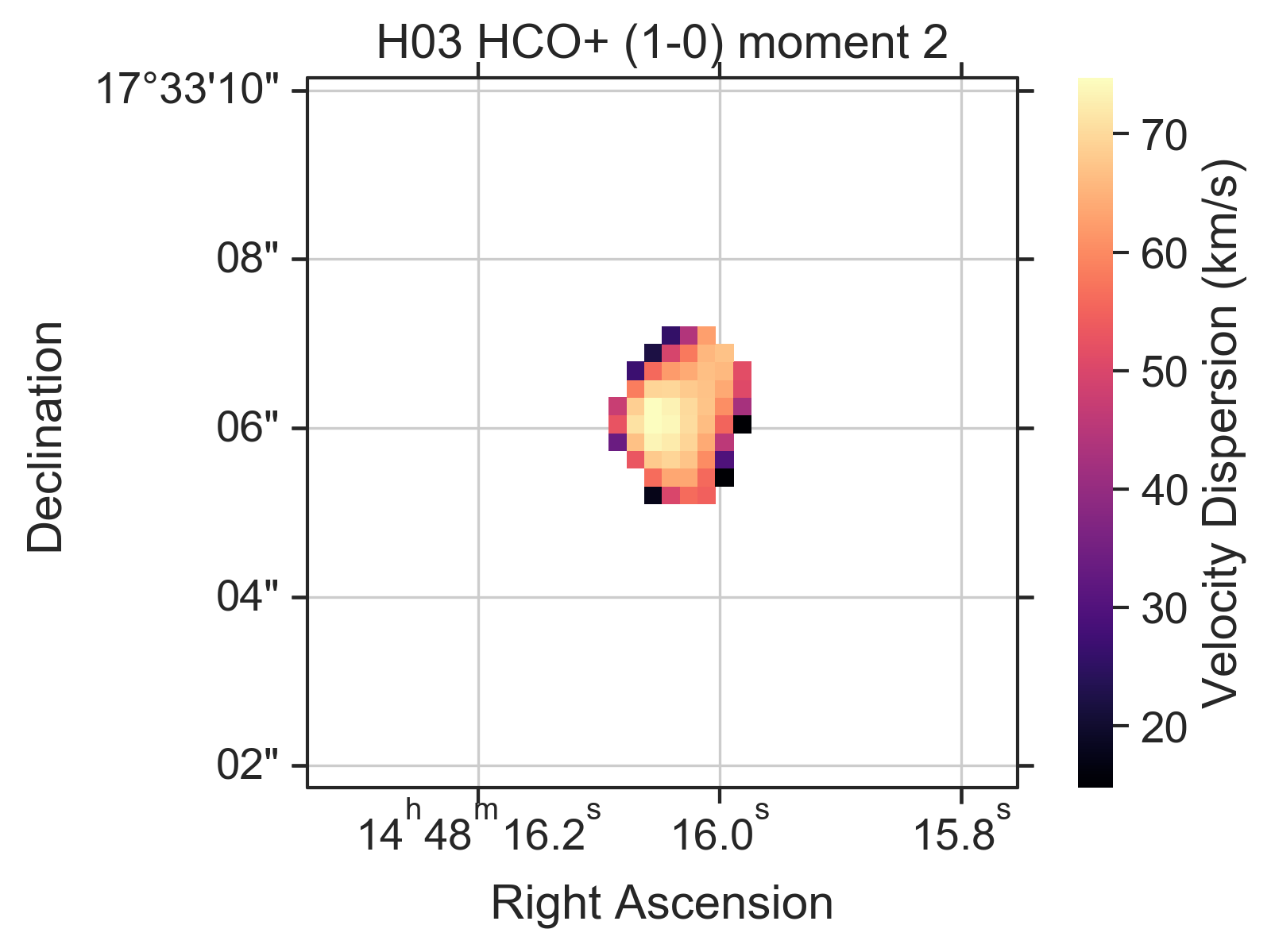}
\includegraphics[width=0.33\textwidth]{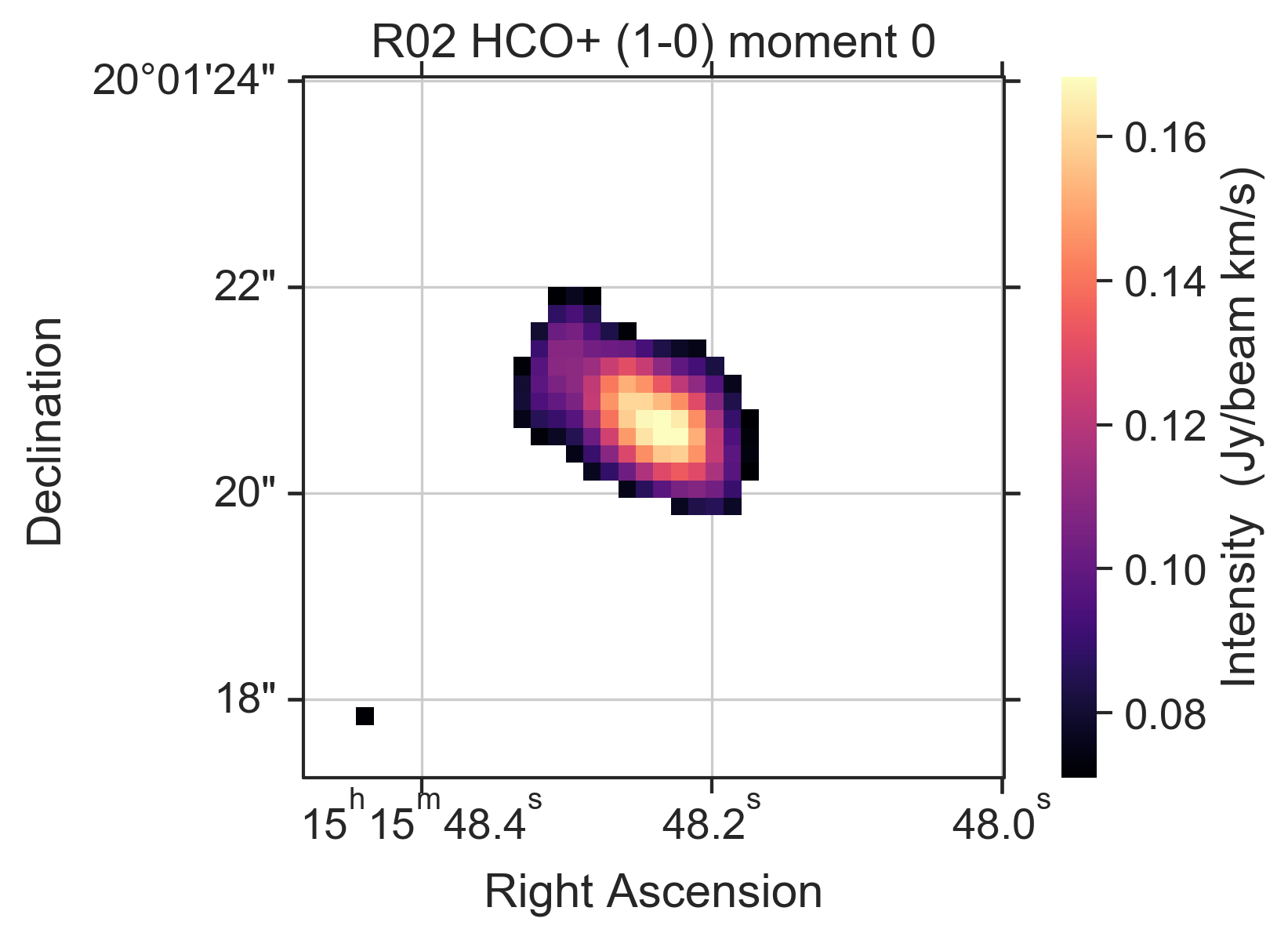}
\includegraphics[width=0.33\textwidth]{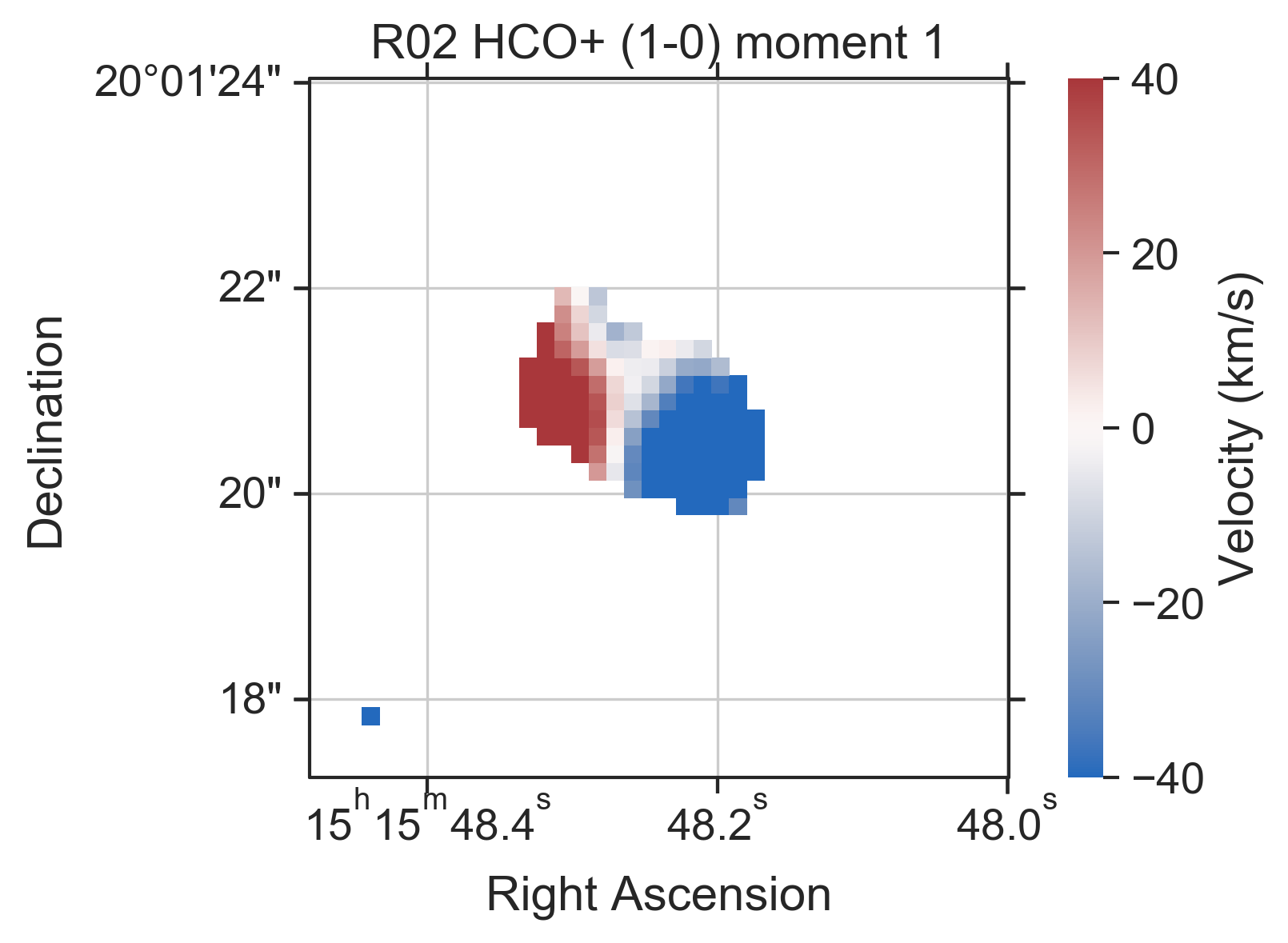}
\includegraphics[width=0.33\textwidth]{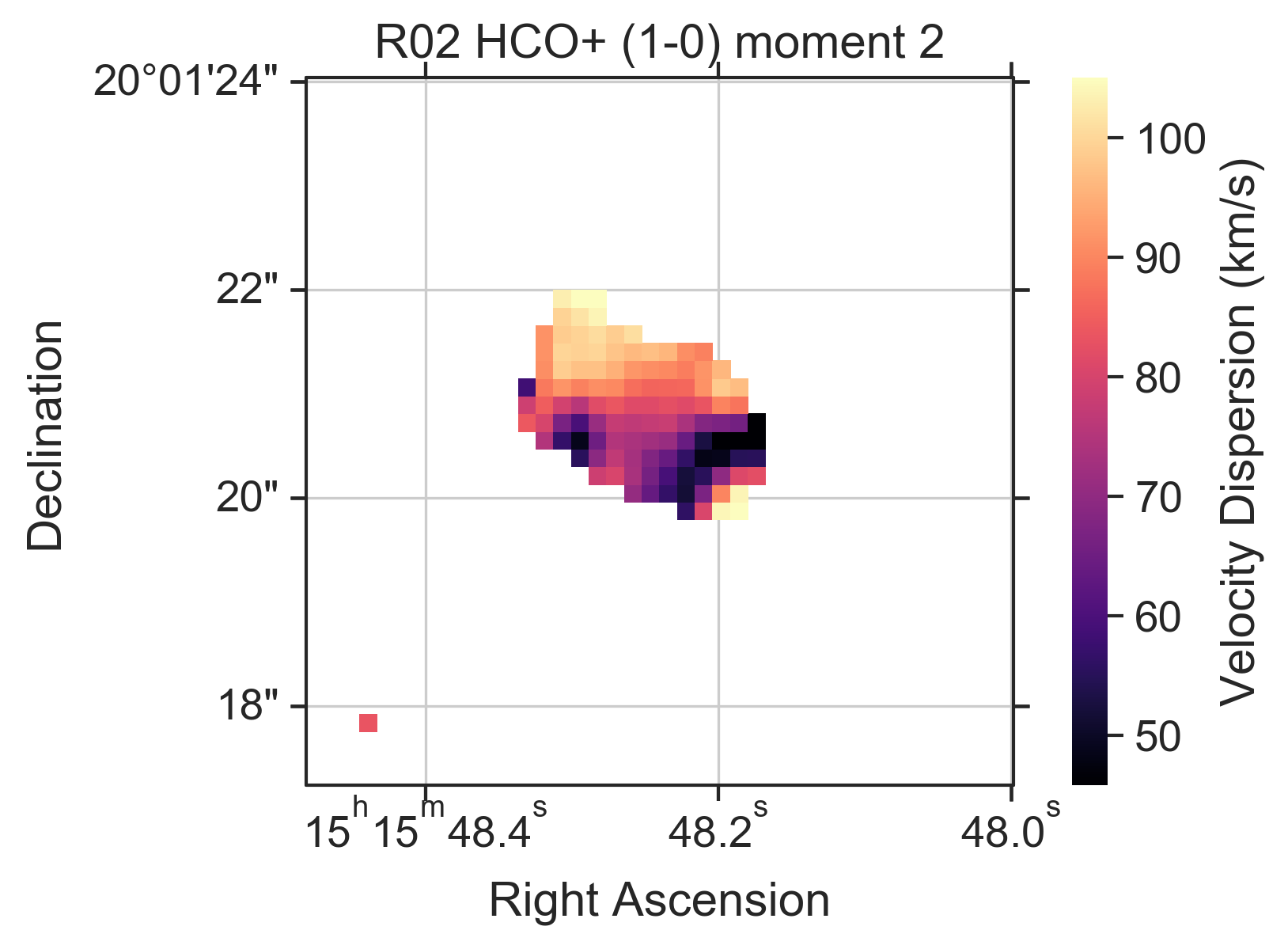}
\includegraphics[width=0.33\textwidth]{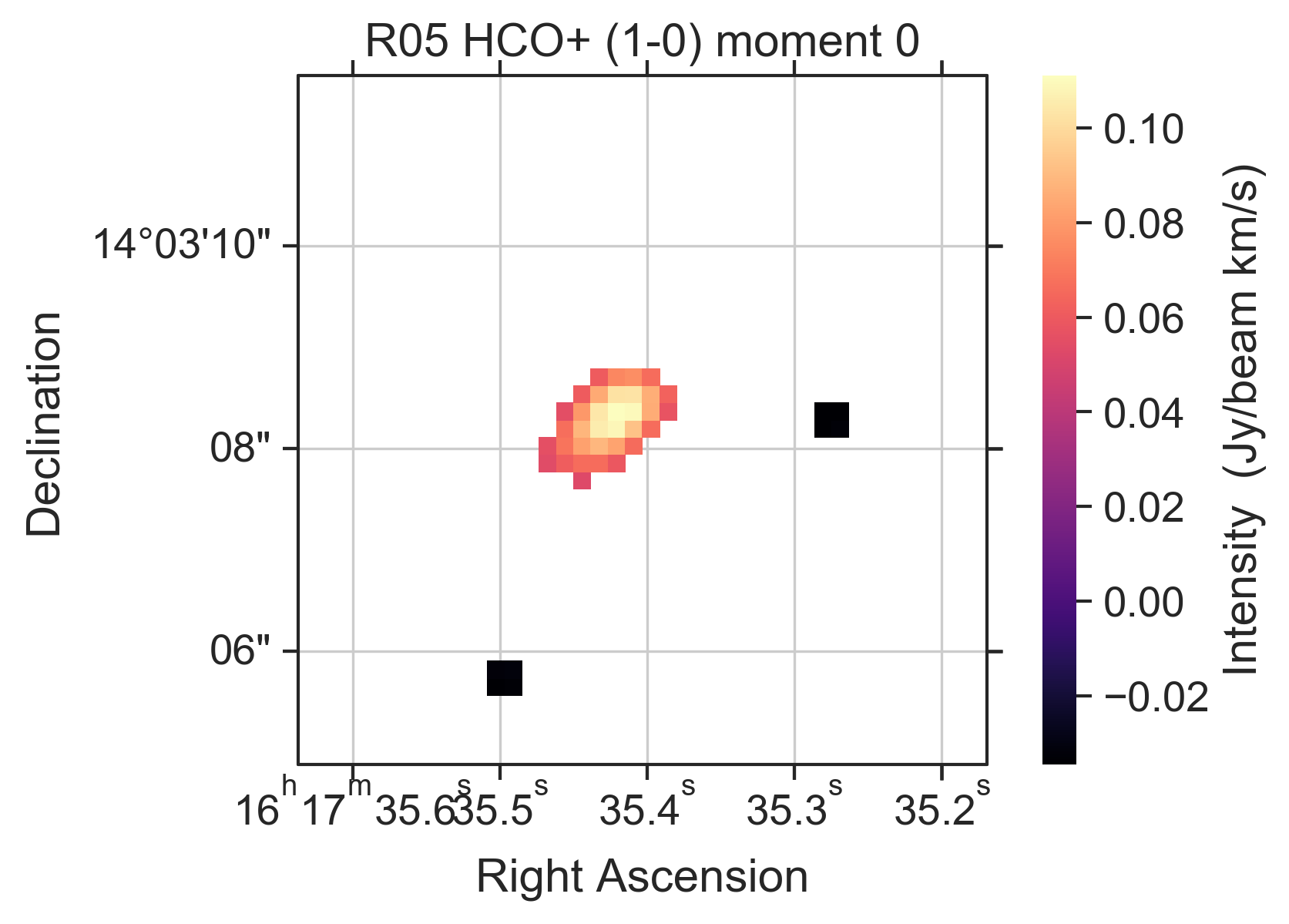}
\includegraphics[width=0.33\textwidth]{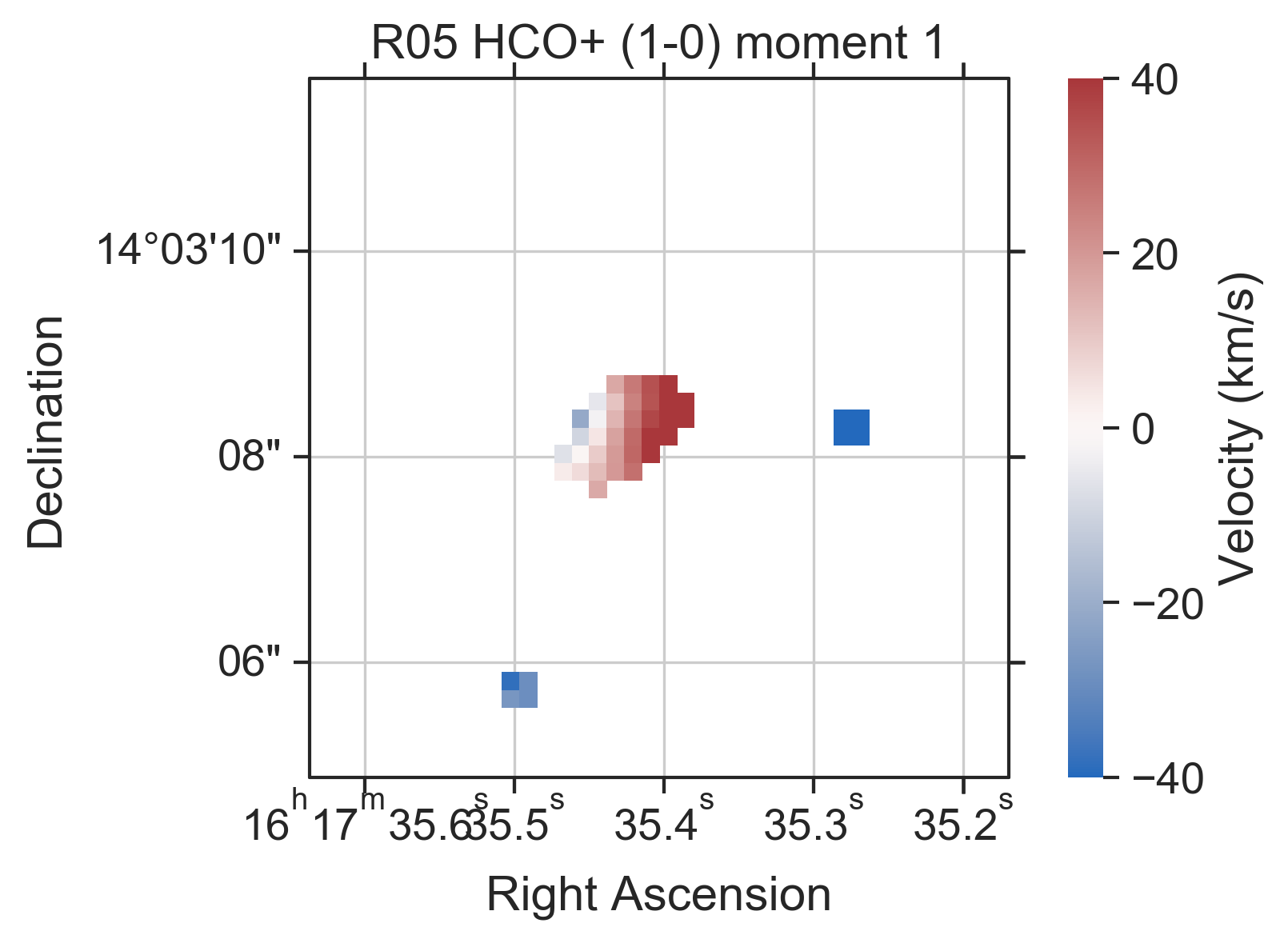}
\includegraphics[width=0.33\textwidth]{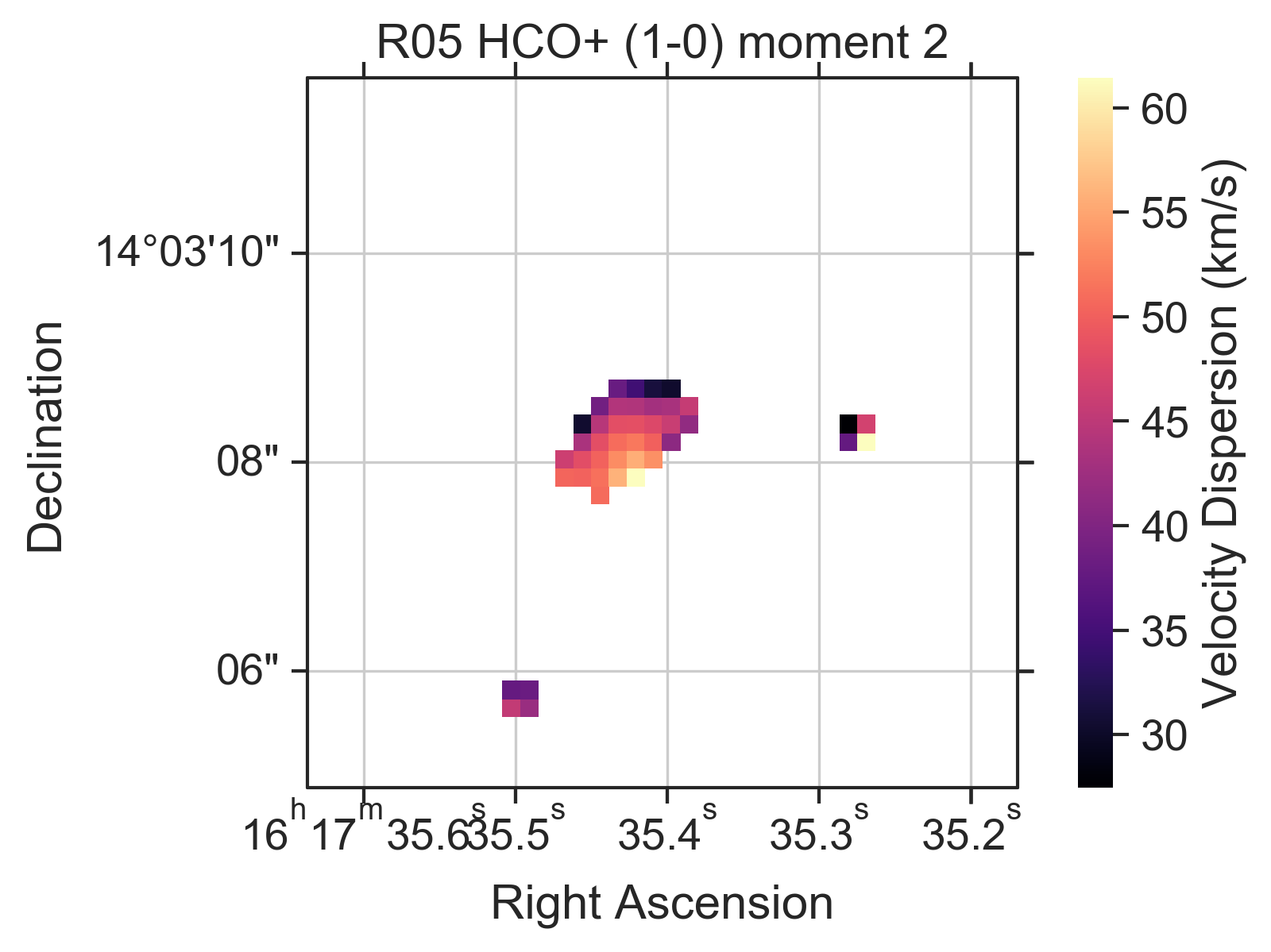}
\caption{Moment 0 (left), moment 1(middle), and velocity dispersion (right) maps for the HCO$^+$ (1--0) emission for the three galaxies with detected emission. }
\label{fig:dense_moments}
\end{figure*}

\begin{figure*}
\includegraphics[width=0.33\textwidth]{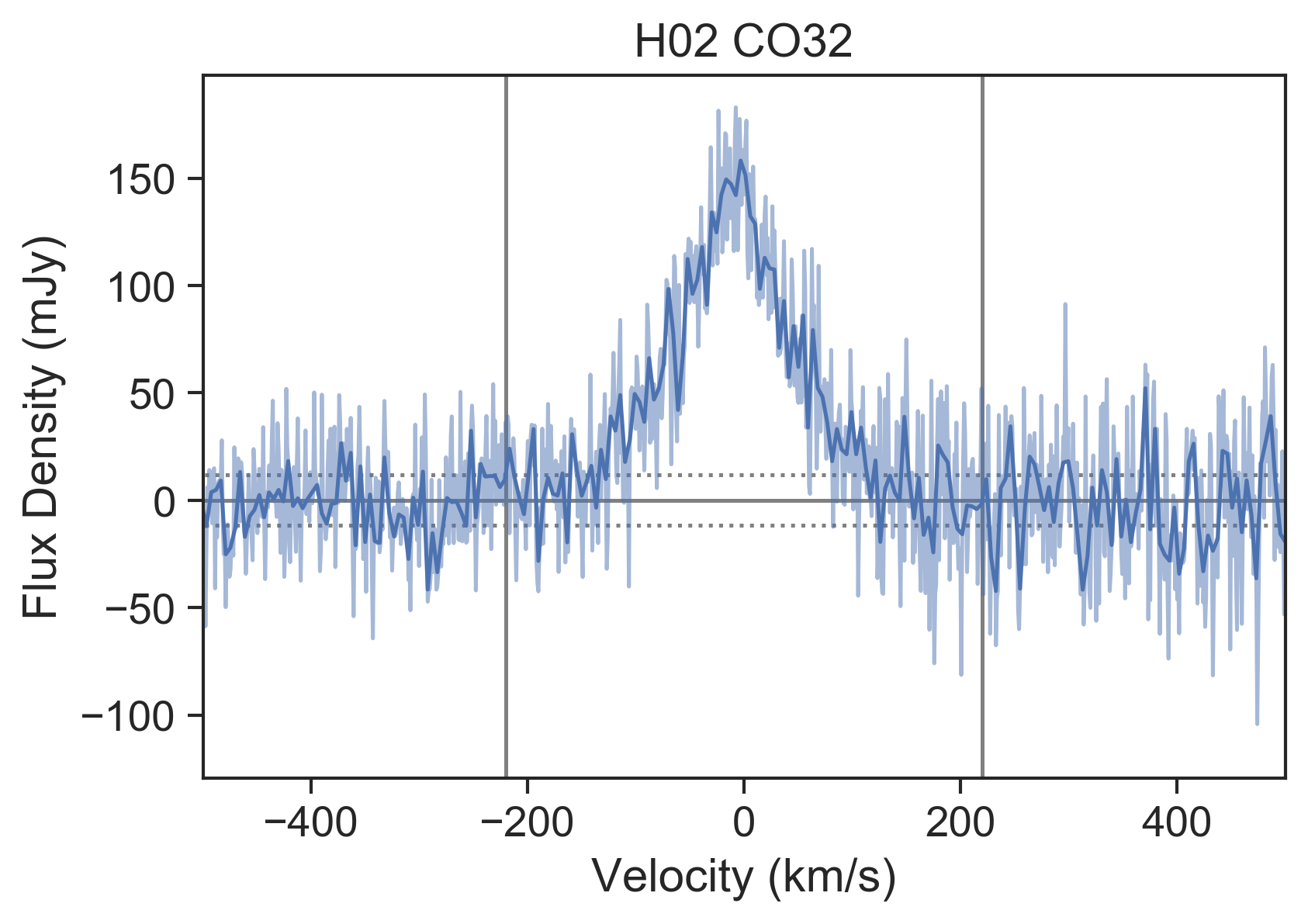}
\includegraphics[width=0.33\textwidth]{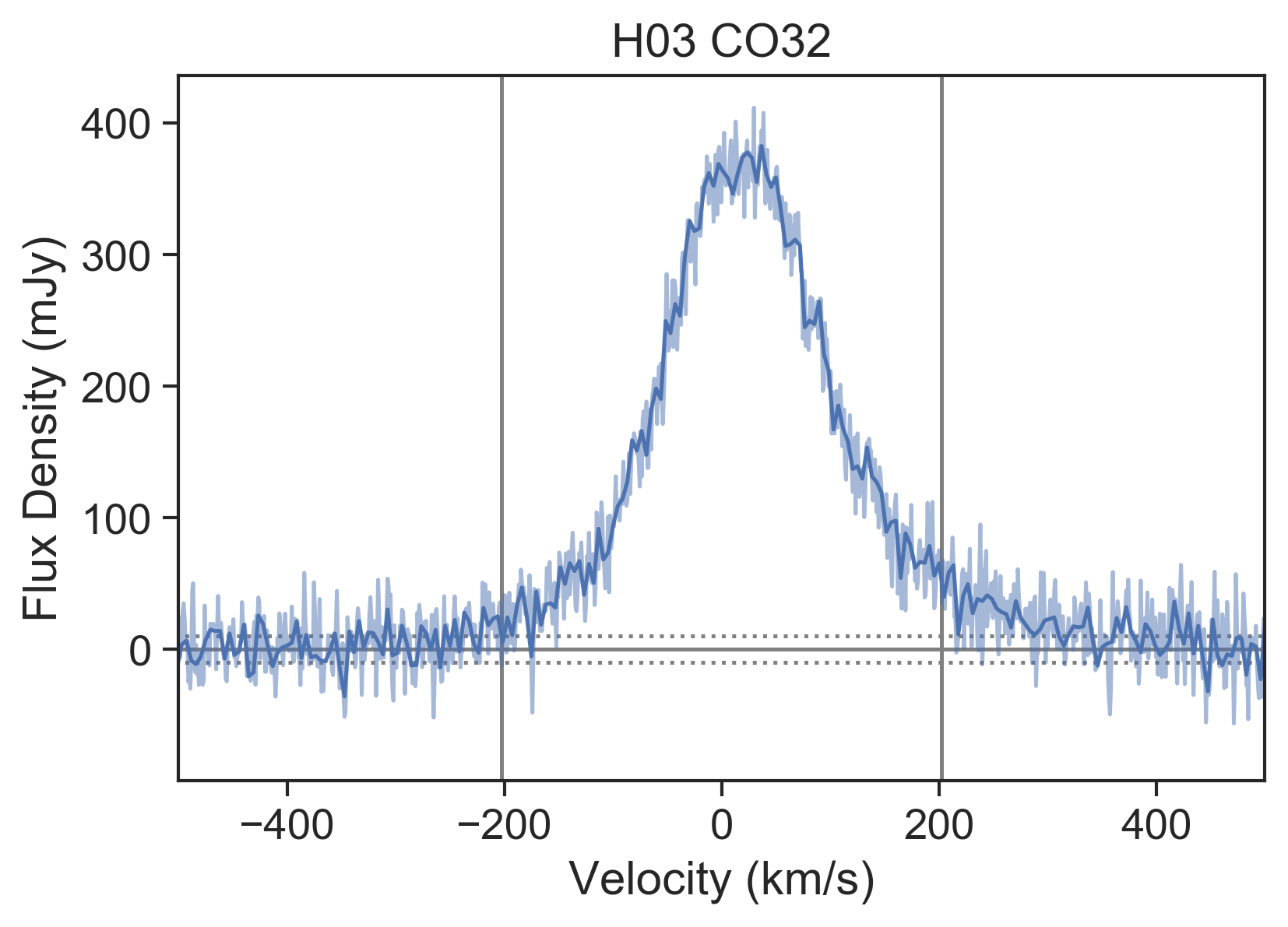}
\includegraphics[width=0.33\textwidth]{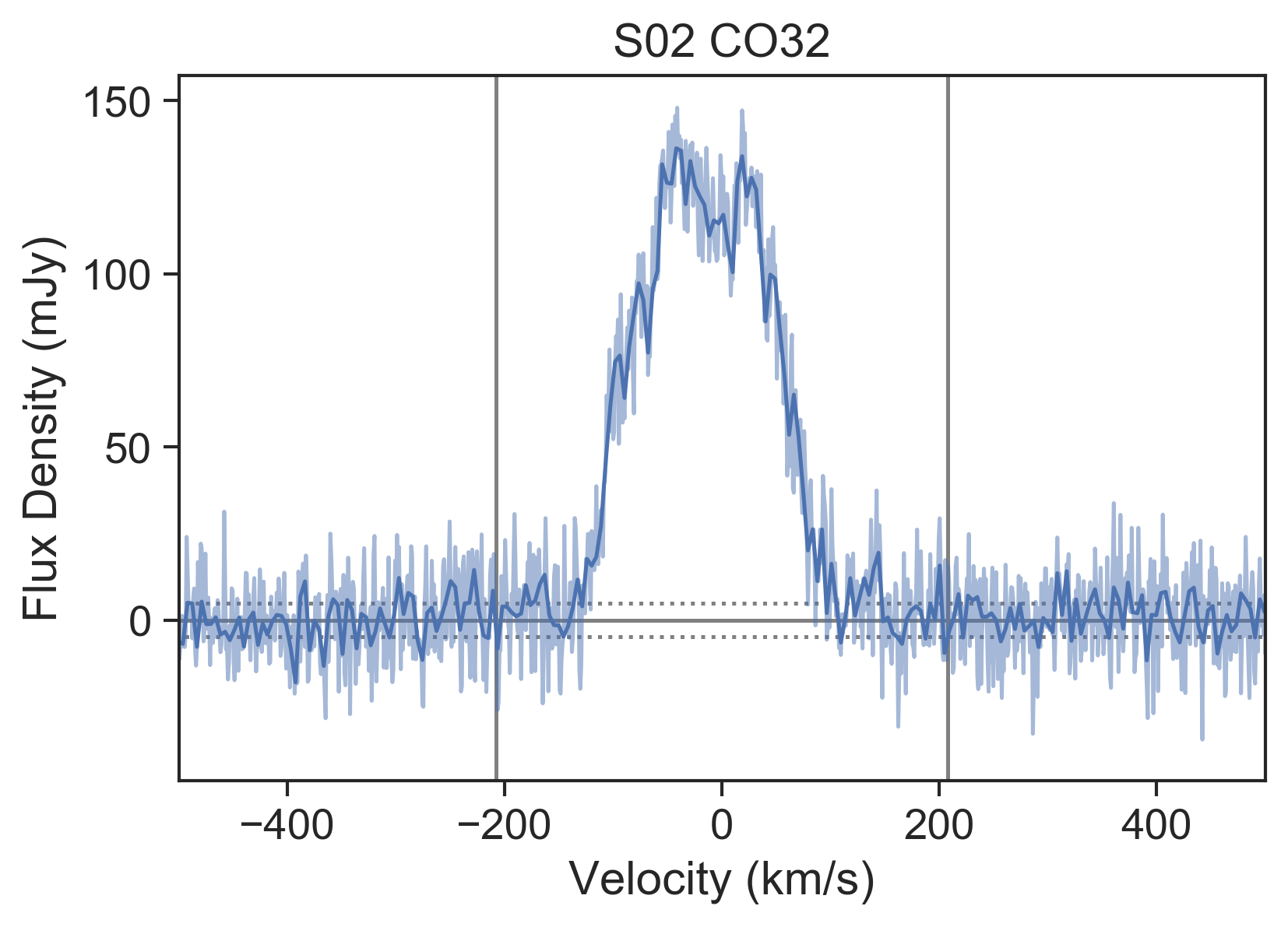}
\caption{Extracted CO (3--2) spectra for each galaxy. For easier visualization, spectra are binned to 5 km/s (dark blue lines). A horizontal center line (grey) and uncertainty bands per 5 km/s channel (dotted grey) are added for comparison. Vertical grey lines represent the integration range for determining the total flux.}
\label{fig:co32_spec}
\end{figure*}

\begin{figure*}
\includegraphics[width=0.33\textwidth]{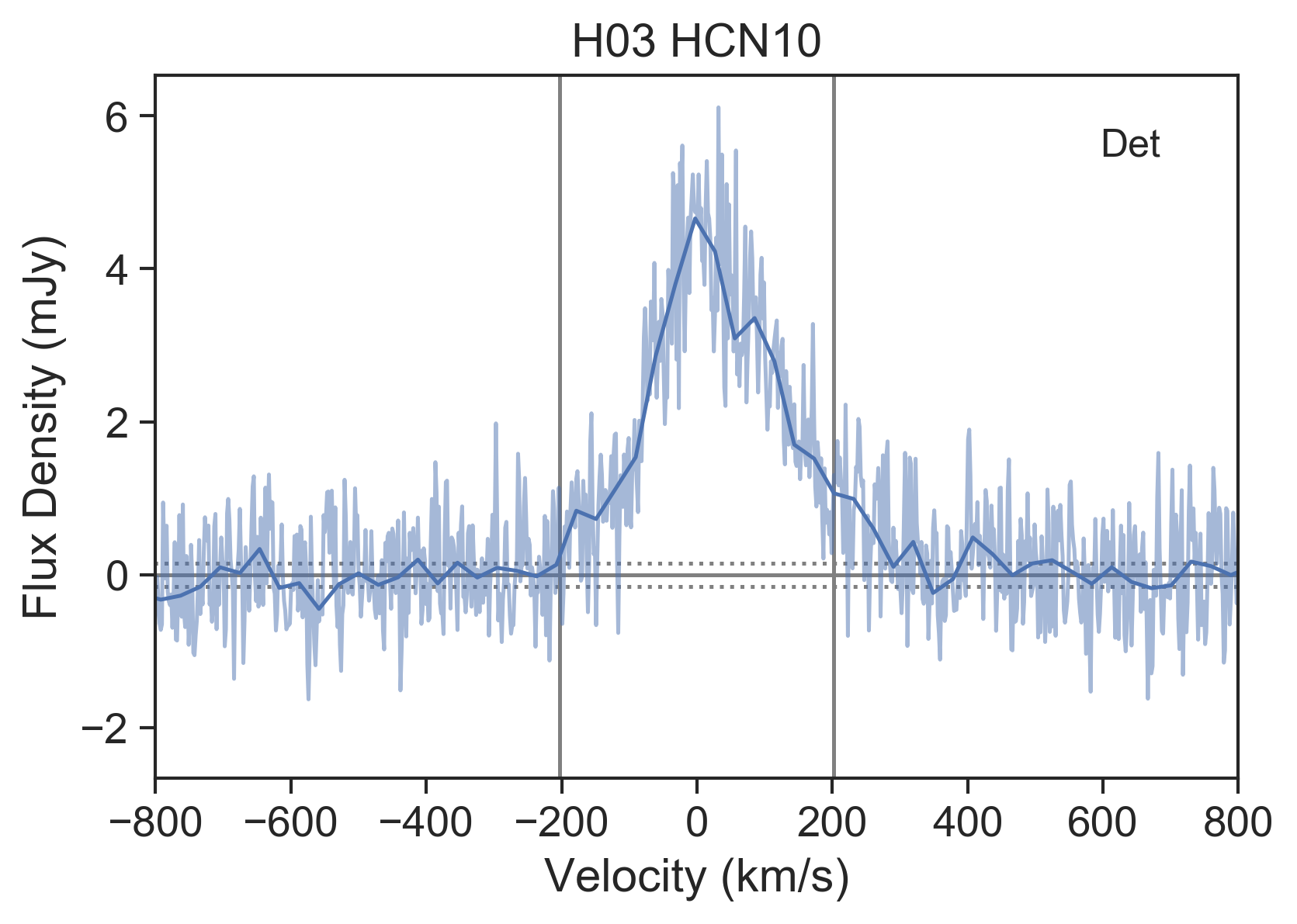}
\includegraphics[width=0.33\textwidth]{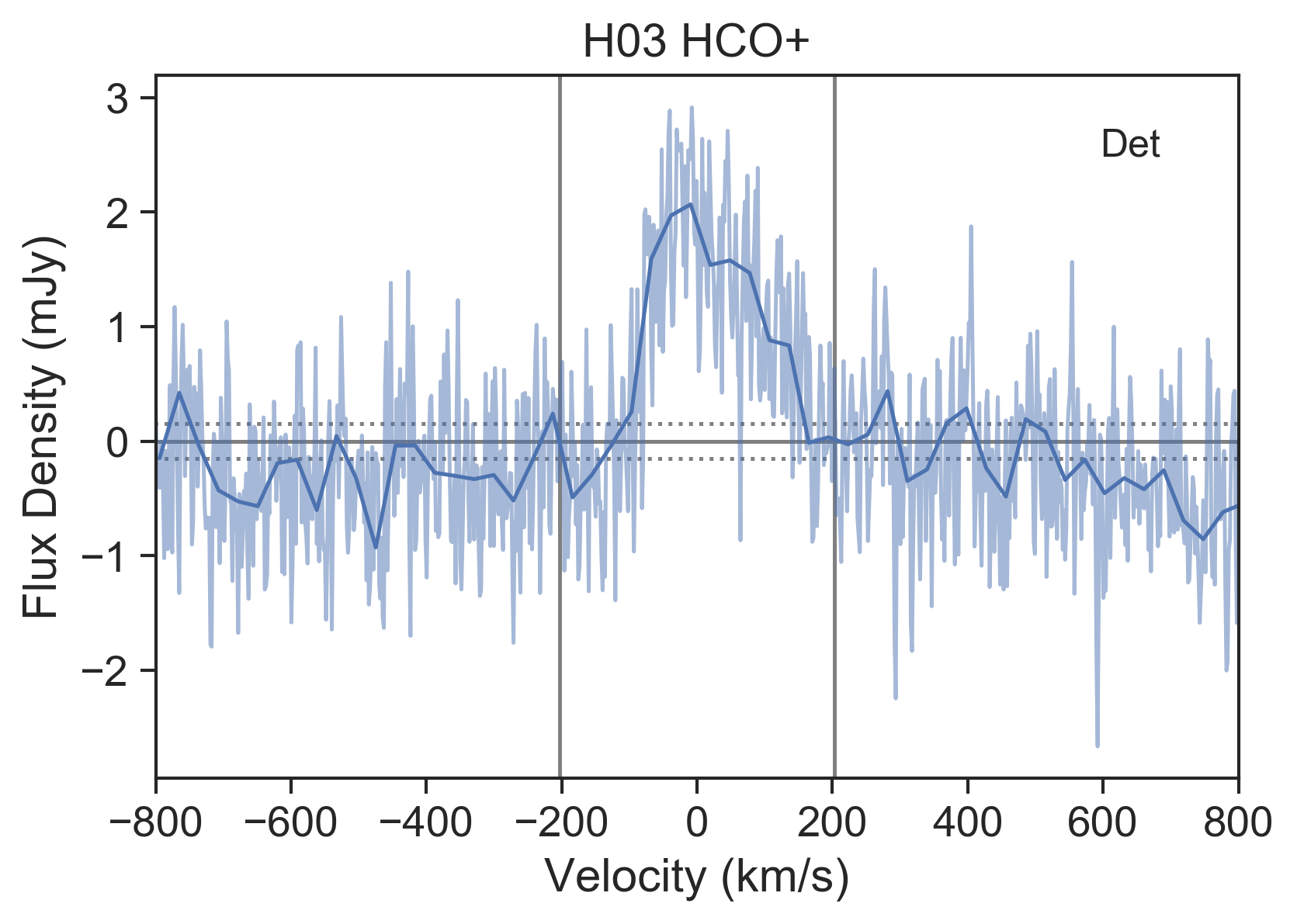}
\includegraphics[width=0.33\textwidth]{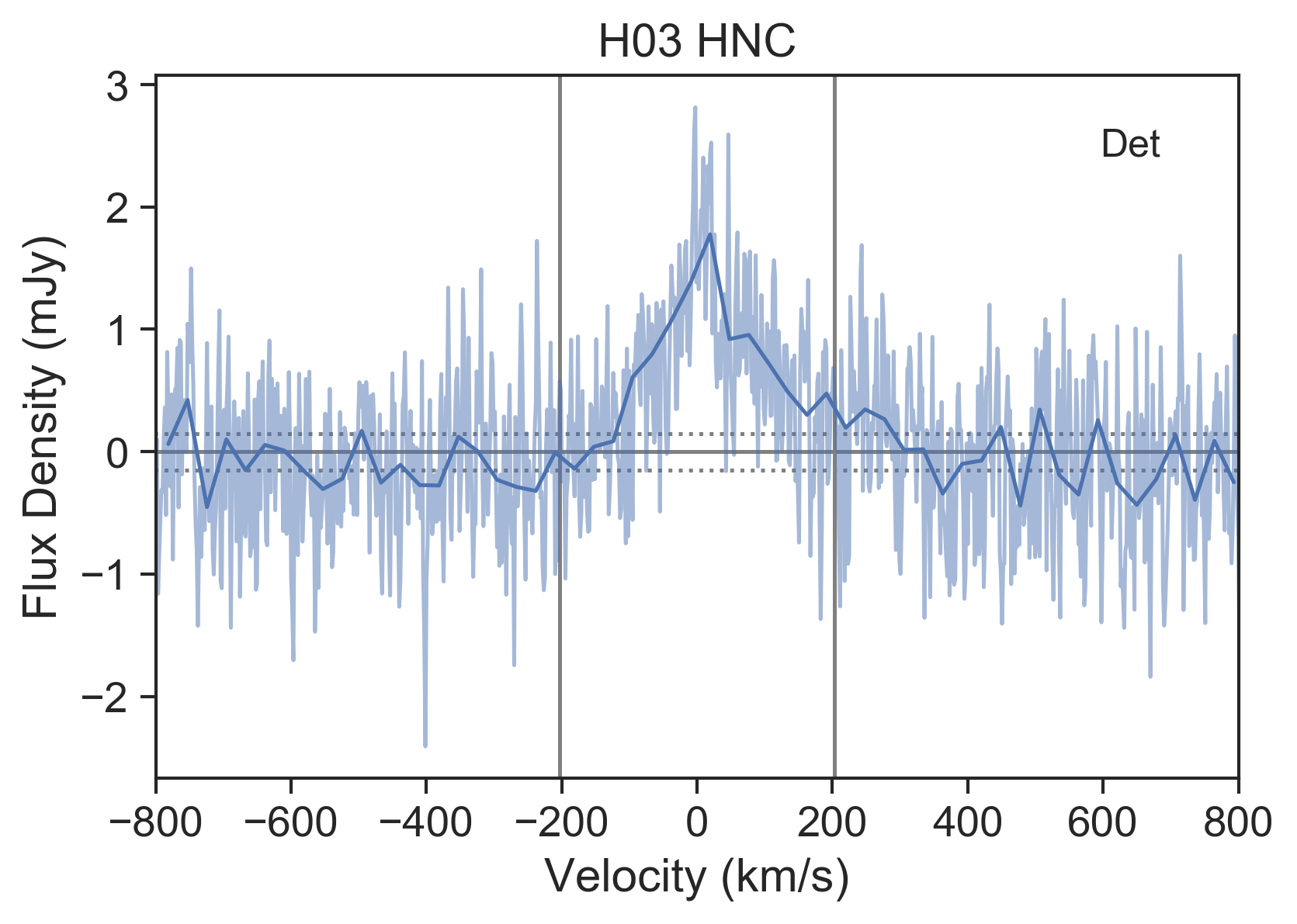}
\includegraphics[width=0.33\textwidth]{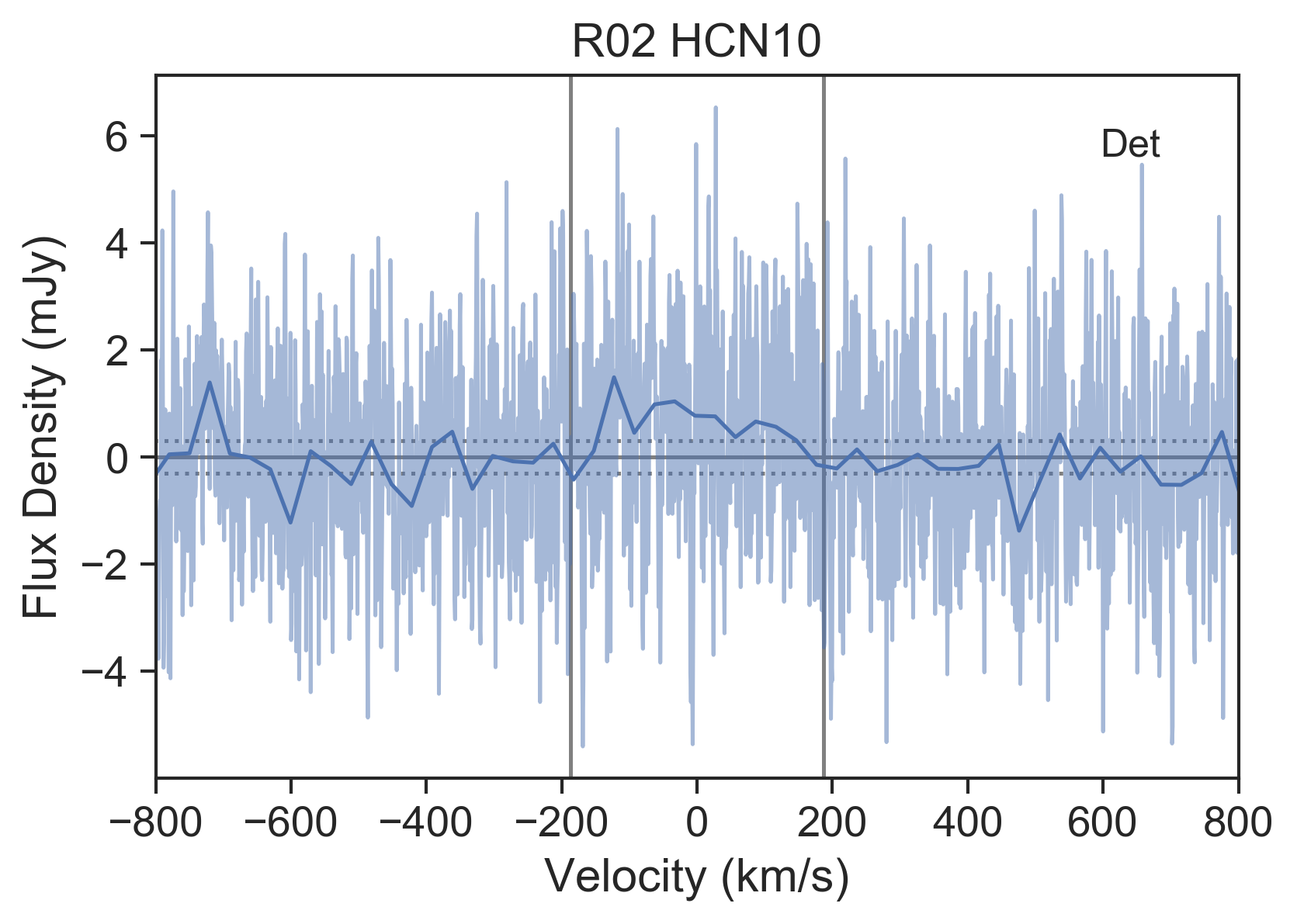}
\includegraphics[width=0.33\textwidth]{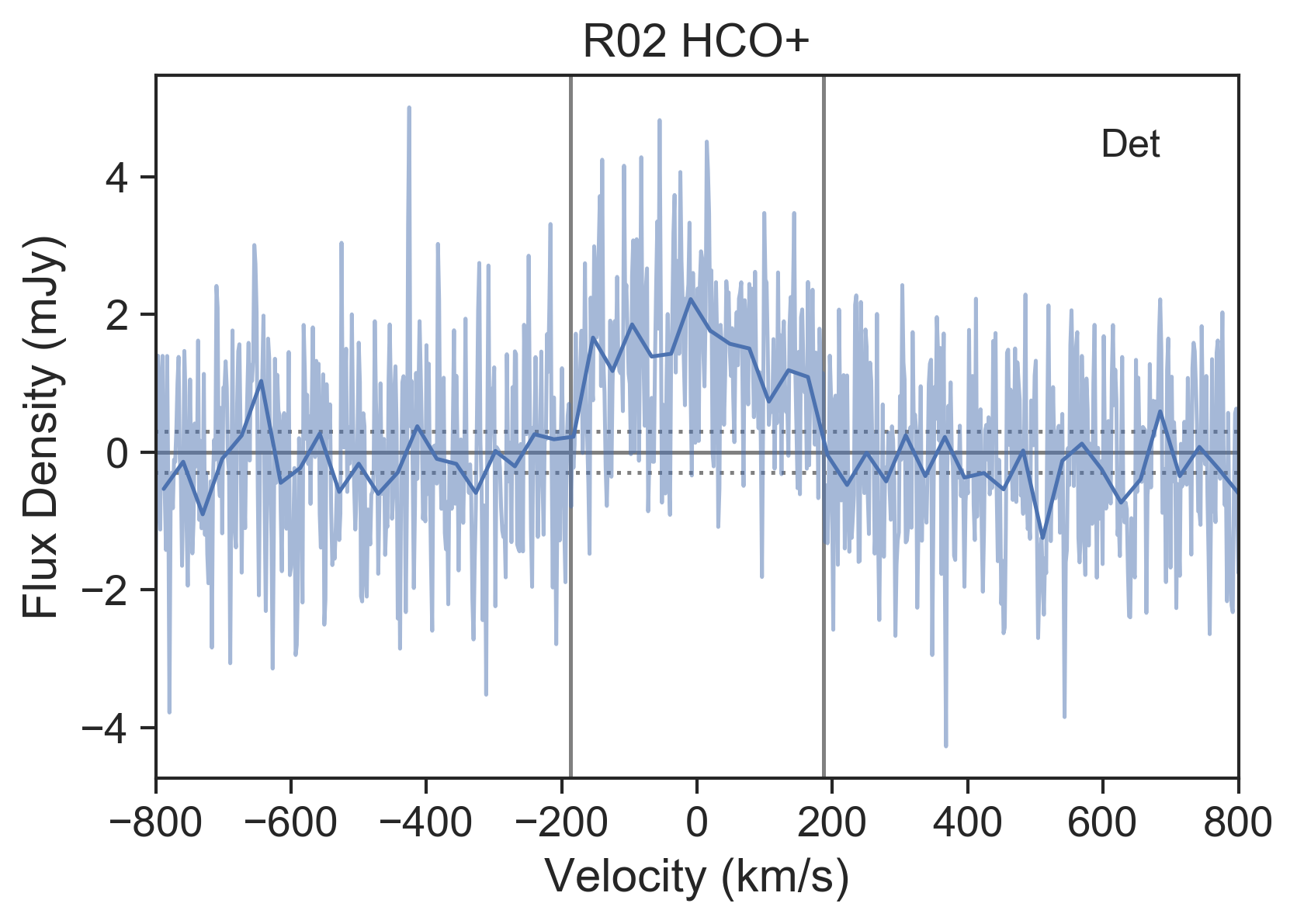}
\includegraphics[width=0.33\textwidth]{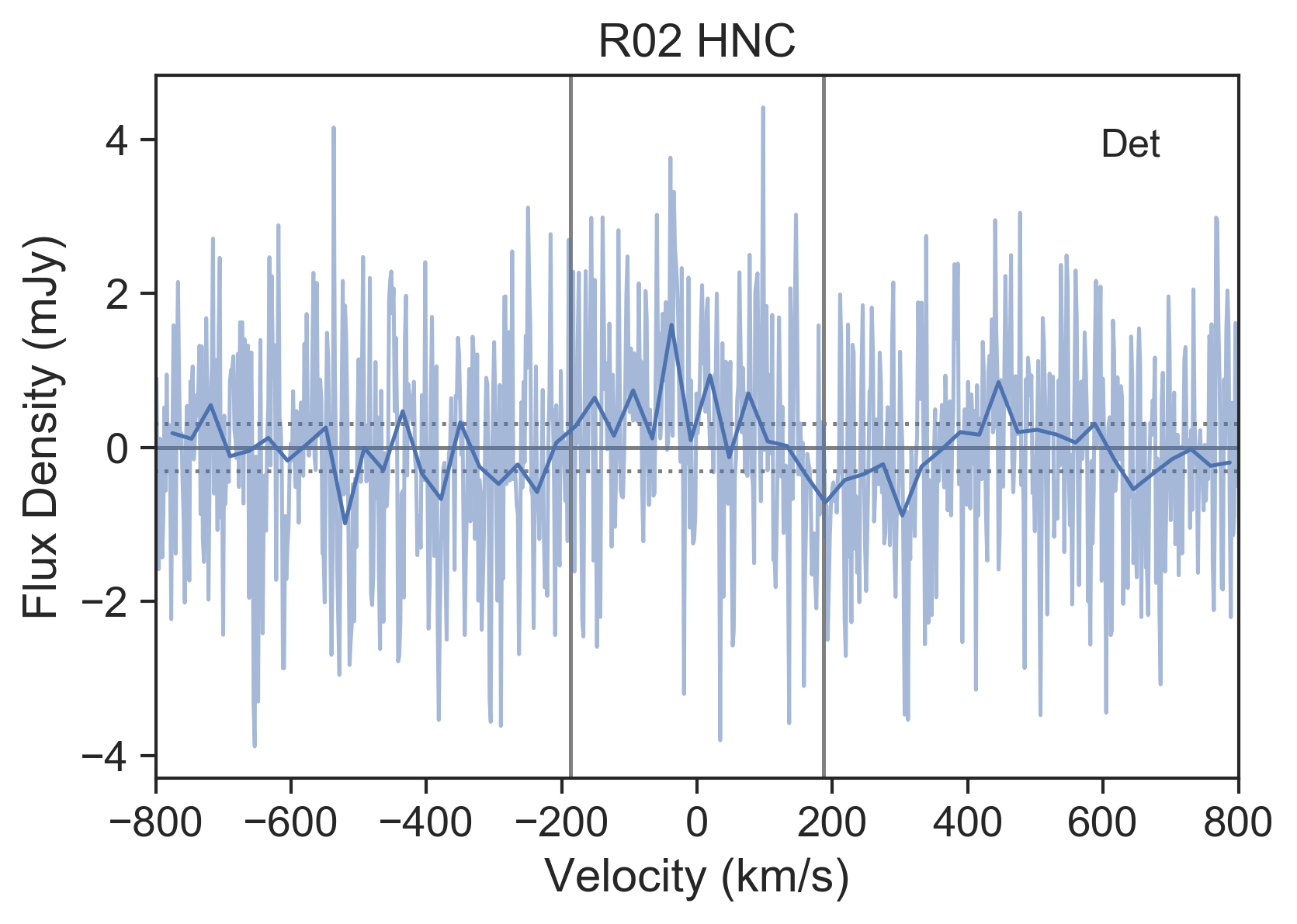}
\includegraphics[width=0.33\textwidth]{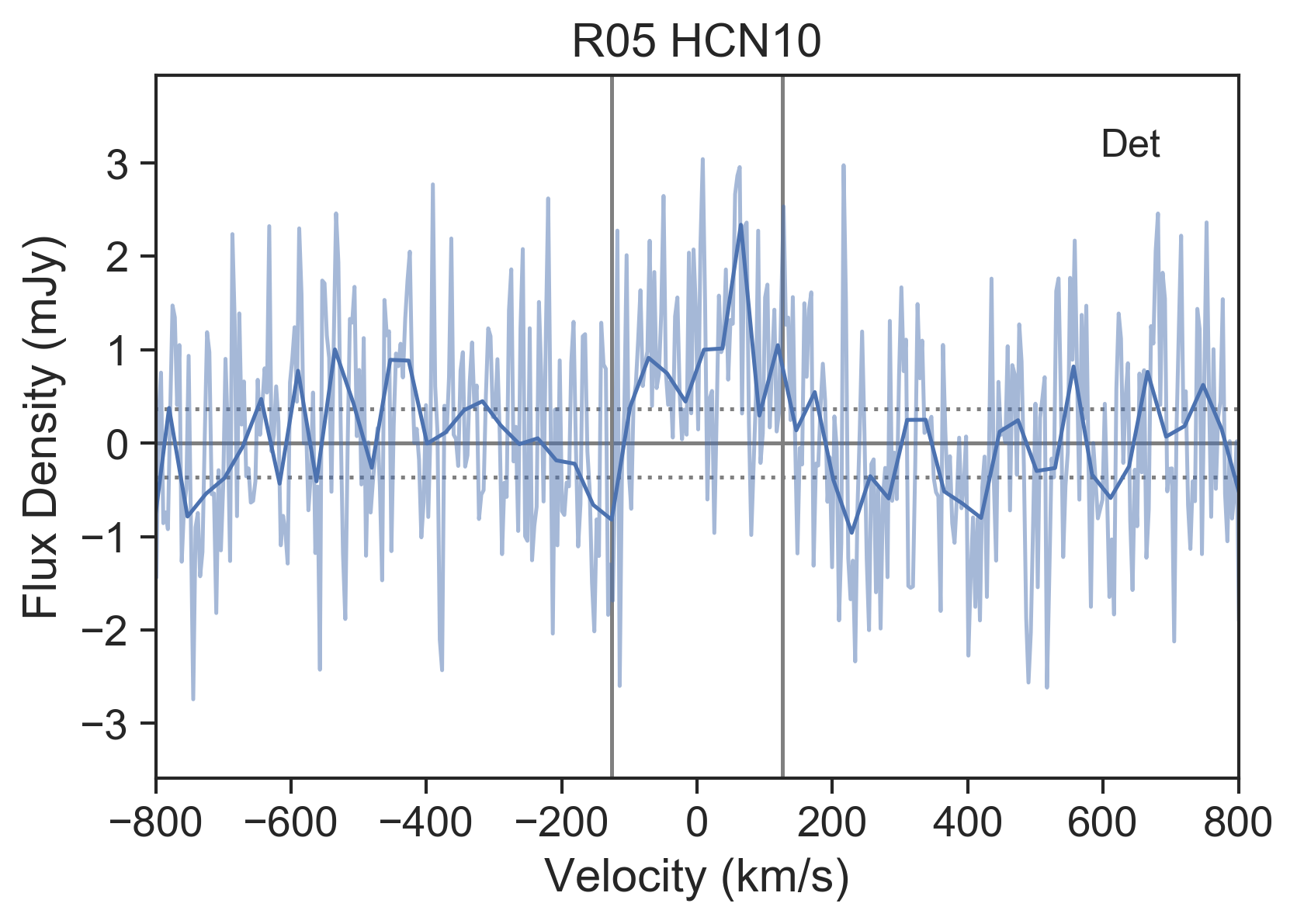}
\includegraphics[width=0.33\textwidth]{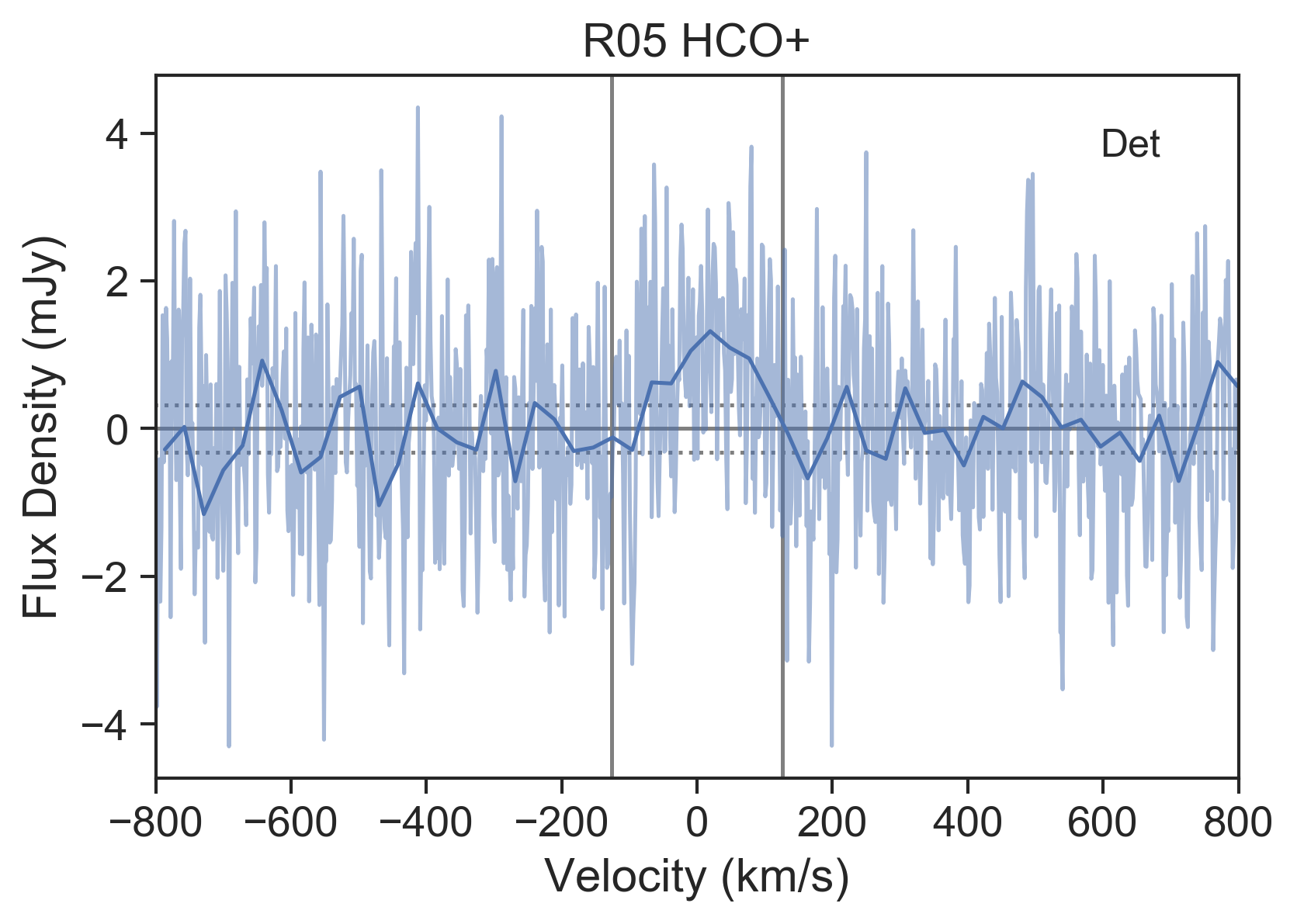}
\includegraphics[width=0.33\textwidth]{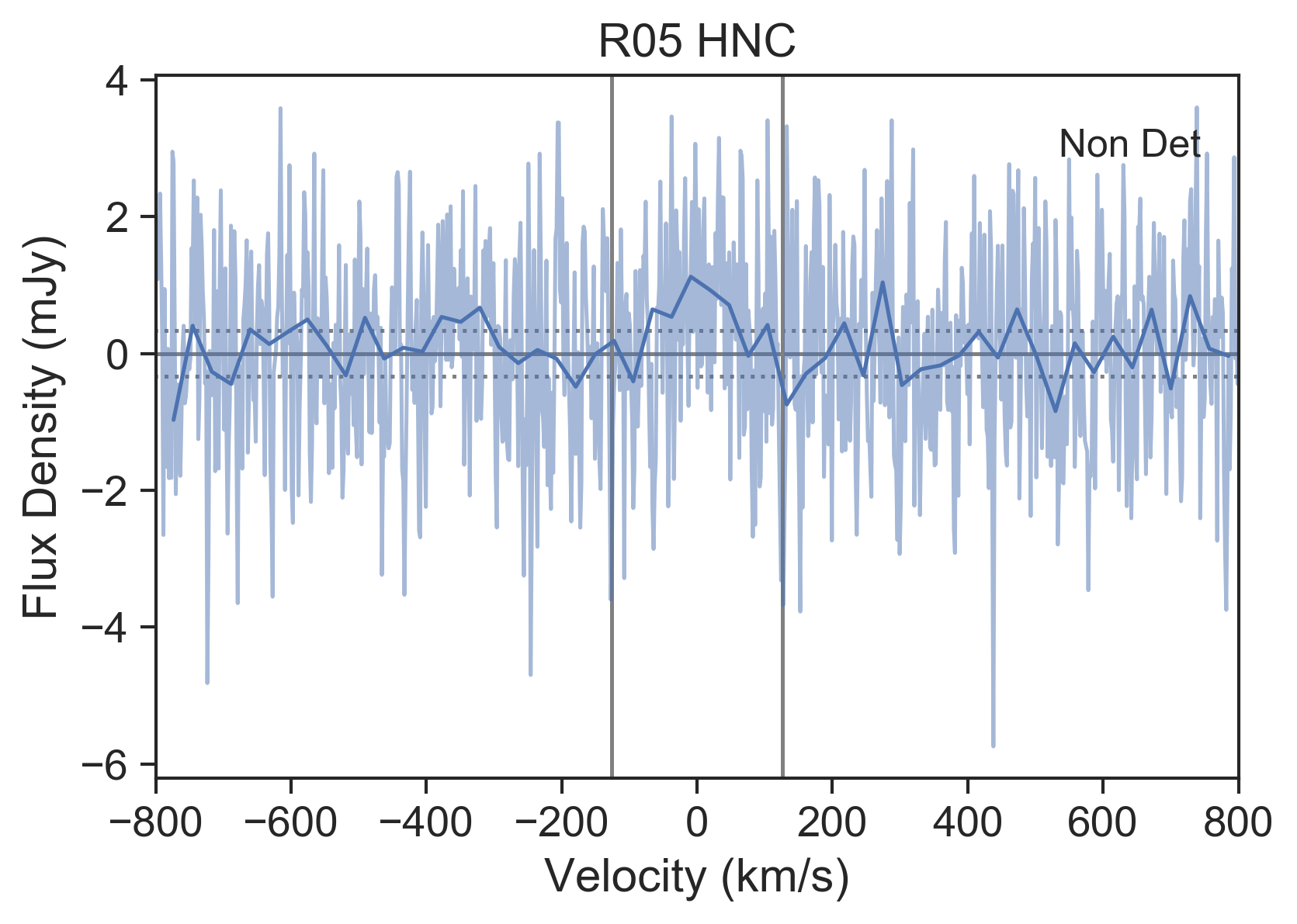}
\includegraphics[width=0.33\textwidth]{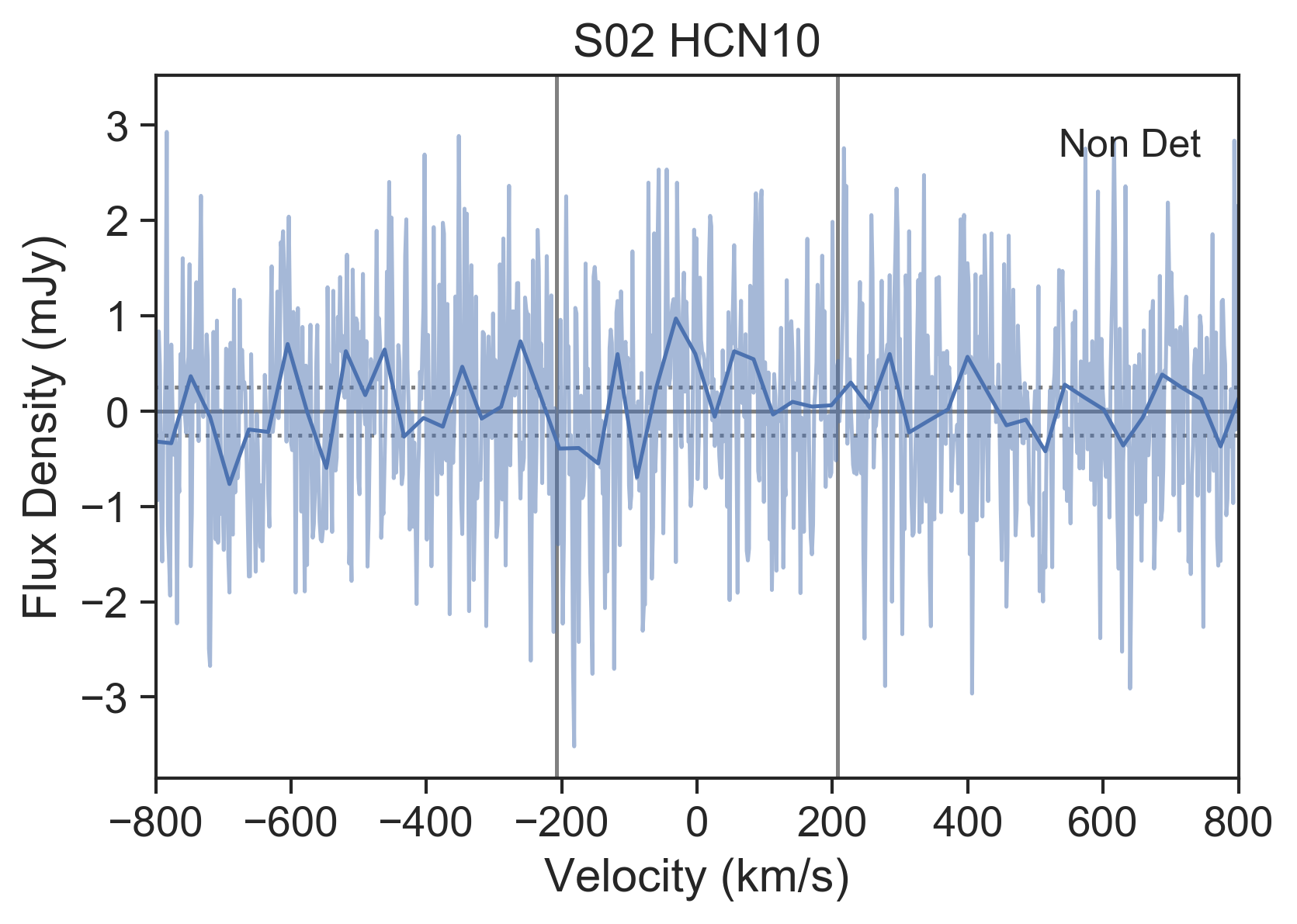}
\includegraphics[width=0.33\textwidth]{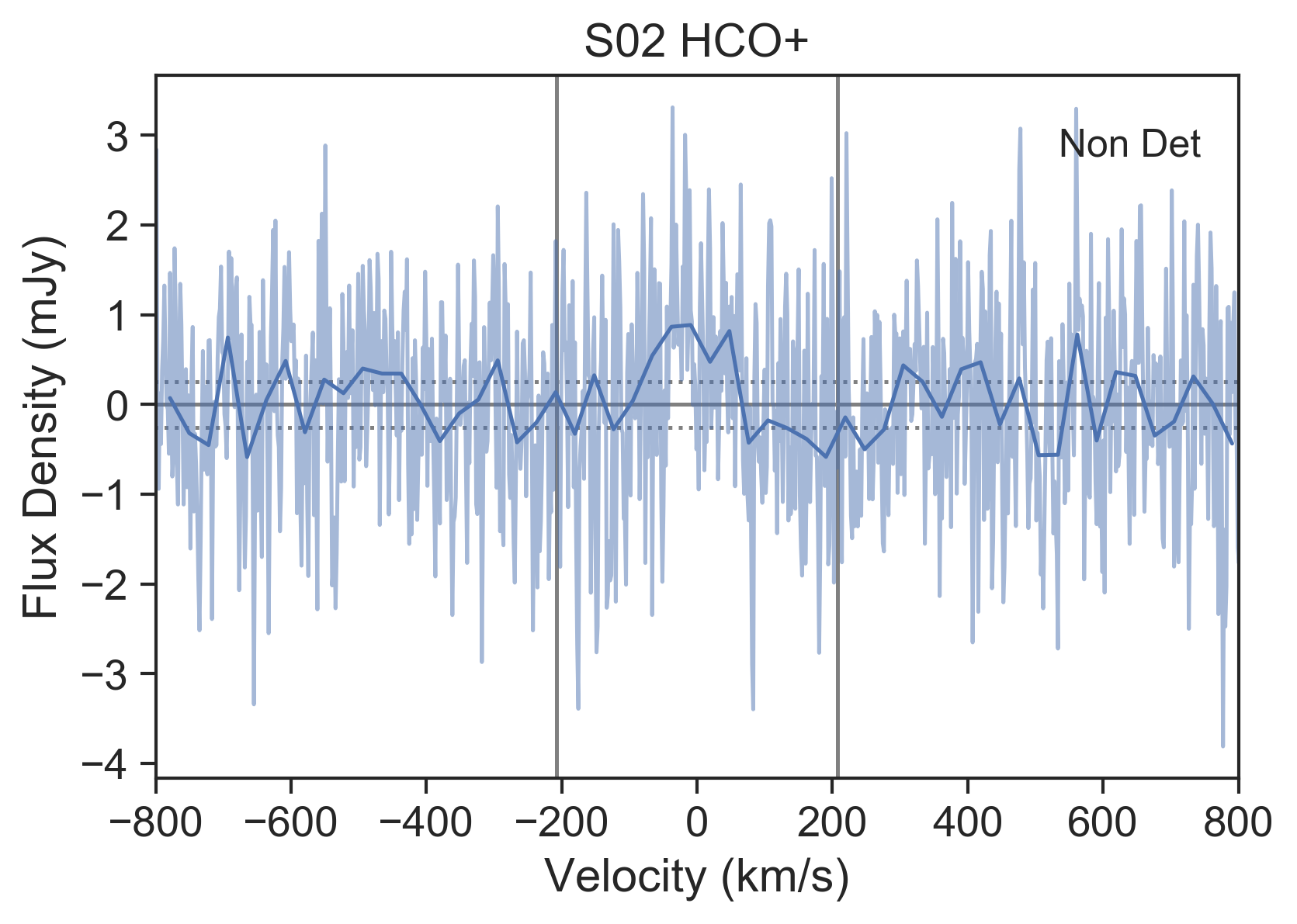}
\includegraphics[width=0.33\textwidth]{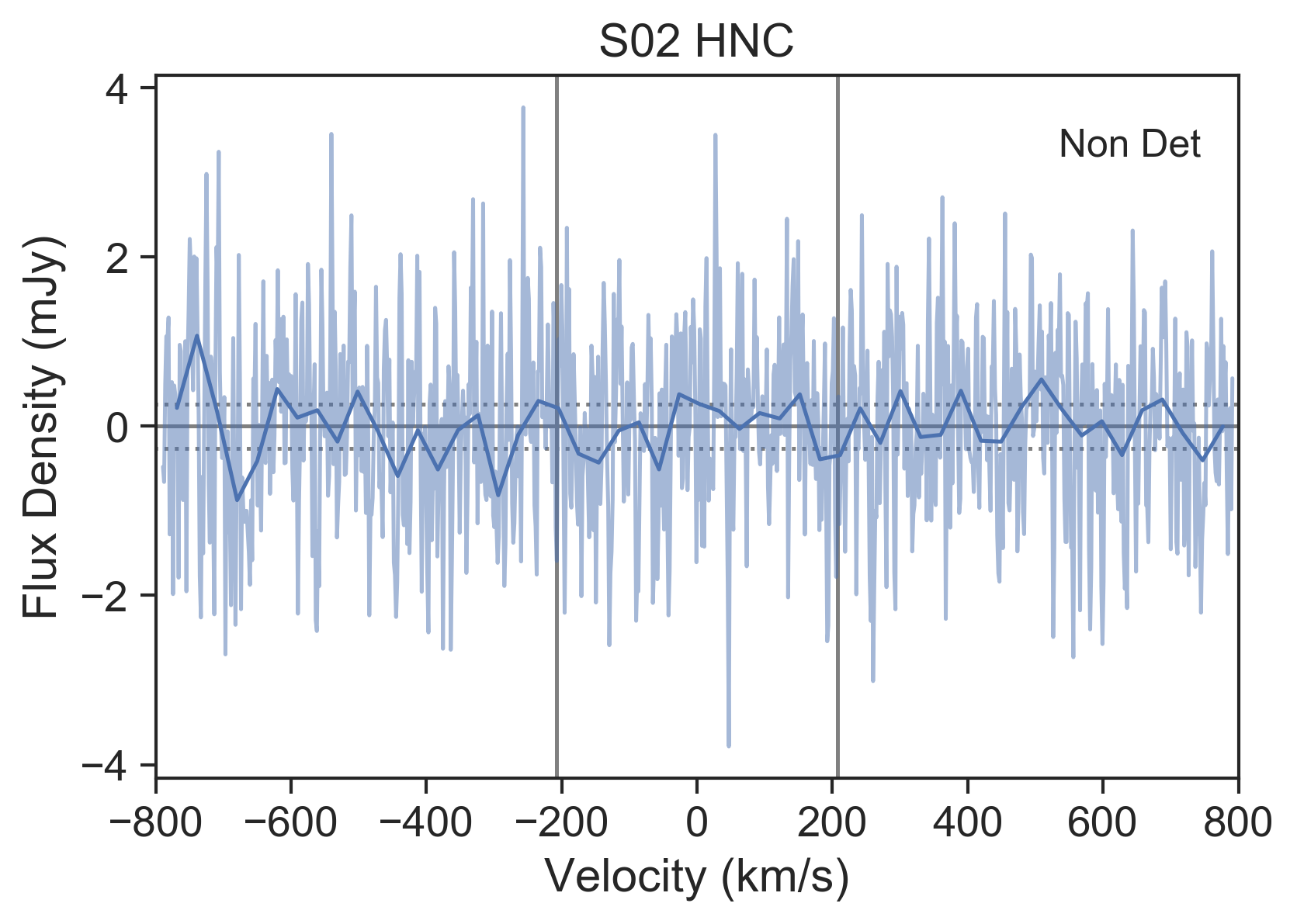}
\caption{Extracted spectra of the three dense gas tracers (HCN (1--0), HCO$^+$(1--0), and HNC (1--0)) for each galaxy. For easier visualization, spectra are binned to 30 km/s (dark blue lines). A horizontal center line (grey) and uncertainty bands per 30 km/s channel (dotted grey) are added for comparison. Vertical grey lines represent the integration range for determining the total flux.}
\label{fig:dense_spec}
\end{figure*}

Moment maps for the both the CO (3--2) and HCN (1--0), HCO$^+$ (1--0), and HNC (1--0) datasets are shown in Figures \ref{fig:moments} and \ref{fig:dense_moments}. Extracted spectra are shown in Figures \ref{fig:co32_spec} and \ref{fig:dense_spec}. More information can be found in \S\ref{sec:almaobs}.

\section{Infrared Star Formation Rates}
\label{sec:appendix_ir}

\begin{figure*}
\includegraphics[width=0.5\textwidth]{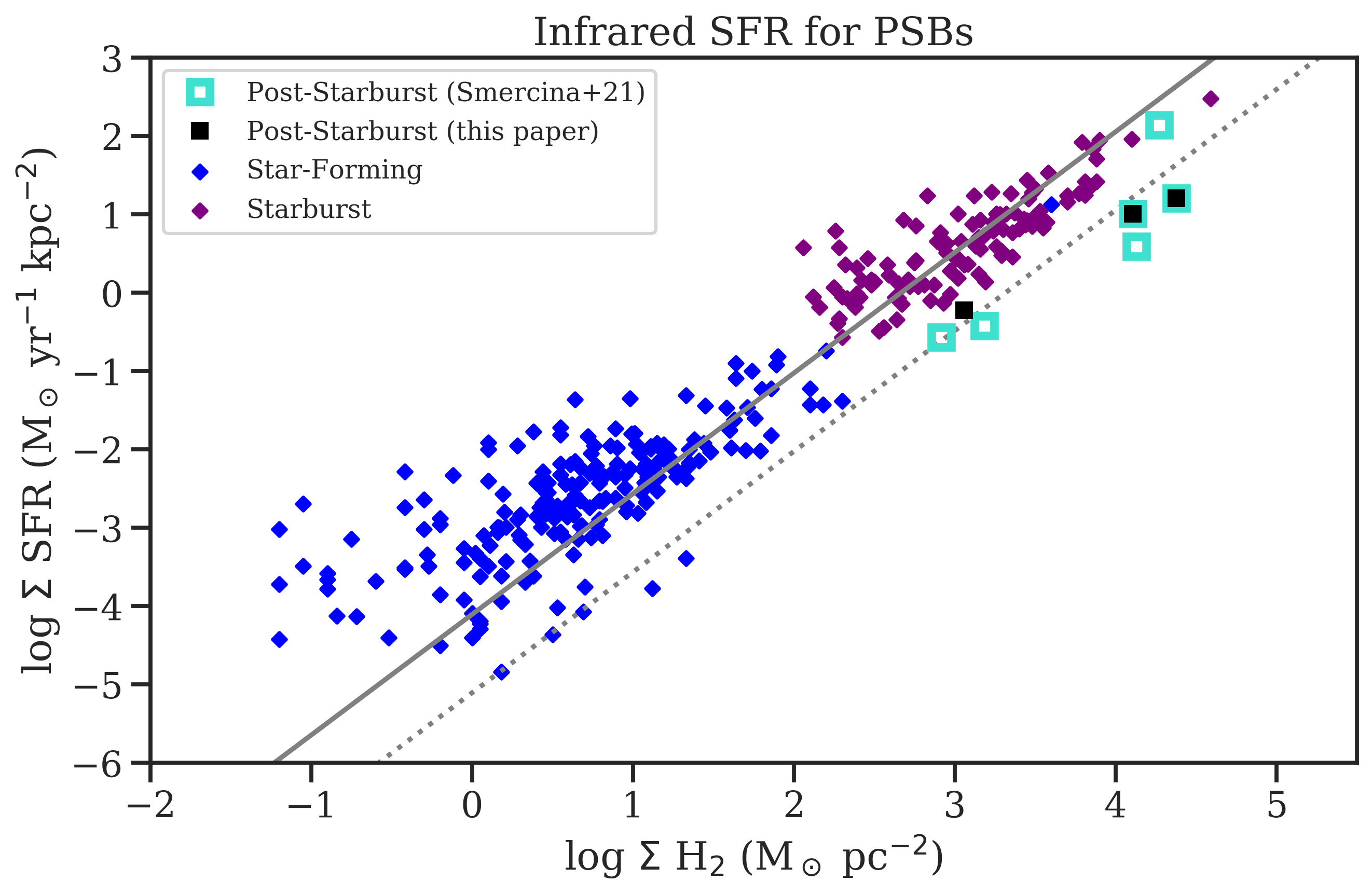}
\includegraphics[width=0.5\textwidth]{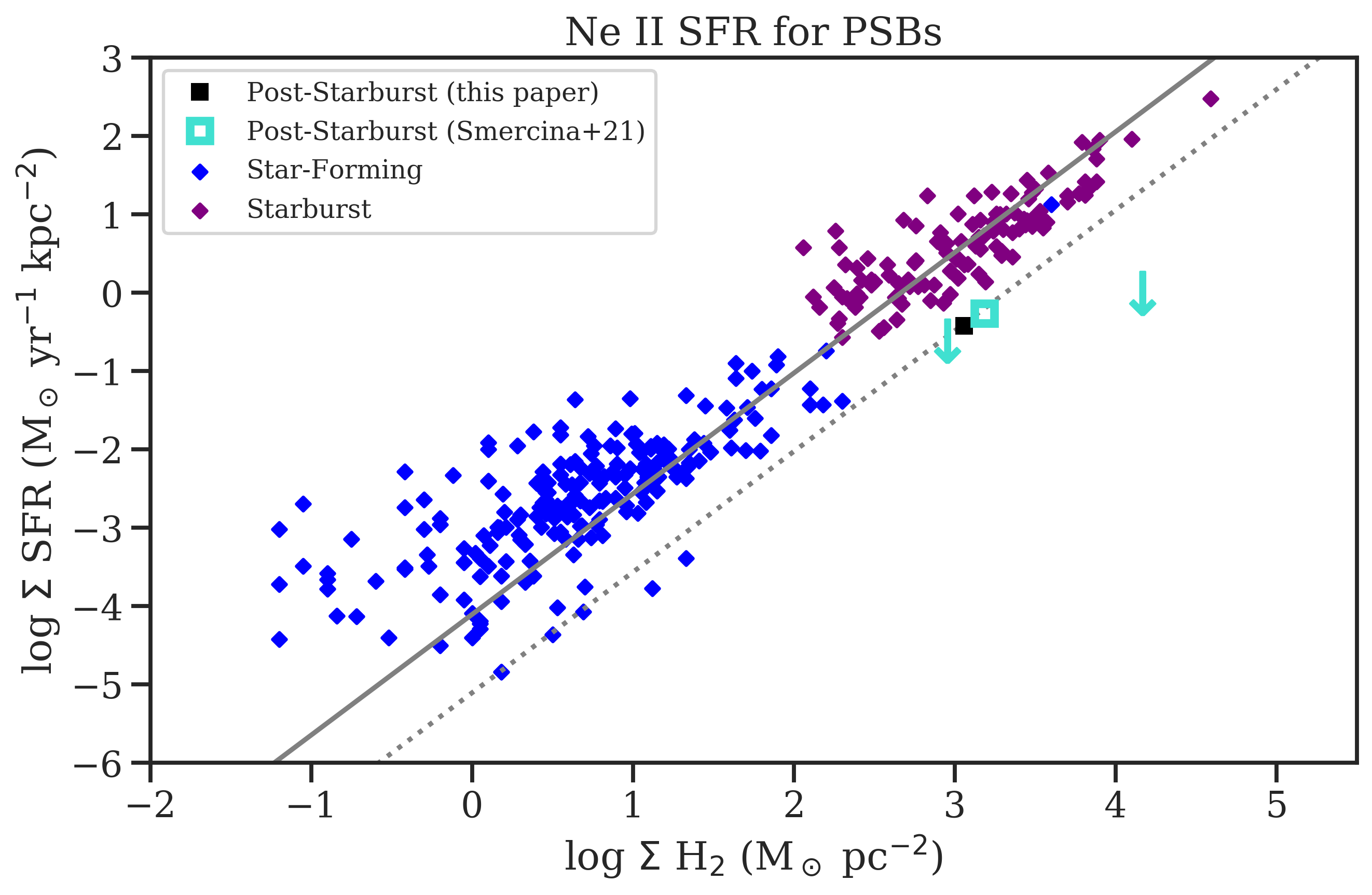}
\caption{Same as Figure \ref{fig:ks}, but with IR-SFR indicators used instead for the post-starburst galaxy sample (left: TIR, right: [Ne II]). Molecular gas surface density vs. SFR surface density for post-starburst galaxies from this work with CO (3--2) sizes and from \citet{Smercina2022} with CO (2--1) sizes, as well as comparison samples of star-forming galaxies from \citet{delosReyes2019} and starbursting galaxies from \citet{Kennicutt2021}. The best-fit relation from \citet{Kennicutt2021} for the total gas density vs. star formation density is plotted in grey, with a dotted line indicating a factor of $10\times$ below the relation. The post-starburst galaxies have very high molecular gas surface densities, yet they lie below the comparison galaxies, with low star formation rate surface densities for their molecular gas surface densities. Our qualitative conclusions here do not depend on the SFR tracer used.}
\label{fig:ks_ir}
\end{figure*}

\begin{figure*}
\includegraphics[width=\textwidth]{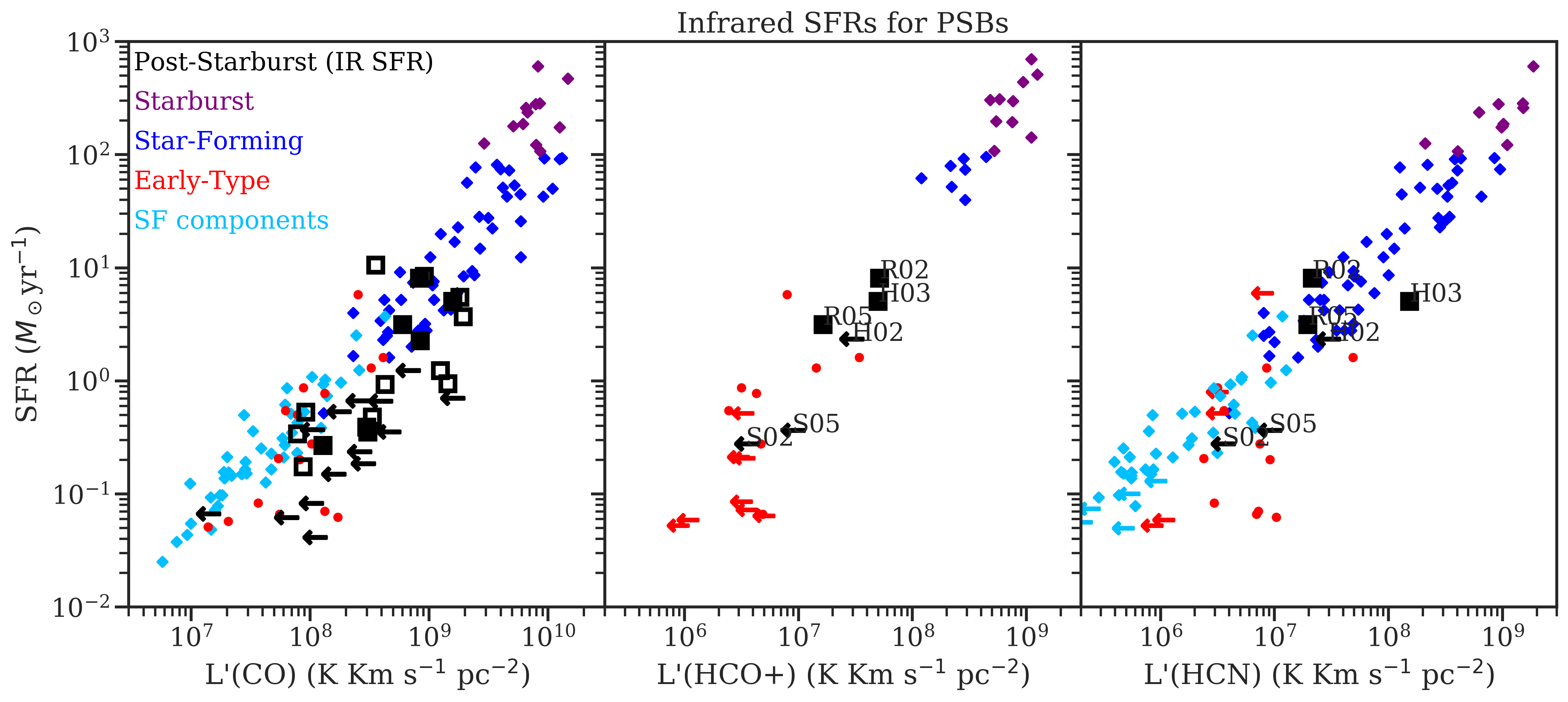}
\caption{Same as Figure \ref{fig:gas_sfr}, but with IR-SFR indicators used instead for the post-starburst galaxy sample. (Left:) \lpco vs. SFR. Early type galaxies \citep{Crocker2012}, starburst and star-forming galaxies \citep{Gao2004}, and star forming galaxy components \citep{Usero2015a} are correlated with low scatter, but post-starburst galaxies \citep{French2015} have low SFRs for their CO luminosities. Black squares indicate post-starburst detections and arrows indicate $3\sigma$ upper limits. Filled squares indicate the ALMA targets considered here (including two galaxies from the \citet{Rowlands2015} sample).  (Middle:) \lpco vs. SFR for the same samples of galaxies. (Right: )\lhcn vs. SFR for the same samples of galaxies. The post-starburst galaxies are more consistent with the comparison samples in their dense gas - star formation relations. The post-starburst galaxies have dense molecular gas properties consistent with either early type galaxies or lying between the star-forming and early type samples.  Our qualitative conclusions here do not depend on the SFR tracer used.
}
\label{fig:gas_sfr_ir}
\end{figure*}

\begin{figure*}
\includegraphics[width=\textwidth]{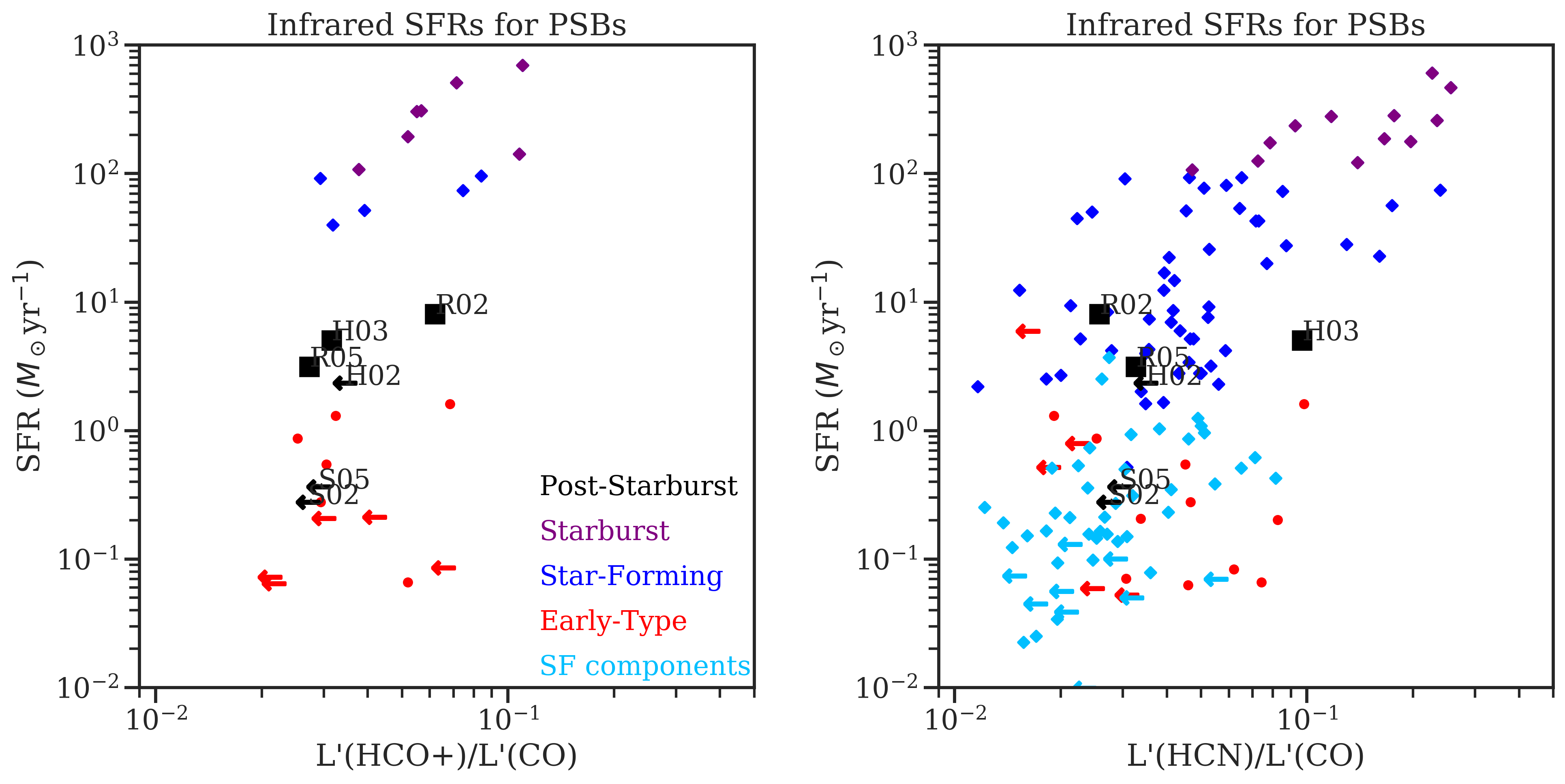}
\caption{Same as Figure \ref{fig:dense_gas_co_ratios}, but with IR-SFR indicators used instead for the post-starburst galaxy sample. Plot colors are the same as \ref{fig:gas_sfr_ir}.  (Left:) Dense gas luminosity ratio  \lhco/\lpco for the same samples as Figure \ref{fig:gas_sfr}. Post-starburst galaxies have low ratios of dense molecular gas to total molecular gas compared to other types of galaxies, although this may evolve rapidly with time during the early post-starburst phase (Figure \ref{fig:age}). (Right:) Dense gas luminosity ratio \lhcn/\lpco for the same samples.  The post-starburst galaxy with high HCN/CO (H03) may have HCN luminosity increased via mechanical heating or cosmic ray heating, as it has a very high HCN/HCO$+$ ratio (see discussion in \S\ref{sec:densegasstate}), and the HCN/CO ratio may overestimate the dense molecular gas fraction.  Our qualitative conclusions here do not depend on the SFR tracer used.}
\label{fig:dense_gas_co_ratios_ir}
\end{figure*}

In Figures \ref{fig:ks_ir}, \ref{fig:gas_sfr_ir} and \ref{fig:dense_gas_co_ratios_ir}, we consider the impact of using the TIR luminosity instead of H$\alpha$ to trace the SFR in the post-starburst sample. A full comparison of the SFR tracers for this sample can be found in \S\ref{sec:sfr}. Our qualitative conclusions do not change, given this sample. We observe the post-starburst galaxies to lie offset in the CO-traced gas vs. SFR plot, while they lie consistent with the comparison samples in the HCO$^+$ vs. SFR plot. Even when using the TIR luminosity as a SFR tracer, the offset observed in the left panel of Figure \ref{fig:gas_sfr_ir} is significant. Even if all of the CO upper limits placed the CO non-detected post-starbursts above the relation, with high star formation efficiencies, the chances of finding as many post-starbursts with low SFE $<10^{9}$ yr$^{-1}$ is low. Using the star-forming, starbursting, and early type galaxies to define the range of expected SFE and scatter, we use a Monte Carlo analysis to determine that we would find the large fraction of low SFE post-starbursts only 1.7\% of the time.

\end{document}